\documentclass[]{mn2e}
\pdfoutput=1
\usepackage{natbib}
\usepackage{times}
\usepackage{epsfig}
\usepackage{amssymb}
\usepackage{amsmath}
\usepackage{setspace}
\usepackage{threeparttable}
\usepackage{rotating}
\usepackage{longtable}
\usepackage{colortbl}
\usepackage{subfigure}

\bibpunct[ ]{(}{)}{;}{a}{}{,}
\def\lsim{\mathrel{\rlap{\lower4pt\hbox{\hskip1pt$\sim$}}
    \raise1pt\hbox{$<$}}}                
\def\gsim{\mathrel{\rlap{\lower4pt\hbox{\hskip1pt$\sim$}}
    \raise1pt\hbox{$>$}}}                

\newcommand\secname{Section}
\newcommand\figurename{Figure}
\newcommand\appname{Appendix}

\newcommand{\eg}{{\em e.g.,}}
\newcommand{\ie}{{\em i.e.,}}

\defcitealias{sahu06}{S06} 
\defcitealias{kotak09}{K09}

\DeclareMathAlphabet{\mathitbf}{OT1}{cmr}{bx}{it}

\newcommand\ion[2]{#1$\,${\sc{#2}}}

\newcommand\eline[2]{\ensuremath{{\rm #1}\,{\textsc{#2}}}}

\newcommand\Feii{\eline{Fe}{ii}}

\newcommand{\ha}{H\ensuremath{\alpha}}

\newcommand\Nai{\eline{Na}{i}}
\newcommand\Oi{\eline{O}{i}}

\newcommand\mcc[1]{\multicolumn{#1}{c}}
\newcommand\mcl[1]{\multicolumn{#1}{l}}

\newcommand\mccc[1]{\multicolumn{#1}{>{\columncolor[gray]{0.95}}c}}

\newcommand{\astjo}{\ensuremath{\ast}}

\newcommand{\lam}{\ensuremath{\lambda}}

\newcommand{\lsun}{L\ensuremath{_{\odot}}}

\newcommand{\msun}{M\ensuremath{_{\odot}}}

\newcommand{\primejo}{\ensuremath{^{\prime}}}

\newcommand{\sigjo}{\ensuremath{\sigma}}
\newcommand{\simjo}{\ensuremath{\sim}}

\newcommand{\hst}{{\em HST}}
\newcommand{\spitzer}{{\em Spitzer}}
\newcommand{\spitzerst}{{\em Spitzer Space Telescope}}

\def\kms{\rm{km\,s$^{-1}$}}

\def\micron{\rm{$\mu$m}}
\def\lam{\rm{$\lambda$}}

\def\mjysr{\rm{MJy\,sr$^{-1}$}}

\def\wm2{\rm{W\,m$^{-2}$}}

\title[The evolution of SN\,2004et]
{The effects of dust on the optical and infrared
evolution of SN\,2004et}
\author[J. Fabbri et al.]
{J. Fabbri$^1$,
M. Otsuka$^2$,
M. J. Barlow$^1$,
Joseph S. Gallagher$^{3,4}$,
R. Wesson$^1$,  
\newauthor
B. E. K. Sugerman$^5$,
Geoffrey C. Clayton$^3$,
M. Meixner$^2$,
J. E. Andrews$^3$,
\newauthor
D. L. Welch$^6$ and
B. Ercolano$^{7,8}$
\\
$^1$Dept. of Physics and Astronomy, University College London, Gower 
Street, London WC1E 6BT, UK\\
$^2$Space Telescope Science Institute,  3700 San Martin Drive, Baltimore, 
MD 21218, USA \\
$^3$Dept. of Physics and Astronomy, Louisiana State University, Baton 
Rouge, LA 70803, USA\\
$^4$Dept. of Mathematics, Physics, and Computer Science, University of
Cincinnati - Raymond Walters College, 9555 Plainfield Rd, Blue Ash, 
OH 45236, USA\\
$^5$Dept. of Physics and Astronomy, Goucher College, 1021 Dulaney 
Valley Road, Baltimore, MD 21204, USA\\
$^6$Dept. of Physics and Astronomy, McMaster University, Hamilton, 
Ontario L8S 4M1, Canada\\
$^7$Universit\"ats-Sternwarte M\"unchen, Scheinerstr. 1, 81679 M\"unchen,
Germany\\
$^8$Cluster of Excellence `Origin and Structure of the Universe',
Boltzmannstr. 2, 85748 Garching, Germany\\
}
\date{Received:}

\begin{document}
\maketitle

\begin{abstract} We present an analysis of multi-epoch observations of the
Type II-P supernova SN~2004et. New and archival optical spectra of
SN~2004et are used to study the evolution of the H$\alpha$ and [O~{\sc i}]
6300~\AA\ line profiles between days 259 and 646. Mid-infrared imaging
with Michelle on Gemini North and with all three instruments of the
\spitzerst\ was carried out between 2004 to 2010, supplemented by archival
\spitzer\ data. We include {\em Spitzer} `warm' mission
photometry at 3.6 and 4.5~$\mu$m obtained on days 1779, 1931 and 2151, 
along with ground-based and {\em HST} optical and near-infrared 
observations obtained between days 79 and 1803. Multi-wavelength light 
curves are presented, as well as optical$-$infrared spectral energy 
distributions (SEDs) for multiple epochs. Starting from about day 300, 
the optical light curves provide evidence for an increasing amount of 
circumstellar extinction attributable to newly formed dust, with
the additional extinction reaching 0.8--1.5 magnitudes in the $V$-band 
by day 690. The overall SEDs were fitted with multiple blackbody 
components, in order to investigate the luminosity evolution of the 
supernova, and then with Monte Carlo radiative transfer models using 
smooth or clumpy dust distributions, in order to estimate how much new 
dust condensed in the ejecta. The luminosity evolution
was consistent with the decay of $^{56}$Co in the ejecta
up until about day 690, after which an additional emission source
is required, in agreement with the findings of \cite{kotak09}.
Clumped dust density distributions consisting of 20\% amorphous carbons 
and 80\% silicates by mass were able to match the observed optical and 
infrared SEDs, with dust masses that increased from 
8$\times10^{-5}$~M$_{\odot}$ on
day 300 to 1.5$\times10^{-3}$~M$_{\odot}$ on day 690, still significantly 
lower than the values needed for core collapse supernovae to make a 
significant contribution to the dust enrichment of galaxies.

\end{abstract}


\section{Introduction}

Theoretical studies by \cite{kozasa89, todini01, nozawa03, bianchi07}
of dust formation in the ejecta of core-collapse supernovae (CCSNe)
produced by massive stars suggested that they could each produce up to
0.1-1.0 solar masses of dust and thereby act as major sources of dust
in galaxies. The discovery at submillimetre wavelengths of large
quantities of dust in very young high redshift galaxies,
\citep[e.g.][]{Bertoldi03, maiolino04, dwek07} seemed to reinforce the
case for massive star CCSNe as major dust contributors. However
studies at infrared wavelengths of Type~II CCSNe in the nearby
Universe have to date not confirmed these dust production predictions.
SN~1987A in the LMC condensed less than $10^{-3}$~M$_{\odot}$ of dust
in its ejecta \citep{wooden93, ercolano07}, while SN~2003gd in M~74
produced less than a few$\times10^{-3}$~M$_{\odot}$ of new dust by day
499 \citep{sugerman06, meikle07}. For a number of other recent Type~II
CCSNe, estimates for the quantities of new dust formed in their ejecta
have not exceeded 10$^{-3}$~M$_{\odot}$ \citep{andrews10, andrews11,
  szalai11, meikle11}.

\cite{kotak09} studied the multi-wavelength evolution of the Type~II-P
SN~2004et, the eighth supernova to be discovered in the past century
in the nearby spiral galaxy NGC~6946. We report here our own
multi-wavelength observations and analysis of the dust production by
this supernova.  Following \cite{li05b}, we adopt an explosion date of
22.0 September 2004 (JD~2,453,270.5) as the day 0 epoch to which our
observation dates are referenced.
{\secname}~\ref{sec:04et_background} provides a brief overview of
previously published optical and infrared (IR) observations of
SN~2004et.  {\secname}~\ref{sec:04et_ha_spec} describes the H$\alpha$
and [O~{\sc i}] 6300~\AA\ line profiles obtained from a sequence of
optical spectra of SN~2004et, while
{\secname}~\ref{sec:04et_midir_phot} describes our mid-IR
observations. The resulting mid-IR fluxes are presented in
{\secname}~\ref{sec:04et_midir_results}, together with a discussion of
the mid-IR light curve evolution.  Complementary photometric data,
obtained by us at optical and near-infrared (NIR) wavelengths using
the \hst\ and Gemini North telescopes, are presented in
{\secname}~\ref{sec:04et_opt_nir_phot}. The spectral energy
distributions (SEDs) of SN~2004et at different epochs are discussed in
{\secname}~\ref{sec:04et_seds}. This section includes blackbody
fitting to the SEDs, allowing limits to be placed on the physical
properties of the SN at the various epochs. Using Monte Carlo
radiative transfer models, we estimate ejecta dust masses in
{\secname}~\ref{sec:04et_rtmodels}. Finally, the results from our
multi-epoch analysis of SN~2004et are summarised in
{\secname}~\ref{sec:04et_discuss}, along with the implications for
dust production by Type~II SNe.

\section{Previous studies of SN~2004et}
\label{sec:04et_background}

\subsection{Optical observations}
\label{ssec:04et_type}

\cite{zwitter04} reported that an echelle spectrum taken on 28
September 2004 showed a relatively featureless spectrum with very
broad, low-contrast \ha\ emission, suggesting SN~2004et to be a Type
II event. This was confirmed by a low-resolution spectrum taken on 1
October 2004 \citep{filippenko04}, which showed the P-Cygni profile of
\ha\ to be dominated by the emission component, but that the other
hydrogen Balmer lines had more typical P-Cygni profiles. The continuum
was reported as quite blue but dropping off steeply shortwards of
4000\,{\AA}.

The {\em V}, {\em R} and {\em I}-band light curves showed SN~2004et to
be a Type II-P (plateau) supernova \citep{li05b} but it exhibited some
differences when compared with the typical Type II-P SN~1999em.
Evolution was slower in the {\em U} and {\em B} bands, consistent with
the ({\em U\,--\,B}) and ({\em B\,--\,V}) colour evolution, leading
\cite{li05b} to conclude that SN~2004et seemed to evolve more slowly
than SN~1999em, especially in the violet part of the spectrum.

Extensive photometric and spectroscopic monitoring in the optical was
carried out by \cite{sahu06} from days 8 to 541 after explosion. They
confirmed the supernova to be of Type II-P based on the plateau
observed in the {\em VRI} bands which lasted for \simjo\,110 days
after the explosion.  From their light curve analysis, they determined
that the SN was caught at a very early stage soon after the shock
breakout, reaching a maximum $B$-band magnitude \simjo\,10 days after
explosion. They also found that the decline rate of the light curve
during the early nebular phase (\simjo\,180--310 days) was similar to
the radioactive decay rate of $^{56}$Co, indicating that $\gamma$-ray
trapping was efficient during this time, and estimated that
0.06~$\pm$~0.02~\msun\ of $^{56}$Ni was synthesised during the
explosion of SN~2004et, in agreement with the subsequent estimates of
\cite{misra07} and \cite{maguire10}.  Based on the plateau luminosity
and duration and the mid-plateau expansion velocity, \cite{sahu06},
\cite{misra07} and \cite{maguire10} have estimated explosion energies
for SN~2004et in the range (0.88-1.20)$\times10^{51}$\,erg.

\subsection{Distance and reddening}
\label{ssec:04et_distance}

There are many estimates in the literature for the distance to the
host galaxy NGC~6946.  \cite{sahu06} summarised a few of these and
included the result from their own analysis of SN~2004et using the
standard candle method of \cite{nugent06} for Type II-P supernovae,
deriving an average distance of 5.6~Mpc.  \cite{herrmann08} estimated
a distance of 6.1$\pm$0.6~Mpc to NGC~6946 using the planetary nebula
luminosity function. For consistency with our previous studies of
SN~2002hh \citep{barlow05, welch07} and SN~2008S \citep{wesson10}, we
adopt the distance of 5.9$\pm$0.4~Mpc to NGC~6946 estimated by
\cite{karachentsev00} from the brightest stars method.

From \cite{schlegel98}, the foreground Galactic reddening towards
SN~2004et is estimated to be \textit{E}(\textit{B\,--\,V})~=~0.34~mag.
\cite{zwitter04} used the equivalent width of \Nai\ \ion{D}{2} lines
from their high-resolution echelle spectra to estimate a total
reddening (Galactic $+$ host) towards the SN of
\textit{E}(\textit{B\,--\,V})~=~0.41~mag. \cite{sahu06} used similar
analysis with their lower-resolution spectra to obtain a comparable
value of \textit{E}(\textit{B\,--\,V})~=~0.43 mag. We adopt a total
reddening to SN~2004et of \textit{E}(\textit{B\,--\,V})~=~0.41~mag.

\subsection{Evidence for dust production by SN~2004et}
\label{ssec:04et_dust}

From their spectroscopic observations of the temporal evolution of the 
\ha\ and [\Oi] 6300, 6363\,{\AA} line profiles from days 277 to 465, 
\cite{sahu06} noted a blueshifting after day 300 of the emission peak of 
both features and a flattening of the \ha\ emission peak. This, together 
with a steepening of the light curve after day \simjo~320, they 
interpreted as indications of early dust formation in the ejecta of 
SN~2004et. \cite{misra07} also found that the rate of decline in the 
optical light curves accelerated between \simjo\,320--386 days.

Fabbri, Sugerman \& Barlow (2005) reported the day 64 detection of
SN~2004et in all four {\em Spitzer} Infrared Array Camera (IRAC)
bands, from 3.6 to 8.0~$\mu$m, in {\em Spitzer} Infrared Nearby Galaxy
Survey (SINGS) Legacy program archival images of NGC~6946 (see
{\secname}~\ref{ssec:04et_midir_obs}).  \cite{kotak09} presented
archival and their own \spitzer\ mid-IR observations of SN~2004et
obtained up to day 1406, together with late-time optical spectra. They
concluded that between days 300--795, the spectral energy distribution
was comprised of three components -- hot, warm and cold -- each
respectively due to emission from: optically thick gas; newly-formed,
ejecta-condensed dust; and an IR echo from the interstellar medium of
the host galaxy. They estimated that the mass of dust formed in the
ejecta grew to a few times $10^{-4}$\,\msun, located in co-moving
clumps of fixed size. From their \spitzer\ Infrared Spectrometer (IRS)
spectra, they reported the first spectroscopic evidence of silicate
dust formed in the ejecta of a supernova, supported by the detection
of strong but declining molecular SiO emission in the 8-$\mu$m region.
They interpreted the appearance of broad, box-shaped optical emission
line profiles about 2 years post-explosion as due to the impact of the
ejecta on the circumstellar medium of the progenitor star, resulting
in the formation of a cool, dense shell to which they attributed
responsibility for a later rise in the mid-IR emission from SN~2004et.

\cite{maguire10} reported optical and NIR photometric and
spectroscopic observations of SN~2004et carried out from just after
explosion to $+$500 days. Their NIR spectrum at day 306 showed a clear
detection of the first overtone band of CO at $\sim$2.3~$\mu$m, which
they interpreted as being a signature of dust formation. By analysing
the optical light curves in the early nebular phase, they found that
the $BVR$ decline rates between days \simjo\,136 and 296 were
consistent with those expected from light curves powered by the
radioactive decay of $^{56}$Co (assuming complete $\gamma$-ray
trapping). However, between days \simjo\,296 and 414 they found that
these decline rates had steepened.  They also noted a significant
blueshift in the peak of the \ha\ emission line from days \simjo\,300
to 464, in agreement with the results of \cite{sahu06}. These results
were interpreted as signatures of dust formation occurring post 300
days. Their very late time ($>$\,1000 days) photometry showed a
flattening of the optical and NIR light curves, which they mainly
attributed to the interaction of the SN ejecta with the circumstellar
medium (CSM), following the work of \cite{kotak09}.

\section{Optical spectroscopy of SN~2004et}
\label{sec:04et_ha_spec}

\begin{table*}\centering
  \parbox[t]{14.5cm}{\protect\caption[Log of optical spectroscopic observations]{Log of optical spectroscopic observations of SN~2004et from the SEEDS collaboration (GMOS-N, PI: Clayton), from the authors \cite{sahu06} and from online archives (TNG and Subaru).\label{tab:04et_opt_spectra}}}
\setlength{\tabcolsep}{1mm}
{\footnotesize
\begin{threeparttable}
\begin{tabular}{lcllcll}
  \hline\hline
  & & & & & & \\    
 Date & Age & Telescope/ & \mcc1{Wavelength} & Exp. & Program ID & Principal \\
  & [days] & instrument &  \mcc1{range [\AA]} & time &            & Investigator \\
  & & & & & & \\ \hline
  & & & & & & \\
2005-06-07 & 259  & HCT HFOSC\tnote{a} & 3500--7000; 5200--9200 & 1$\times$900\,s &--&\cite{sahu06} \\
2005-08-01 & 314  & HCT HFOSC\tnote{a} & 3500--7000; 5200--9200 & 1$\times$900\,s &--&\cite{sahu06} \\
2005-08-05 & 317  & Gemini  GMOS-N  & 3500--10000 & 3$\times$900\,s & GN-2005B-Q-54 & G. Clayton \\
2005-08-29 & 336  & TNG LRS\tnote{b}  & 3890-8000 & 1$\times$1800\,s & AOT12 CAT-G109 & E. de la Rosa \\
2005-10-17 & 391  & HCT HFOSC\tnote{a} & 3500--7000; 5200--9200 & 1$\times$900\,s &--&\cite{sahu06} \\
2005-10-31 & 404  & Gemini  GMOS-N  & 3500--10000 & 3$\times$900\,s & GN-2005B-Q-54 & G. Clayton \\
2005-11-23 & 428  & HCT HFOSC\tnote{a} & 3500--7000; 5200--9200 & 1$\times$900\,s &--&\cite{sahu06} \\
2006-06-30 & 646  & Subaru FOCUS\tnote{c}  & 4670-8970 & 2$\times$900\,s & S06A-152 & K. Kawabata \\
  & & & & & & \\ \hline
\end{tabular}
\begin{tablenotes}
 \scriptsize
 \item [a] Raw data, including calibration frames and flux standards, provided by the authors \citep{sahu06}.
 \item [b] Raw and calibration data were downloaded from the online TNG archive at http://ia2.oats.inaf.it/.
 \item [c] Raw and calibration data downloaded from the online Subaru Mitaka Okayama Kiso Archive (SMOKA) at  http://smoka.nao.ac.jp/index.jsp.
 \end{tablenotes}
\end{threeparttable}
}
\end{table*}

As part of the SEEDS program, Gemini North spectroscopic observations
of SN~2004et were obtained on 5 August and 31 October 2005,
corresponding to 317 and 404 days after explosion. A log of these and
the other H$\alpha$-region observations that we analyse here can by
found in {\tablename}~\ref{tab:04et_opt_spectra}. We utilised GMOS-N
equipped with a 0\farcs75 slit width and the B600-G5303 grating in
long-slit mode with a position angle of 296\degr. Three spectra were
obtained during each epoch with identical exposure times of 900\,s.
The central wavelengths of the images were 5950, 5970, and
5990\,{\AA}, respectively, to allow for gap removal of the combined
spectra. A 2$\times$2 binning of the CCD pixels in the low gain
setting was employed.  All spectra were taken with adjacent GMOS
baseline calibration flat exposures to correct for sensitivity
gradients across the CCD, and CuAr spectra were utilised for the
initial calibration of the dispersion solution.

Our GMOS-N spectra were reduced using the Gemini IRAF package.
Pipeline processed calibration images were obtained from the Gemini
Science Archive. The spectra were trimmed and then overscan, bias and
flat-field corrected using the task {\tt gsreduce}. Wavelength
calibration solutions were determined from the CuAr lamp spectra using
{\tt gswavelength}, and the solution was applied to the SN~2004et
spectra via {\tt gstransform}.  Object spectra were extracted using
{\tt gsextract}.  The observations were not flux calibrated since our
primary goal was to monitor the evolution of the line profiles.

\begin{figure*}
\includegraphics[scale=0.55]{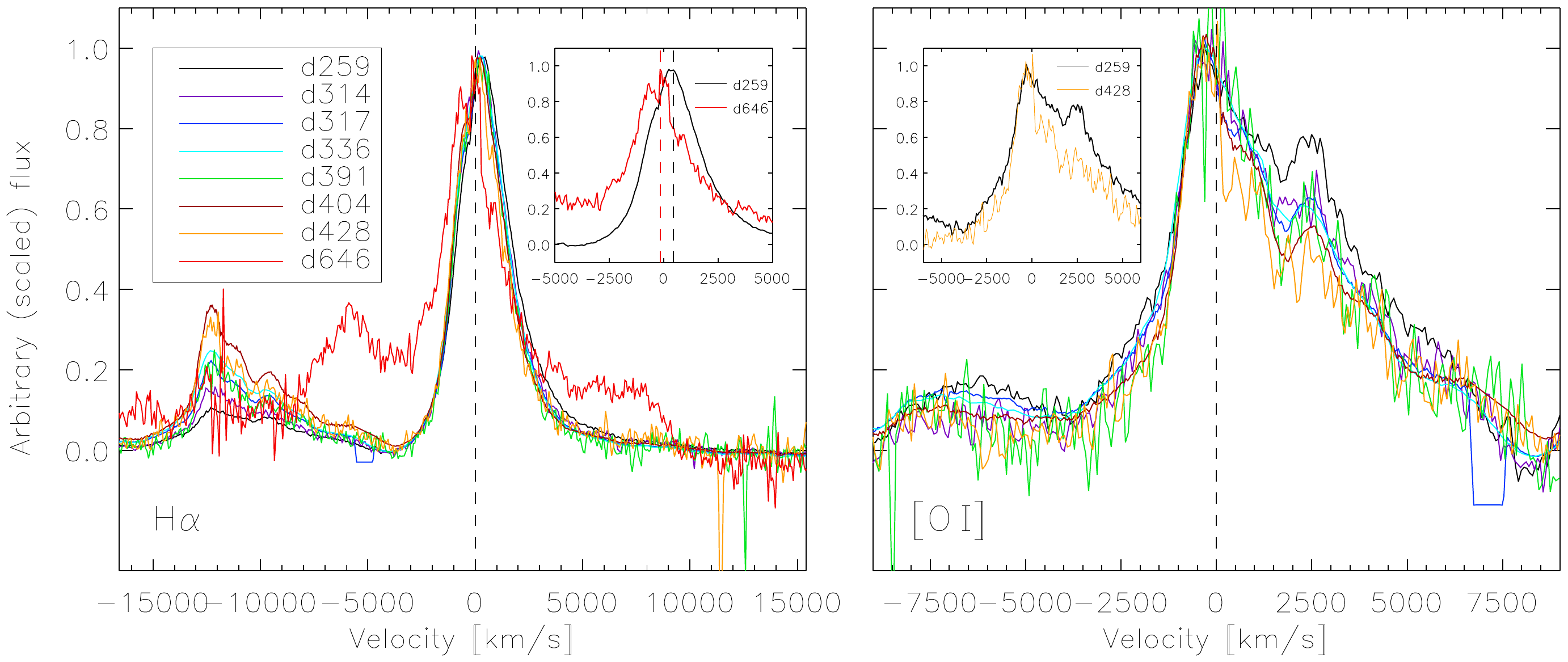}
\hspace{0.2cm}
\caption[\ha\ profile evolution]{The \ha\ 
  and [O~{\sc i}] profile evolution. 
  The left hand panel shows the \ha\ spectra from days 259 to 646,
  with the normalised continua subtracted to a zero-level and
  the peaks of the emission line normalised to unity.
  The dashed vertical line in the main plot corresponds to the rest 
  wavelength of \ha\ (NGC~6946 has a radial velocity of +40~km~s$^{-1}$). 
  The inset shows a comparison of the
  earliest and latest \ha\ spectra, obtained just over a year 
  apart, to highlight the blueshifting (\simjo\,600\,\kms) during this
  interval. The dashed lines in the inset plot indicate the measured line
  centres.  The right hand panel shows the evolution of the [O~{\sc i}] 
  6300~\AA\ profile from days 259 to 428. The dashed line in the main 
  plot indicates the rest velocity of the 6300~\AA\ line. The emission 
  feature in the red wing corresponds to the [O~{\sc i}] 6363~\AA\ line.
  The spectra from days 259, 314, 391 and 428 are from
  \cite{sahu06} (raw data kindly provided by the authors and calibrated 
  by us).  The spectra
  from days 317 and 404 are SEEDS GMOS-N observations, and those
  from days 336 and 646 are from archival TNG-LRS and Subaru-FOCAS spectra
  respectively.}
\label{fig:04et_halpha}   
\end{figure*}

\cite{sahu06} presented photometric and spectroscopic data for
SN~2004et from approximately 8 to 541 days after the explosion. Their
results showed a shift to the blue of the central peak of both \ha\
and [OI] 6300,6363{\,\AA} at late times. They concluded that this was
indicative of new dust formed in the ejecta of SN~2004et. However,
comparisons between the \ha\ profiles in our GMOS-N spectra and those
at similar epochs in the archived calibrated spectra of \cite{sahu06},
obtained from the \textit{Online Supernova Spectrum Archive
  (SUSPECT)}\footnote{http://bruford.nhn.ou.edu/$\sim$suspect}, showed
some inconsistencies, with their spectra showing clear blueshifting of
the \ha\ emission line profile between days 314 and 391, while our own
spectra showed little change between days 317 and 404.  In order to
help resolve this discrepancy, the \cite{sahu06} authors generously
provided their raw data for SN~2004et for several of the epochs
presented in their paper. This allowed us to reduce the respective
sets of data in the same way.  A description of the observational
setup used by \citet{sahu06} can be found in their paper.

We also obtained two archival spectroscopic observations of SN~2004et,
taken on days 336 and 646. The earlier epoch corresponds to an observation
on 29 August 2005 taken with the Low Resolution Spectrograph on the 3.58-m 
Telecopio Nazionale Galileo (TNG) on La Palma. The latter
epoch corresponded to an observation on 30 June 2006 with the Faint Object
Camera and Spectrograph (FOCAS) on the 8.2-m Subaru telescope on Mauna Kea
in Hawaii. The reduction procedure matched that carried out by us for the
GMOS-N and \citet{sahu06} spectra.

Since our goal was to self-consistently align a number of spectra
taken with four different instrument/telescope setups, we did not rely
solely on the initial wavelength calibrations. During the extraction
of each SN spectrum, a sky spectrum was also extracted that was
ultimately subtracted from the SN spectrum. Since the sky and SN
spectra possessed identical wavelength calibrations, we derived
corrections to the initial wavelength calibrations of each of the SN
spectra using the strong and narrow [\Oi] 5577{\,\AA} and 6300{\,\AA}
sky emission lines.

The temporal evolution of the \ha\ profile between days 259 and 646,
following our re-analysis of the combined set of spectra, is depicted
in the left-hand panel of {\figurename}~\ref{fig:04et_halpha}.  The
spectra are displayed in velocity space, with the continua subtracted
and the peaks of the emission line normalised to approximately unity.
The inset shows a close-up view of the earliest and latest \ha\
profiles, obtained just over a year apart at days 259 and 646, to
highlight the overall blueshifting of the profile during this time.
The dashed lines in the inset indicate the line peak centres measured
on days 259 and 646, with the blueshifting of the latter profile
providing strong evidence for the formation of dust in the ejecta
during the intervening period, as described by \citet{lucy89} for
SN~1987A.

The \ha\ line profiles presented by \cite{sahu06} 
showed a significant blueshifting with time of the emission peak,
amounting to a few hundred \kms, with the
largest shift appearing to occur between days 277 and 314.
\cite{sahu06} did not quantify the shifts in their
\ha\ profiles, beyond stating that a blueshift in
the emission peak was clearly seen after day 300 in their 
day 277-465 \ha\ (and [\Oi] 6300,6363{\,\AA}) profiles. 
\cite{kotak09} confirmed this from their own analysis of the
\citet{sahu06} spectra, reporting a shift of $-$400\,\kms\ in the
whole \ha\ profile between days 301 and 314, but little sign of a
progressive blueshifting during the subsequent day 314--465 period.
Using the \citet{sahu06} spectra, along with additional
spectra, \cite{maguire10} found that the peak of the \ha\ emission
line was at +280 $\pm$ 50\,\kms\ between days 163 and 300, but 
from days 314 to 464 showed a constant blueshift, to $-$137\,\kms. 

\begin{figure}\centering
\includegraphics[scale=0.55]{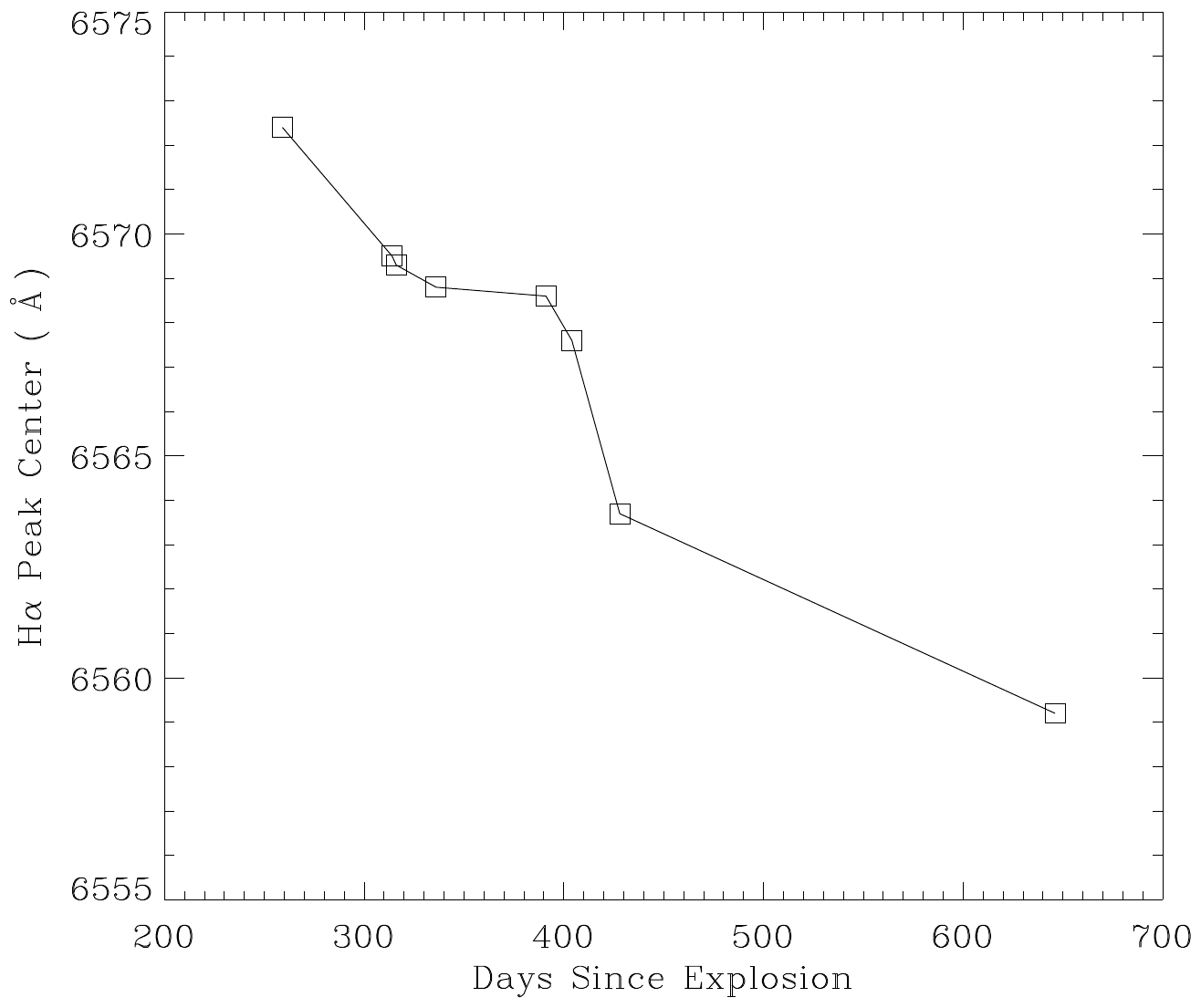}
{\caption[The wavelength evolution of the \ha\ emission 
peak]{The evolution with time of the measured wavelength
    of the \ha\ emission peak of SN~2004et.}
  \label{fig:04et_halpha_peak_blueshift}}
\end{figure}

We have used our recalibrated spectral dataset to measure the
wavelength of peak H$\alpha$ emission in each of the profiles plotted
in {\figurename}~\ref{fig:04et_halpha}.  The evolution of the peak
wavelength is shown in
{\figurename}~\ref{fig:04et_halpha_peak_blueshift}. The line peaks
show an overall blueshifting between days 259 and 646 of
$\sim$13.2{\,\AA}, corresponding to a velocity shift of
$\sim-$600\,\kms. Between days 259 and 314 we measure an initial
blueshift of \simjo\,3{\,\AA} ($\approx$ 140\,\kms), not as large as
the 400\,\kms\ shift between days 301 and 314 measured by
\cite{kotak09} from the \cite{sahu06} spectra in the {\em SUSPECT}
archive. While we find little change between days 314/317 and day 391
in the measured emission line peaks in the recalibrated spectral
dataset (Figure~\ref{fig:04et_halpha_peak_blueshift}), between days
404 and 428 we measure a blueshifting of the emission peak by
\simjo\,4{\,\AA} (185 km~s$^{-1}$). followed by a further blueshifting
of \simjo\,4.5{\,\AA} (205 km~s$^{-1}$) between the spectra obtained
at days 428 and 646. We therefore find that the majority of the
blueshifting occurred after day 391.

Whilst there is no strong evidence for a developing line asymmetry in the
earlier epoch H$\alpha$ profiles ({\figurename}~\ref{fig:04et_halpha}), a
diminution of the red wing can be discerned by days 428 and 646. By day
646 a significant blueshifting (\simjo\,4.5\,\AA) of the whole profile
since day 428 is evident, with multiple peaks evident at the centre of the
profile, although there are earlier inflections in the blue wing of the
\ha\ profiles. Two broad features either side of the main emission line
(at \simjo\,$+$7000 and $-$6000\,\kms) have appeared by day 646.
\cite{kotak09} presented three later optical spectra of SN~2004et,
obtained with the Keck telescopes on days 823, 933 and 1146, which show
similarities to, and a development of, the Subaru day 646 \ha\ profile.
They described the \ha\ profile from the late-time Keck spectra as having
a steep-sided, box-like component, with a half width at zero intensity
(HWZI) of 8500\,\kms and noted that the characteristic ejecta profile of
\ha\ seen in the \cite{sahu06} spectra may still be present at days 823
and 933, with a HWZI of \simjo\,2000\,\kms, but that its presence at day
1146 was less certain.

The 6300 and 6363~\AA\ lines of [O~{\sc i}] can be seen in emission on
the blue side of the H$\alpha$ profiles plotted in
{\figurename}~\ref{fig:04et_halpha}. In the right-hand panel of
{\figurename}~\ref{fig:04et_halpha}, we have plotted the [O~{\sc i}]
6300-\AA\ velocity profiles for days 259 through to 428 (the day 646
profile had too low a signal to noise). The inset compares the day 259
and day 428 [O~{\sc i}] profiles, showing a clear diminution of the
red wing of the day 428 profile relative to day 259.

\begin{figure*}\centering
  \includegraphics[scale=0.9,angle=0,clip=true]
  {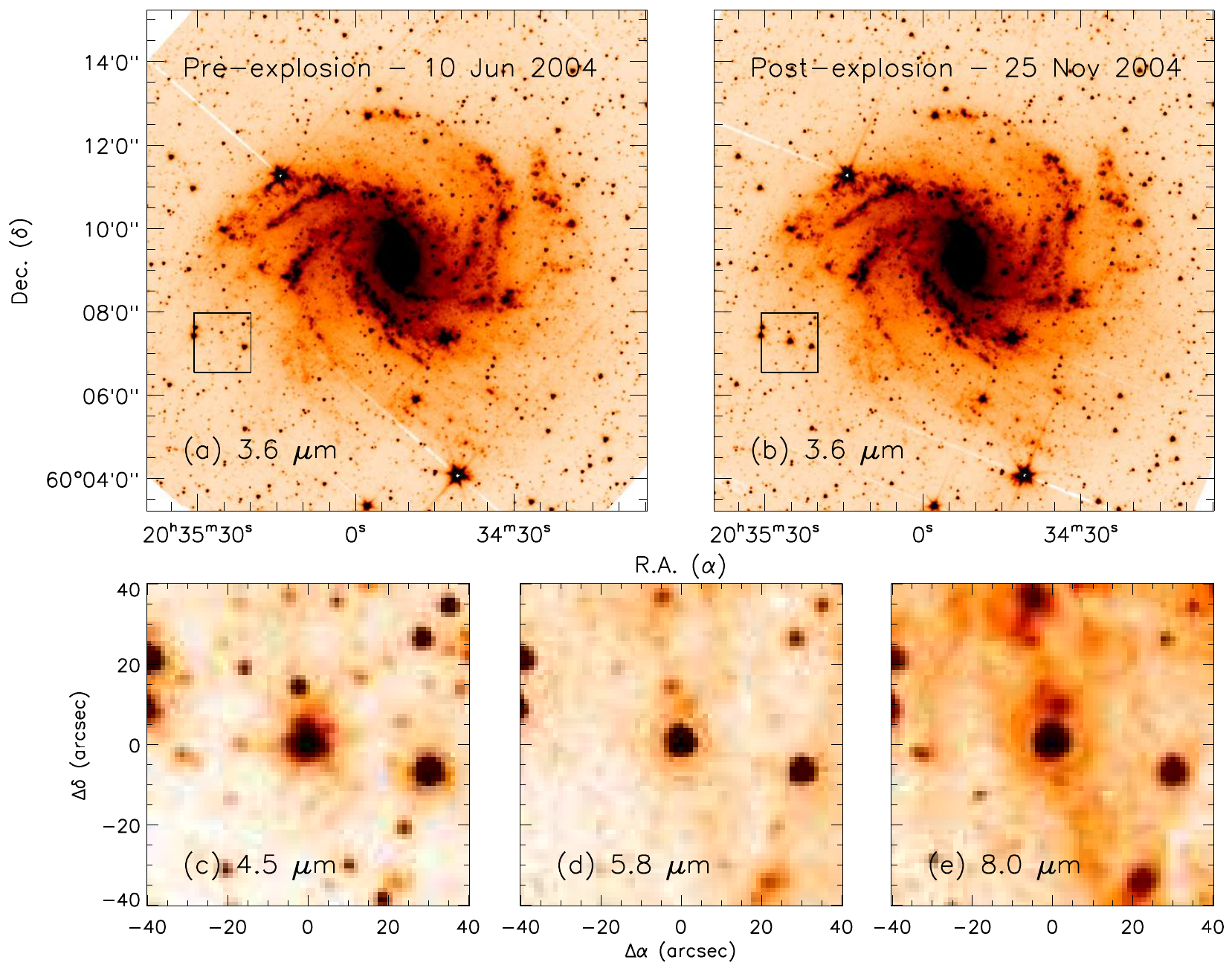} \vspace{2ex} 
   {\caption[The
    position of SN~2004et in SINGS Legacy IRAC images of NGC~6946.]{The
    position of SN~2004et in SINGS Legacy IRAC images of NGC~6946.
    Panels (a) and (b) show the whole galaxy at 3.6-$\mu$m pre- (day
    $-$104) and post-explosion (day 64) respectively. The square is
    centred on the SN coordinates, with the SN clearly evident at day
    64.  Panels (c), (d) and (e) zoom in on the square region of the
    SN field at 4.5, 5.8 and 8.0~$\mu$m respectively, at day
    64.}\label{fig:04et_irac_sings_field}}
\end{figure*}

\section{Mid-infrared observations}
\label{sec:04et_midir_phot}

\subsection{Gemini-Michelle and \spitzer\ photometry}
\label{ssec:04et_midir_obs}

NGC~6946, the host galaxy of SN~2004et, was observed with \spitzer\ by 
the SINGS Legacy program \citep{kennicutt03} between June and November 
2004, such that the region of the SN was serendipitously imaged pre- and 
post-explosion. {\figurename}~\ref{fig:04et_irac_sings_field} shows the
position of the SN in relation to its host galaxy NGC~6946 in the pre-
and post-explosion SINGS IRAC images, together with a closer view of
the SN region in the different IRAC wavebands. 

SN~2004et was clearly detected in the SINGS IRAC image at day 64 and
was subsequently monitored with IRAC, MIPS and the IRS Peak-Up 
Imaging (PUI) module
via our \spitzer\ GO programs during Cycles 2, 3, 4, 6 and 7
(Cycle 6 was the beginning of the \spitzer\ `warm' mission, in which
the observatory operates using only the 3.6 and 4.5\,\micron\ IRAC
channels).  Archival \spitzer\ data for the SN from
Cycles 2 and 3 were also downloaded to provide a more complete
time-sample of the SN's mid-IR evolution. In summary, post-explosion
mid-IR observations of SN~2004et were taken between November 2004 and
August 2010, corresponding to an age range of 64 to 2151 days. With
the \spitzerst, there are 12 epochs of observations in each of the
four IRAC wavebands, with a further three epochs of IRAC observations
during the `warm' mission in just the 3.6- and 4.5-$\mu$m channels.
There are 8 epochs of IRS PUI observations at 16~$\mu$m, and 9
epochs of MIPS 24-$\mu$m observations. Of the 33 individual
\spitzer\ observations listed in {\tablename}~\ref{tab:04et_midir_obs}, 
the first 3 were obtained by the SINGS Legacy program, 5 were obtained by 
programs led by PIs Meikle and Kotak and 25 were obtained by our
SEEDS program. 

In addition to the \spitzer\ mid-IR observations, broad-band N\primejo\ 
photometry at 11.2~$\mu$m was obtained with Michelle on Gemini-North 
during 2005, 2006, 2007 and 2008, consisting of six observations at the 4 
epochs.  A time-ordered list of all the \spitzer\ and Michelle mid-IR 
observations of 
SN~2004et is provided in {\tablename}~\ref{tab:04et_midir_obs}.  The 
listed exposure time is the total time spent on-source. The final column 
provides a key to the list of observing programs from which these data 
were obtained; the program number and principal investigator are detailed 
in a footnote to the table.  Data retrieved from the archive are marked as 
such. {\appname}~\ref{ssec:04et_midir_reduce} describes how the mid-IR
data were processed.

\begin{table*}\centering
\begin{spacing}{0.90}
  \parbox[t]{14.5cm}{\protect\caption[Summary of mid-IR observations of SN~2004et]{Summary of the mid-infrared imaging observations of SN~2004et with the \spitzerst\ and Gemini North.\label{tab:04et_midir_obs}}}
{\small
\begin{threeparttable}
\begin{tabular}{lrlccccc}
  \hline\hline

  & & & & & & & \\
  \mcc1{UT date} & \mcc1{Age} & \mcc1{Detector} & \mcc1{\lam$_{eff}$} & \mcc1{FoV} & \mcc1{Pixel scale} & \mcc1{Exp. time} & \mcc1{Ref.}\\

                 & [days]     &                 & \mcc1{[\micron]}         & \mcc1{[\arcmin~$\times$~\arcmin]} & \mcc1{[\arcsec/pixel]} & \mcc1{[s]} & \\
  & & & & & & & \\ 
\hline
  & & & & & & & \\
2004-06-10 & -104 & IRAC & 3.6/4.5/5.8/8.0 & 5.2\,$\times$\,5.2 & 1.2 & 107.2 & \tnotens{[1]}\\
2004-07-09 &  -75 & MIPS & 24.0            & 5.4\,$\times$\,5.4 & 1.5 & 161.5 & \tnotens{[1]}\\
2004-11-25 &   64 & IRAC & 3.6/4.5/5.8/8.0 & 5.2\,$\times$\,5.2 & 1.2 & 107.2 & \tnotens{[1]}\\
2005-07-13 &  294 & IRS-PUI & 16.0            &   1.0\,$\times$\,1.2 & 1.2 & 629.2 & \tnotens{[2]}\\
2005-07-19 &  300 & IRAC & 3.6/4.5/5.8/8.0 & 5.2\,$\times$\,5.2 & 0.75 & 14.4 & \tnotens{[3]}\\
2005-07-30 &  311 & Michelle & 11.2 (N\primejo) & 0.5\,$\times$\,0.4 & 0.1 & 1081.9 & \tnotens{[4]}\\
2005-08-03 &  315 & MIPS & 24.0            & 5.4\,$\times$\,5.4 & 0.75 & 140.0 & \tnotens{[3]}\\
2005-09-17 &  360 & IRAC & 3.6/4.5/5.8/8.0 & 5.2\,$\times$\,5.2 & 0.75 & 14.4 & \tnotens{[3]}\\
2005-09-24 &  367 & MIPS & 24.0            & 5.4\,$\times$\,5.4 & 0.75 & 140.0 & \tnotens{[3]}\\
2005-11-02 &  406 & IRAC & 3.6/4.5/5.8/8.0 & 5.2\,$\times$\,5.2 & 0.75 & 536.0 & \tnotens{[2]}\\
2005-12-22 &  456 & IRS-PUI & 16.0            &   1.0\,$\times$\,1.2 & 1.2 & 629.2 & \tnotens{[2]}\\
2005-12-30 &  464 & IRAC & 3.6/4.5/5.8/8.0 & 5.2\,$\times$\,5.2 & 0.75 & 14.4 & \tnotens{[3]}\\
2006-01-11 &  476 & MIPS & 24.0            & 5.4\,$\times$\,5.4 & 0.75 & 140.0 & \tnotens{[3]}\\
2006-05-12 &  597 & Michelle & 11.2 (N\primejo) & 0.5\,$\times$\,0.4 & 0.1 & 811.4 & \tnotens{[5]}\\
2006-05-14 &  599 & Michelle & 11.2 (N\primejo) & 0.5\,$\times$\,0.4 & 0.1 & 376.3 & \tnotens{[5]}\\
2006-08-04 &  681 & IRS-PUI & 16.0            &   1.0\,$\times$\,1.2 & 1.2 & 629.2 & \tnotens{[6]}\\
2006-08-13 &  690 & IRAC & 3.6/4.5/5.8/8.0 & 5.2\,$\times$\,5.2 & 0.75 & 14.4 & \tnotens{[7]}\\
2006-09-01 &  709 & MIPS & 24.0            & 5.4\,$\times$\,5.4 & 0.75 & 140.0 & \tnotens{[7]}\\
2006-09-10 &  718 & IRS-PUI & 16.0            &   1.0\,$\times$\,1.2 & 1.2 & 56.6 & \tnotens{[7]}\\
2006-12-29 &  828 & IRAC & 3.6/4.5/5.8/8.0 & 5.2\,$\times$\,5.2 & 0.75 & 124.8 & \tnotens{[7]}\\
2007-01-21 &  851 & MIPS & 24.0            & 5.4\,$\times$\,5.4 & 0.75 & 420.0 & \tnotens{[7]}\\
2007-01-27 &  857 & IRS-PUI & 16.0            &   1.0\,$\times$\,1.2 & 1.2 & 132.1 & \tnotens{[7]}\\
2007-06-26 & 1007 & IRS-PUI & 16.0            &   1.0\,$\times$\,1.2 & 1.2 & 283.1 & \tnotens{[7]}\\
2007-07-03 & 1015 & IRAC & 3.6/4.5/5.8/8.0 & 5.2\,$\times$\,5.2 & 0.75 & 321.6 & \tnotens{[7]}\\
2007-07-09 & 1020 & Michelle & 11.2 (N\primejo) & 0.5\,$\times$\,0.4 & 0.1 & 1999.2 & \tnotens{[8]}\\
2007-07-10 & 1021 & MIPS & 24.0            & 5.4\,$\times$\,5.4 & 0.75 & 420.0 & \tnotens{[7]}\\
2007-08-02 & 1044 & IRS-PUI & 16.0            &   1.0\,$\times$\,1.2 & 1.2 & 283.1 & \tnotens{[9]}\\
2007-08-12 & 1054 & IRAC & 3.6/4.5/5.8/8.0 & 5.2\,$\times$\,5.2 & 0.75 & 321.6 & \tnotens{[9]}\\
2007-08-27 & 1069 & MIPS & 24.0            & 5.4\,$\times$\,5.4 & 0.75 & 420.0 & \tnotens{[9]}\\
2007-12-09 & 1173 & IRS-PUI & 16.0          &   1.0\,$\times$\,1.2 & 1.2 & 283.1 & \tnotens{[9]}\\
2007-12-27 & 1191 & IRAC & 3.6/4.5/5.8/8.0 & 5.2\,$\times$\,5.2 & 0.75 & 321.6 & \tnotens{[9]}\\
2008-01-07 & 1202 & MIPS & 24.0            & 5.4\,$\times$\,5.4 & 0.75 & 420.0 & \tnotens{[9]}\\
2008-01-17 & 1212 & IRS-PUI & 16.0          &   1.0\,$\times$\,1.2 & 1.2 & 1258.4 & \tnotens{[10]}\\
2008-06-21 & 1368 & Michelle & 11.2 (N\primejo) & 0.5\,$\times$\,0.4 & 0.1 & 1928.6 & \tnotens{[11]}\\
2008-07-09 & 1386 & Michelle & 11.2 (N\primejo) & 0.5\,$\times$\,0.4 & 0.1 & 2257.9 & \tnotens{[12]}\\
2008-07-18 & 1395 & IRAC & 3.6/4.5/5.8/8.0 & 5.2\,$\times$\,5.2 & 0.75 & 321.6 & \tnotens{[9]}\\
2008-07-29 & 1406 & MIPS & 24.0            & 5.4\,$\times$\,5.4 & 0.75 & 420.0 & \tnotens{[9]}\\
2009-08-06 & 1779 & IRAC & 3.6/4.5        & 5.2\,$\times$\,5.2 & 0.75 & 1161.6 & \tnotens{[13]}\\
2010-01-05 & 1931 & IRAC & 3.6/4.5        & 5.2\,$\times$\,5.2 & 0.75 & 1161.6 & \tnotens{[13]}\\
2010-08-13 & 2151 & IRAC & 3.6/4.5        & 5.2\,$\times$\,5.2 & 0.75 & 1161.6 & \tnotens{[14]}\\
  & & & & & & & \\ \hline

\end{tabular}
\begin{tablenotes}
\scriptsize
\item [{[1]}] \spitzer\ Cycle 1 SINGS Legacy program 00159, PI: Kennicutt.
\item [{[2]}] Archival data, \spitzer\ Cycle 2 GO program 20256, PI: Meikle
\item [{[3]}] This paper, \spitzer\ Cycle 2 GO program 20320, PI: Sugerman.
\item [{[4]}] This paper, Gemini semester 05A program GN-2005A-Q-20, PI: Barlow.
\item [{[5]}] This paper, Gemini semester 06A program GN-2006A-Q-1, PI: Barlow.
\item [{[6]}] Archival data, \spitzer\ Cycle 3 GO program 30292, PI: Meikle.
\item [{[7]}] This paper, \spitzer\ Cycle 3 GO program 30494, PI: Sugerman.
\item [{[8]}] This paper, Gemini semester 07A program GN-2007A-Q-5, PI: Barlow.
\item [{[9]}] This paper, \spitzer\ Cycle 4 GO program 40010, PI: Meixner.
\item [{[10]}] Archival data, \spitzer\ Cycle 4 GO program 40619, PI: Kotak.
\item [{[11]}] This paper, Gemini semester 07B program GN-2007B-Q-4, PI: Barlow.
\item [{[12]}] This paper, Gemini semester 08B program GN-2008B-Q-44, PI: Barlow.
\item [{[13]}] This paper, \spitzer\ Cycle 6 GO program 60071, PI:
  Andrews. Cycle 6 took place during the \spitzer\ `warm' mission
  following completion of the cryogenic mission. Only IRAC 3.6 and
  4.5\,\micron\ channels were available, with expected sensitivity
  unchanged from performance in the cryogenic mission.
\item [{[14]}] This paper, \spitzer\ Cycle 7 GO program 70008, PI:
  Andrews, \spitzer\ `warm' mission. 
\end{tablenotes}
\end{threeparttable}
}
\end{spacing}
\end{table*}

\begin{table*}\centering
\footnotesize
  \setlength{\tabcolsep}{0.4mm}
  \setlength{\LTcapwidth}{22cm}
  \begin{spacing}{1}
\caption[Mid-infrared photometry of SN~2004et]{Mid-infrared photometry of SN~2004et. The shaded rows indicate those fluxes across different wavebands considered as the same epoch. \label{tab:04et_midir_fluxes_psffit_diff}}
\begin{threeparttable}
\begin{tabular}{crccccccccccccccrcccrcccrc}
  \hline\hline
  \mcc{26}{} \\ 

  UT date & \mcc1{Age} & \mcc{24}{Flux density\tnote{1} / upper limits\tnote{2} ~[$\mu$Jy]} 
  \\ \cline{3-26}

  & \mcc1{[days]} & \mcc{12}{} &  & \mcc3{} &   & \mcc3{} &   & \mcc3{}\\
  \mcc{2}{} & \mcc{12}{IRAC\tnote{3}} &  & \mcc3{Michelle N\primejo} &  & \mcc3{IRS-PUI\tnote{3}} &  & \mcc3{MIPS\tnote{3}} \\

  \mcc{2}{} & \mcc3{3.6 \micron} & \mcc3{4.5 \micron} & \mcc3{5.8 \micron} & 
  \mcc3{8.0 \micron} &  & \mcc3{11.2 \micron} &  & \mcc3{16 \micron} &  & \mcc3{24 \micron} \\
\hline
 2004-06-10 & -104 & & 83.8 $\pm$ 10.2 & & & 45.4 $\pm$ 8.3 & & & 178 $\pm$ 25 & & & 412 $\pm$ 51 & &  & \mcc3{$\cdots$} &  & \mcc3{$\cdots$} &  & \mcc3{$\cdots$}\\

 2004-07-09 &  -75 & \mcc3{$\cdots$} & \mcc3{$\cdots$} & \mcc3{$\cdots$} & \mcc3{$\cdots$} &  & \mcc3{$\cdots$} &  & \mcc3{$\cdots$} & & & 376 $\pm$ 25 &  \\  \hline

 2004-11-25 &   64 & & 17490 $\pm$ 532 & & & 13038 $\pm$ 745 & & & 10046 $\pm$ 179 & & & 6020 $\pm$ 91 & &  & \mcc3{$\cdots$} &  & \mcc3{$\cdots$} &  & \mcc3{$\cdots$} \\

 \rowcolor[gray]{0.95}
 \rowcolor[gray]{0.95}
 2005-07-13 &  294 & \mccc3{$\cdots$} & \mccc3{$\cdots$} & \mccc3{$\cdots$} & \mccc3{$\cdots$} &  & \mccc3{$\cdots$} &  & & 930 $\pm$ 33 & &  & \mccc3{$\cdots$} \\
 \rowcolor[gray]{0.95}
 2005-07-19 &  300 & & 726 $\pm$ 61 & & & 3151 $\pm$ 97 & & & 1291 $\pm$ 218 & & & 2162 $\pm$ 154 & &  & \mccc3{$\cdots$} &  & \mccc3{$\cdots$} &  & \mccc3{$\cdots$} \\
 \rowcolor[gray]{0.95}
 2005-07-30 &  311 & \mccc3{$\cdots$} & \mccc3{$\cdots$} & \mccc3{$\cdots$} & \mccc3{$\cdots$} &  & & 1700 $\pm$ 200 & &  & \mccc3{$\cdots$} &  & \mccc3{$\cdots$} \\
 \rowcolor[gray]{0.95}
 2005-08-03 &  315 & \mccc3{$\cdots$} & \mccc3{$\cdots$} & \mccc3{$\cdots$} & \mccc3{$\cdots$} &  & \mccc3{$\cdots$} &  & \mccc3{$\cdots$} &  & &  832 $\pm$ 75 & \\

 2005-09-17 &  360 & & 430 $\pm$ 71 & & & 1728 $\pm$ 88 & & & 935 $\pm$ 285 & & & 1731 $\pm$ 156 & &  & \mcc3{$\cdots$} &  & \mcc3{$\cdots$} &  & \mcc3{$\cdots$} \\
 2005-09-24 &  367 & & \mcc3{$\cdots$} & & \mcc3{$\cdots$} & & \mcc3{$\cdots$} &  & \mcc3{$\cdots$} &  & \mcc3{$\cdots$} &  & & 735 $\pm$ 74 & \\

 \rowcolor[gray]{0.95}
 2005-11-02 &  406 & & 315 $\pm$ 21 & & & 1045 $\pm$ 21  & & & 707 $\pm$ 79 & & & 1500 $\pm$ 77 & &  & \mccc3{$\cdots$} &  & \mccc3{$\cdots$} &  & \mccc3{$\cdots$} \\

 2005-12-22 &  456 & \mcc3{$\cdots$} & \mcc3{$\cdots$} & \mcc3{$\cdots$} & \mcc3{$\cdots$} &  & \mcc3{$\cdots$} &  & & 890 $\pm$ 32 & &  & \mcc3{$\cdots$} \\
 2005-12-30 &  464 & & 174 $\pm$ 67 & & & 656 $\pm$ 77 & & & 606 $\pm$ 293 & & & 952 $\pm$ 162 & &  & \mcc3{$\cdots$}&  & \mcc3{$\cdots$} &  & \mcc3{$\cdots$} \\
 2006-01-11 &  476 & \mcc3{$\cdots$} & \mcc3{$\cdots$} & \mcc3{$\cdots$} & \mcc3{$\cdots$} &  & \mcc3{$\cdots$} &  & \mcc3{$\cdots$} &  & & 686 $\pm$ 89 & \\

 \rowcolor[gray]{0.95}
 2006-05-12/14 &  597/599 & \mccc3{$\cdots$} & \mccc3{$\cdots$} & \mccc3{$\cdots$} & \mccc3{$\cdots$} &  & \mccc3{$\leq$ 650} &  & \mccc3{$\cdots$} &  & \mccc3{$\cdots$} \\
 \rowcolor[gray]{0.95}
 2006-08-04 &  681 & \mccc3{$\cdots$} & \mccc3{$\cdots$} & \mccc3{$\cdots$} & \mccc3{$\cdots$} &  & \mccc3{$\cdots$} &  & & 670 $\pm$ 34 & &  & \mccc3{$\cdots$} \\
 \rowcolor[gray]{0.95}
 2006-08-13 &  690 & \mccc3{$\leq$ 49} & & 45.9 $\pm$ 15.3 & & & 115 $\pm$ 38 & & & 342 $\pm$ 114 & &  & \mccc3{$\cdots$} &  & \mccc3{$\cdots$} &  & \mccc3{$\cdots$} 
\\
 \rowcolor[gray]{0.95}
 2006-09-01 &  709 & \mccc3{$\cdots$} & \mccc3{$\cdots$} & \mccc3{$\cdots$} & \mccc3{$\cdots$} &  & \mccc3{$\cdots$} &  & \mccc3{$\cdots$} &  & & 663 $\pm$ 77 & \\
  \rowcolor[gray]{0.95}
2006-09-10 &  718 & \mccc3{$\cdots$} & \mccc3{$\cdots$} & \mccc3{$\cdots$} & \mccc3{$\cdots$} &  & \mccc3{$\cdots$} &  & & 562 $\pm$ 30 & &  & \mccc3{$\cdots$} \\
 2006-12-29 &  828 & \mcc3{$\leq$ 45} & & 29.4 $\pm$ 9.8 & & \mcc3{$\leq$ 87} & & 145 $\pm$ 48 & &  & \mcc3{$\cdots$}&  & \mcc3{$\cdots$} &  & \mcc3{$\cdots$} \\
 2007-01-21 &  851 & \mcc3{$\cdots$} & \mcc3{$\cdots$} & \mcc3{$\cdots$} & \mcc3{$\cdots$} &  & \mcc3{$\cdots$} &  & \mcc3{$\cdots$} &  & & 513 $\pm$ 54 & \\
 2007-01-27 &  857 & \mcc3{$\cdots$} & \mcc3{$\cdots$} & \mcc3{$\cdots$} & \mcc3{$\cdots$} &  & \mcc3{$\cdots$} &  & & 446 $\pm$ 31 & &  & \mcc3{$\cdots$} \\

 \rowcolor[gray]{0.95}
 2007-06-26 & 1007 & \mccc3{$\cdots$} & \mccc3{$\cdots$} & \mccc3{$\cdots$} & \mccc3{$\cdots$} &  & \mccc3{$\cdots$} &  & & 324 $\pm$ 29 & &  & \mccc3{$\cdots$} \\
 \rowcolor[gray]{0.95}
 2007-07-03 & 1015 & \mccc3{$\leq$ 23} & & 14.1 $\pm$ 4.7 & & \mccc3{$\leq$ 83} & & 113 $\pm$ 37 & &  & \mccc3{$\cdots$} &  & \mccc3{$\cdots$} &  & \mccc3{$\cdots$} \\
 \rowcolor[gray]{0.95}
 2007-07-09 & 1020 & \mccc3{$\cdots$} & \mccc3{$\cdots$} & \mccc3{$\cdots$} & \mccc3{$\cdots$} &  & \mccc3{$\leq$ 400} &  & \mccc3{$\cdots$} &  & \mccc3{$\cdots$} \\
 \rowcolor[gray]{0.95}
 2007-07-10 & 1021 & \mccc3{$\cdots$} & \mccc3{$\cdots$} & \mccc3{$\cdots$} & \mccc3{$\cdots$} &  & \mccc3{$\cdots$} &  & \mccc3{$\cdots$} &  & & 644 $\pm$ 48 & \\

 2007-08-02 & 1044 & \mcc3{$\cdots$} & \mcc3{$\cdots$} & \mcc3{$\cdots$} & \mcc3{$\cdots$} &  & \mcc3{$\cdots$} &  & & 366 $\pm$ 35 & &  & \mcc3{$\cdots$} \\
 2007-08-12 & 1054 &  \mcc3{$\leq$ 34} & & 15.1 $\pm$ 5.0 & & \mcc3{$\leq$ 78} & & 115 $\pm$ 38 & &  & \mcc3{$\cdots$} &  & \mcc3{$\cdots$} &  & \mcc3{$\cdots$} \\
 2007-08-27 & 1069 & \mcc3{$\cdots$} & \mcc3{$\cdots$} & \mcc3{$\cdots$} & \mcc3{$\cdots$} &  & \mcc3{$\cdots$} &  & \mcc3{$\cdots$} &  & & 610 $\pm$ 49 & \\

 \rowcolor[gray]{0.95}
 2007-12-09 & 1173 & \mccc3{$\cdots$} & \mccc3{$\cdots$} & \mccc3{$\cdots$} & \mccc3{$\cdots$} &  & \mccc3{$\cdots$} &  & & 1023 $\pm$ 33 & &  & \mccc3{$\cdots$} \\
 \rowcolor[gray]{0.95}
 2007-12-27 & 1191 & & 52.5 $\pm$ 17.5 & & & 189 $\pm$ 10 & & & 345 $\pm$ 75 & & & 458 $\pm$ 73 & &  & \mccc3{$\cdots$} &  & \mccc3{$\cdots$} &  & \mccc3{$\cdots$} \\
 \rowcolor[gray]{0.95}
 2008-01-07 & 1202 & \mccc3{$\cdots$} & \mccc3{$\cdots$} & \mccc3{$\cdots$} & \mccc3{$\cdots$} &  & \mccc3{$\cdots$} &  & \mccc3{$\cdots$} &  & & 1276 $\pm$ 42 & \\
 \rowcolor[gray]{0.95}
 2008-01-17 & 1212 & \mccc3{$\cdots$} & \mccc3{$\cdots$} & \mccc3{$\cdots$} & \mccc3{$\cdots$} &  & \mccc3{$\cdots$} &  & & 1055 $\pm$ 31 & &  & \mcc3{$\cdots$}\\

 2008-06-21 & 1368 & \mcc3{$\cdots$} & \mcc3{$\cdots$} & \mcc3{$\cdots$} & \mcc3{$\cdots$} &  & & 1036 $\pm$ 212 & &  & \mcc3{$\cdots$} &  & \mcc3{$\cdots$} \\

 2008-07-09 & 1386 & \mcc3{$\cdots$} & \mcc3{$\cdots$} & \mcc3{$\cdots$} & \mcc3{$\cdots$} &  & & 1016 $\pm$ 224 & &  & \mcc3{$\cdots$} &  & \mcc3{$\cdots$} \\

 2008-07-18 & 1395 & & 76.3 $\pm$ 22.6 & & & 258 $\pm$ 11 & & & 465 $\pm$ 49 & & & 578 $\pm$ 82 & &  & \mcc3{$\cdots$} &  & \mcc3{$\cdots$} &  & \mcc3{$\cdots$} \\
 2008-07-29 & 1406 & \mcc3{$\cdots$} & \mcc3{$\cdots$} & \mcc3{$\cdots$} & \mcc3{$\cdots$} &  & \mcc3{$\cdots$} &  & \mcc3{$\cdots$} &  & & 1563 $\pm$ 54 & \\
2009-08-06\tnote{4} & 1779 & & 17.3 $\pm$ 7.6 & & & 113 $\pm$ 8 & & \mcc3{$\cdots$} & \mcc3{$\cdots$}  & & \mcc3{$\cdots$} &  & \mcc3{$\cdots$} &  & \mcc3{$\cdots$} \\
2010-01-05\tnote{4} & 1931 & & 14.4 $\pm$ 6.9 & & & 90.3 $\pm$ 16.8 & & \mcc3{$\cdots$} & \mcc3{$\cdots$}  & & \mcc3{$\cdots$} &  & \mcc3{$\cdots$} &  & \mcc3{$\cdots$} \\
2010-08-13\tnote{4} & 2151 & & 7.2 $\pm$ 3.0 & & & 76.7 $\pm$ 5.3 & & \mcc3{$\cdots$} & \mcc3{$\cdots$}  & & \mcc3{$\cdots$} &  & \mcc3{$\cdots$} &  & \mcc3{$\cdots$} \\

\hline
\mcl2{Pre-explosion} & & 83.8 $\pm$ 10.2 & & & 45.4 $\pm$ 8.3 & & & 178 $\pm$ 25 & & & 412 $\pm$ 51 & & & \mcc3{$\cdots$} & & & 221 $\pm$ 22 & & & & 376 $\pm$ 25 & \\

\hline

\end{tabular}
\begin{tablenotes}
\scriptsize
\item []
\item [1] All post-explosion flux densities were measured with PSF-fitted
  photometry (using IRAF {\tt daophot}). Pre-explosion IRAC and
  MIPS fluxes were measured in an aperture of radius 5\arcsec\ with
  sky annuli at inner radius 7\farcs5 and outer radius 10\arcsec\
  respectively, using a 2-\sigjo\ clipped-mean sky algorithm (using
  IRAF {\tt phot}).
\item [2] Flux upper limits for the non-detections in the
  Gemini-Michelle data are 3-\sigjo\ values based on the standard
  deviation of the background in the region of the SN position scaled
  to a diffraction-limited size aperture. 
\item [3] \spitzer\ IRAC, MIPS and IRS-PUI data have had pre-explosion
  flux levels subtracted: IRAC and MIPS by use of difference imaging
  techniques to subtract pre-explosion SINGS images, and IRS-PUI by
  estimating the pre-explosion level from blackbody fits to the SED at
  day 1015 (see text).  Measured (IRAC and MIPS) and estimated
  (IRS-PUI) pre-explosion fluxes at the position of the SN are
  summarised in the last row of the table.
\item [4] Observations at days 1779, 1931 and 2151 were obtained
  during the \spitzer\ `warm' mission, where only IRAC 3.6 and
  4.5\,\micron\ channels were available.

\end{tablenotes}
\end{threeparttable}
\end{spacing}
\end{table*}

 \begin{figure*}\centering
   \includegraphics[scale=0.75,angle=0,clip=true]
   {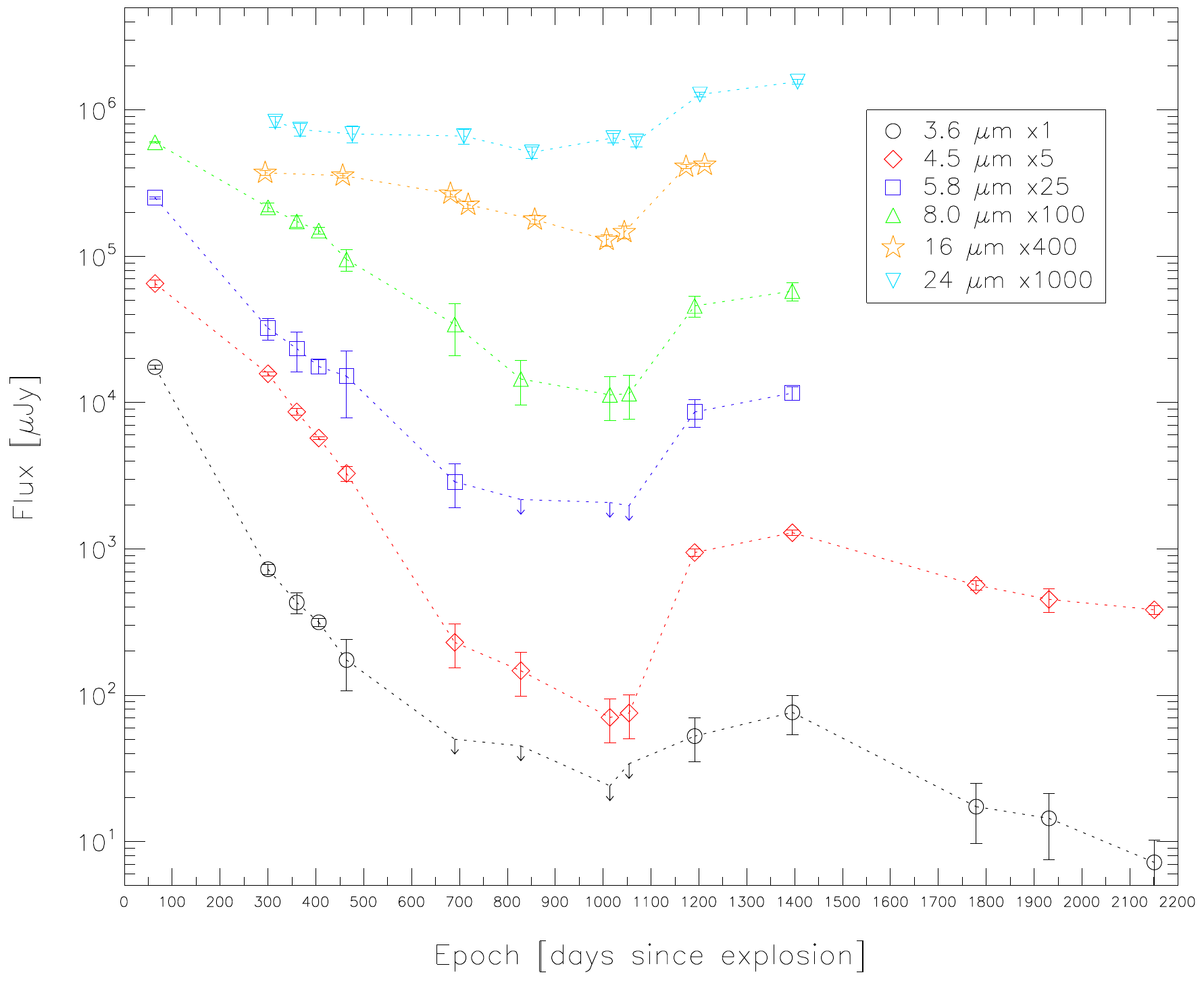} \vspace{1ex}
  {\caption[\spitzer\ mid-infrared light curves of SN
     2004et]{\spitzer\ mid-infrared light curves of SN 2004et: 3.6,
       4.5, 5.8 and 8.0\,\micron\ (IRAC), 16\,\micron\ (IRS Peak-Up
       Imaging) and 24\,\micron\ (MIPS). The IRAC and MIPS flux
       densities are from PSF-fitted photometry carried out on
       difference images which use the pre-explosion image (day -104
       for IRAC and day -75 for MIPS) as the reference image. The
       16\,\micron\ flux densities are from PSF-fitted photometry of
       the original images from which an estimated pre-explosion flux
       was subtracted (see text for details). For non-detections,
       upper limits to the flux densities are indicated by the
       downward-pointing arrows. 
       For clarity the light curves have been shifted vertically by
       the factors indicated.}\label{fig:04et_irac_pui_mips_lcurves}}
 \end{figure*}

\subsection{Evolution of the mid-IR emission}
\label{sec:04et_midir_results}

{\tablename}~\ref{tab:04et_midir_fluxes_psffit_diff} lists the
complete set of \spitzer\ and Gemini mid-infrared flux densities and
associated uncertainties/upper-limits of SN~2004et from days 64 to
2151 as determined from the PSF-fitting techniques described by
\cite{fabbri11}.  \spitzer\ IRAC, MIPS and IRS-PUI data have had
pre-explosion flux levels subtracted.  Measured (IRAC and MIPS) and
estimated (PUI) pre-explosion flux densities at the position of the SN
are summarised in the last row of the table.

The IRAC, PUI 16-$\mu$m and MIPS 24-$\mu$m light curves (with
pre-explosion levels subtracted) are shown in
{\figurename}~\ref{fig:04et_irac_pui_mips_lcurves}.  For clarity, the
different light curves have been arbitrarily shifted by the factors
shown.  Upper limits to the flux densities at 3.6~$\mu$m and
5.8~$\mu$m are indicated by the downward pointing arrows.  Each
waveband demonstrates the decline in brightness from the earliest
epochs to around day 800 when the SN has faded or is fading to its
faintest levels. At 3.6~$\mu$m and 5.8~$\mu$m, the upper limits
measured from the difference images indicate that the SN faded to
below background levels for \simjo~200 days. The distinctive rise in
brightness after this time ($>$~1000 days) is evident in all
wavebands. The latest \spitzer\ data at days 1779, 1931 and 2151 were
obtained during the post-cryogenic phase of the mission, where only
the shortest wavelength IRAC channels at 3.6 and 4.5~$\mu$m were
available. They show that sometime between days 1395 and 1779, the
mid-IR brightness of the SN at 3.6 and 4.5~$\mu$m began to decline
again, continuing with a slower decline to day 2151. The 3.6~$\mu$m
flux at day 2151 is about 9\,\% higher than the pre-explosion level,
whilst the 4.5-$\mu$m flux at the latest epoch is a factor of 2.7
brighter than the pre-explosion level.  Whilst the 3.6\,\micron\ flux
densities at days 1779 to 2151 are lower than the upper limits between
days 690 and 1015, the on-source integration times for the later
observations were between factors of 3 and 80 longer than those for
the earlier observations. The corresponding increase in
signal-to-noise for the IRAC `warm' images, together with difference
imaging techniques, allowed the SN to be reliably detected at deeper
levels than for previous observations with shorter exposure times when
the SN was also faint.

\begin{figure*}\centering
  \includegraphics[scale=0.54,angle=0,clip=true]
  {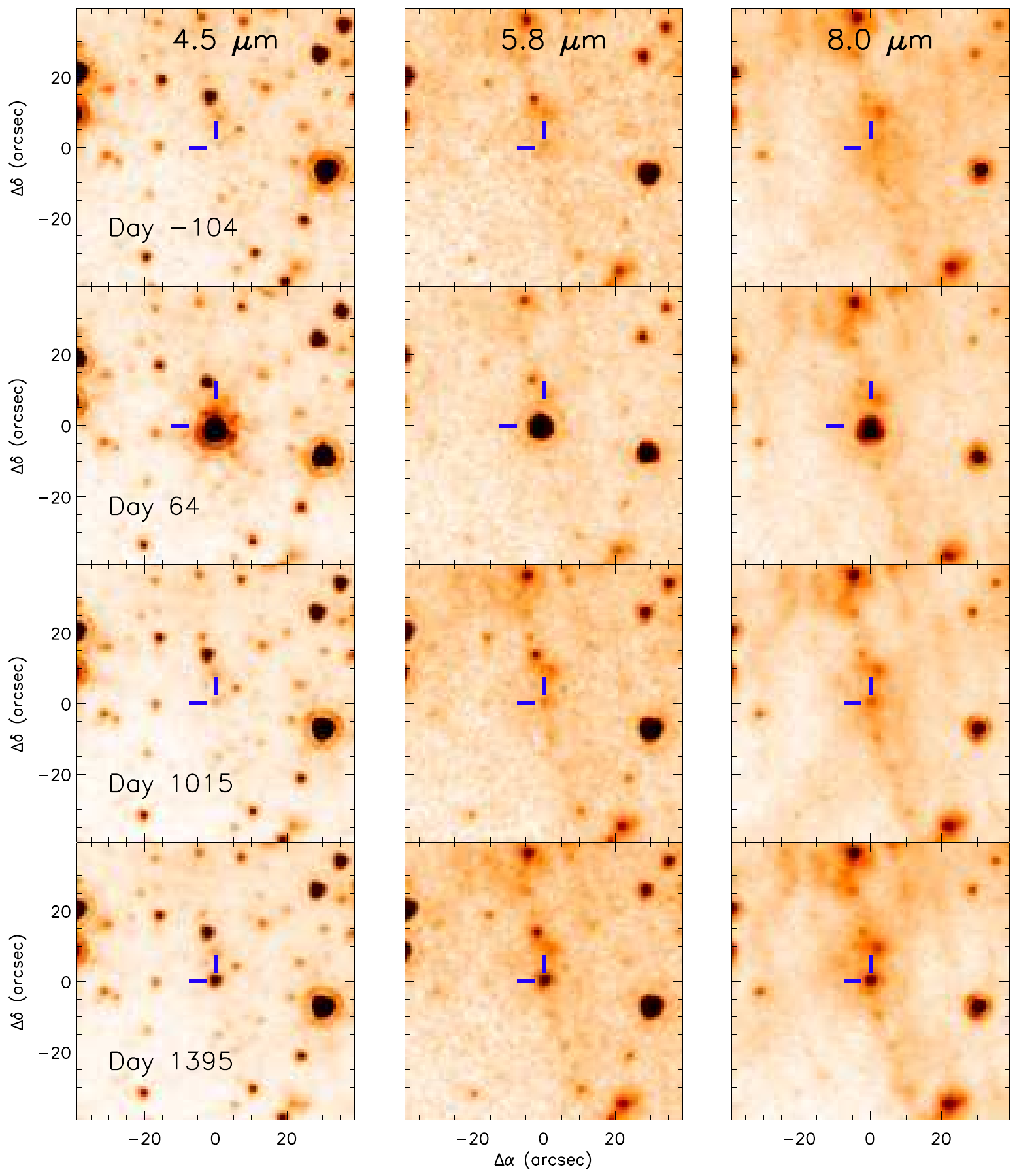} \vspace{2ex}
  \includegraphics[scale=0.54,angle=0,clip=true]
  {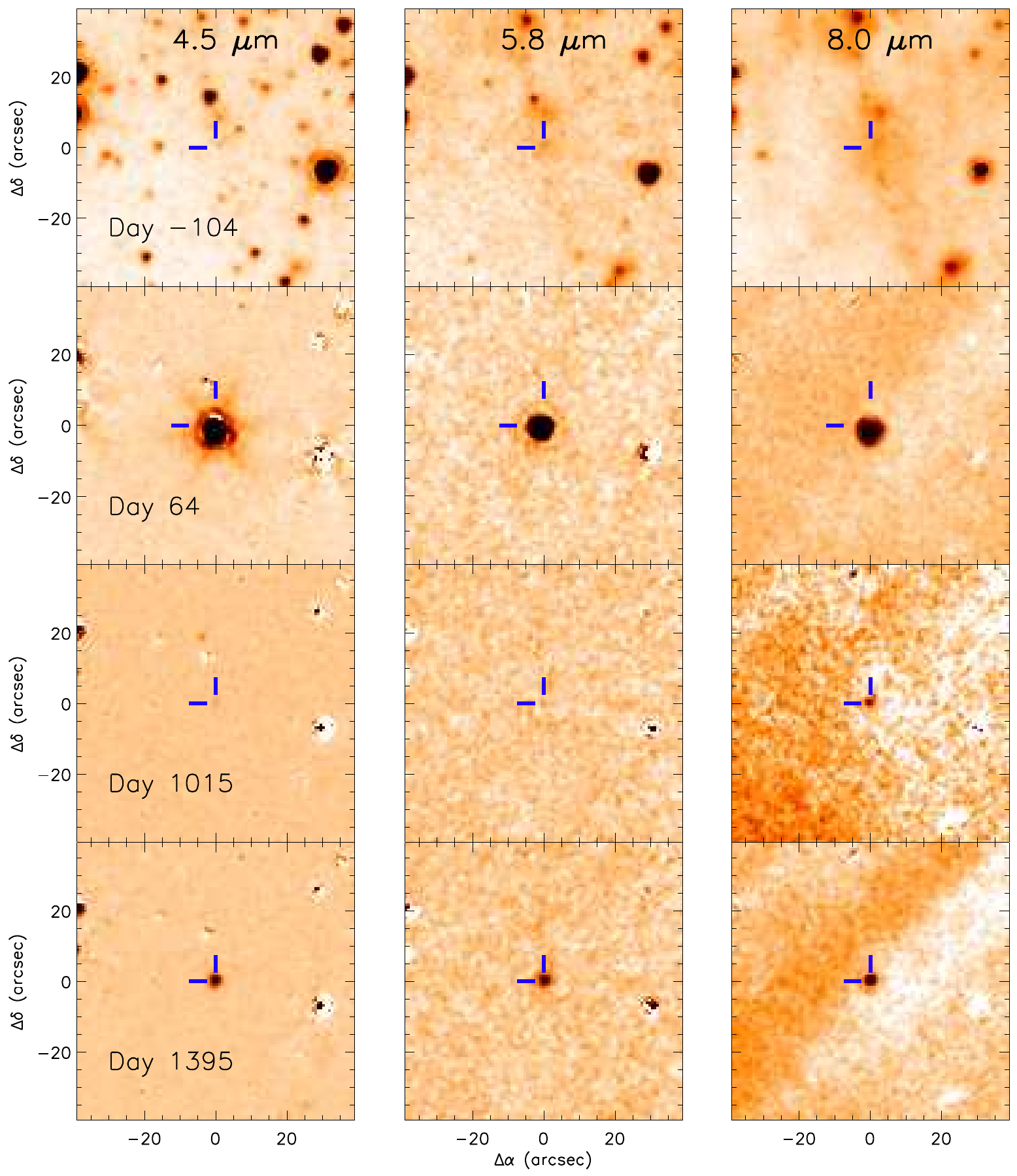} \vspace{2ex}
    {\caption[The SN position in \spitzer\ images at
    selected epochs, showing the mid-IR evolution in IRAC bands 4.5,
    5.8 and 8.0~$\mu$m.]{Leftmost three columns: the SN position in 
      \spitzer\ images at
      selected epochs, showing the mid-IR evolution in IRAC bands 4.5,
      5.8 and 8.0$\mu$m.  The first row shows the SN field in the
      pre-explosion SINGS images at day -104. The first mid-IR
      detections were obtained 64 days after explosion when the SN was
      still very bright (second row). The third row shows that the SN
      had faded to almost pre-explosion levels by day 1015, yet by day
      1395 the SN has brightened again (fourth 
      row). Rightmost three columns: difference images for the same epochs
      and wavelengths.  The pre-explosion images
      at day $-$104 (top row) have been registered to, and 
      subtracted
      from, the post-explosion images using PSF-matched difference
      imaging techniques (see
      {\secname}~\ref{ssec:04et_midir_reduce}). Note the strong,
      uneven background residuals at
      8.0~$\mu$m.}\label{fig:04et_irac_orig_diff}}
\end{figure*}

A selection of pre- and post-explosion IRAC images at 4.5~$\mu$m,
5.8~$\mu$m and 8.0~$\mu$m are shown in
{\figurename}~\ref{fig:04et_irac_orig_diff}, depicting the
mid-infrared evolution of SN~2004et. The supernova position is shown
in the pre-explosion SINGS images at day -104 revealing evidence of
extended emission in this region.  The first mid-IR images of the SN,
obtained 64 days after explosion (second row of figure), showed the SN
to be very bright. This was during the photospheric plateau phase
which characterises Type II-P supernovae, where hot blackbody emission
dominates the optical emission and its Rayleigh-Jeans tail extends
into the infrared. By day 1015 (third row), it can be seen that the SN
has faded to almost pre-explosion levels, but a late rise in
brightness is clearly evident by day 1395.  Comparable difference
images depicting the net mid-IR emission at the SN position for the
same epochs are also shown in
{\figurename}~\ref{fig:04et_irac_orig_diff}. The rightmost three boxes
in the top row again show the pre-explosion images at day -104, which
were used as the reference images for subtraction from the
post-explosion images, to yield the difference images shown in the
remaining panels. At day 1015 when the SN has faded to its faintest
levels, a detection at 4.5~$\mu$m can just be discerned. For the same
epoch at 5.8~$\mu$m, whilst there is positive emission coincident with
the position of the SN, this is at a similar level to the average
noise levels in the residual background of the difference image and is
therefore considered to be a non-detection for which an upper limit to
the flux is derived. The SN is much more clearly detected in the
8.0-$\mu$m difference image at day 1015, although the irregular
residual background, seen as diagonal bands across all of the
difference images as this wavelength, creates relatively large
uncertainties in the final measured flux.

For the 16-$\mu$m \spitzer-PUI data, the pre-explosion flux estimated
by \cite{kotak09} was \simjo\,56\,\% higher than that used here and so
we obtain generally higher post-explosion fluxes than those presented
by \cite{kotak09}. The factor by which our post-explosion fluxes
exceed those of \cite{kotak09} varies from \simjo\,1.1 to 3.3. The
largest differences in the photometry occur for days 1007--1044, when
the SN was close to its faintest levels in this waveband. During the
final epochs observed (days 1212--1173), when the rebrightening was
strongest, the fluxes agree to within 10\%.

For the 24-$\mu$m \spitzer-MIPS data, \cite{kotak09} measured the flux
in the two SINGS pre-explosion images observed at days -75 and -73, as
processed with the standard \spitzer\ pipeline. They found the flux at
day -75 to be a factor of almost 1.3 higher than that at day -73. From
these they measured an average pre-explosion flux which is almost
40\,\% lower than the pre-explosion flux measured here.  As discussed
in {\secname}~\ref{ssec:04et_midir_reduce}, the pre-explosion MIPS 
image used here for the analysis of SN~2004et was the SINGS enhanced
mosaic from their 5th data delivery, which combined data from both
dates.  We found a flux consistent with that of \cite{kotak09} at day
-73 but at day -75 we measured the flux to be about a factor of 1.7
higher.  We also measured the fluxes of a number of nearby isolated
point sources that were present in both images. No systematic offset
was found, with the photometry differing by no more than 1-10\%.

Despite measuring a higher pre-explosion flux than \cite{kotak09} at
24~$\mu$m, we find that photometry from our difference images yielded
generally \textit{higher} post-explosion fluxes than theirs.
Different techniques/software were used to produce the final
difference images, which may have resulted in different background
residuals affecting the photometry.  In addition, \cite{kotak09} used
aperture photometry to measure their fluxes, whereas we used
PSF-fitting. A comparison between results from aperture photometry and
PSF-fitting is presented by \cite{fabbri11}.  In addition, different
techniques may have been used to interpolate the 16- and 24-$\mu$m
data to the IRAC epochs used for the SED analysis.

\section{Optical and near-infrared photometry}
\label{sec:04et_opt_nir_phot}

\subsection{The observations}
\label{ssec:04et_opt_nir_obs}

Optical and NIR photometric observations of SN~2004et were obtained as
part of the SEEDS program over the years 2004 to 2009, spanning
79--1803 days after explosion.

The first two epochs of optical photometry were obtained during the
plateau phase of the SN at days 79 and 89 with the 32-inch Tenagra II
telescope in Arizona. A further three epochs of optical photometry
were obtained during the nebular phase between days 317 and 664 with
the Gemini Multi-Object Spectrograph on Gemini-North (GMOS-N) in
Hawaii. In addition, an archival flux-calibrated Subaru spectrum was
used to obtain optical photometry at day 646 by integrating over the
$BVRI$ filter transmission curves. Two epochs of late-time, high
resolution observations of the SN field were obtained with the \hst\
Wide Field Planetary Camera 2 (WFPC2) around three years after
explosion at days 1054 and 1215. A final epoch of optical photometry,
almost 4 years after explosion, was obtained with GMOS-N at day 1412.

The first epoch of NIR photometry was obtained at day 268 with the
2.3-m Bok telescope, part of the Steward Observatory at Kitt Peak,
Arizona. This was followed by three epochs of data from the Near
InfraRed Imager (NIRI) on Gemini-North at dates close in time to the
optical images taken with GMOS-N during the nebular phase. Late-time,
high-resolution data was obtained with the \hst\ Near Infrared Camera
and Multi-Object Spectrometer 2 (NICMOS2), at epochs corresponding to
those of the optical WFPC2 data.  A final $H$-band image was taken
approximately 5 years (day 1803) after explosion with the WIYN
High-resolution InfraRed Camera: WHIRC \citep{meixner10}, on the WIYN
3.5-m telescope at Kitt Peak, Arizona.

{\tablename}~\ref{tab:04et_opt_nir_obslog} provides a complete log of
the optical and NIR photometric observations of SN~2004et taken as
part of the SEEDS project. {\appname}~\ref{ssec:04et_opt_nir_reduce}
describes how the optical and near-IR data were processed, while
{\appname}~\ref{ssec:04et_opt_nir_hst} describes our late-time {\em
  HST} optical and near-IR images which revealed the single point
source seen at the SN position in IRAC images to be comprised of at
least three sources.

\begin{table*}\centering
  \parbox[t]{13.0cm}{\protect\caption[Log of optical and near-infrared photometric observations]{Log of optical and near-infrared photometric observations of SN~2004et from the SEEDS collaboration.\label{tab:04et_opt_nir_obslog}}}
\setlength{\tabcolsep}{1.5mm}
{\footnotesize
\begin{tabular}{lcllcll}
  \hline\hline
  & & & & & & \\    
  Date & Age & Telescope/ & Filters & Exp. time & Program ID & Principal \\
  & [days] & instrument &      &        &            & Investigator \\

  & & & & & & \\ \hline
  & & & & & & \\
2004-12-10 & 79   & Tenagra II 32'' & $VRI$ & 9$\times$100\,s & -- & D. Welch \\
2004-12-20 & 89   & Tenagra II 32'' & $VRI$ & 9$\times$100\,s & -- & D. Welch \\
2005-06-17 & 268  & Bok 2.3m IR Camera & $JHK$ & 20$\times$30\,s& -- & K. Gordon\\ 
2005-08-05 & 317  & Gemini  GMOS-N  & $g^{'}r^{'}i^{'}$ & 1$\times$60\,s & GN-2005B-Q-54 & G. Clayton \\
2005-08-05 & 317  & Gemini NIRI    & $JHK$ & 22$\times$30\,s & GN-2005B-Q-54 & G. Clayton \\
2005-10-17 & 390  & Gemini NIRI    & $JHK$ & 10$\times$30\,s & GN-2005B-Q-54 & G. Clayton \\
2005-10-31 & 404 & Gemini  GMOS-N  & $g^{'}r^{'}i^{'}$ & 1$\times$60\,s & GN-2005B-Q-54 & G. Clayton \\
2006-07-06 & 652  & Gemini NIRI    & $JHK$ & 22$\times$30\,s & GN-2006A-Q-52 & G. Clayton \\
2006-07-18 & 664  & Gemini GMOS-N  & $g^{'}r^{'}i^{'}$ & 1$\times$60\,s & GN-2006A-Q-52 & G. Clayton \\
2007-07-08 & 1019 & \hst\ WFPC2    & {\em F606W,F814W} & 4$\times$400\,s & GO11229 & M. Meixner \\
2007-07-08 & 1019 & \hst\ NICMOS2 & {\em F110W,F205W} & 5$\times$128\,s & GO11229 & M. Meixner \\
2007-07-08 & 1019 & \hst\ NICMOS2 & {\em F160W} & 4$\times$128\,s & GO11229 & M. Meixner \\
2008-01-20 & 1215 & \hst\ WFPC2    & {\em F606W,F814W} & 4$\times$400\,s & GO11229 & M. Meixner \\
2008-01-20 & 1215 & \hst\ NICMOS2 & {\em F110W,F205W} & 5$\times$128\,s & GO11229 & M. Meixner \\
2008-01-20 & 1215 & \hst\ NICMOS2 & {\em F160W} & 4$\times$128\,s & GO11229 & M. Meixner \\
2008-08-04 & 1412 & Gemini GMOS-N  & $g^{'}r^{'}i^{'}$ & 2$\times$600\,s & GN-2008B-Q-44 & M. Barlow \\
2009-08-30 & 1803 & WIYN WHIRC   & $H$ & 4$\times$180\,s & 2009B-0516 & M. Otsuka \\
  & & & & & & \\ \hline
\end{tabular}
}
\end{table*}


\subsection{Photometry and light curve evolution}
\label{ssec:04et_opt_nir_lcurves}

\begin{figure*}\centering
   \includegraphics[scale=0.65,angle=0,clip=true]
   {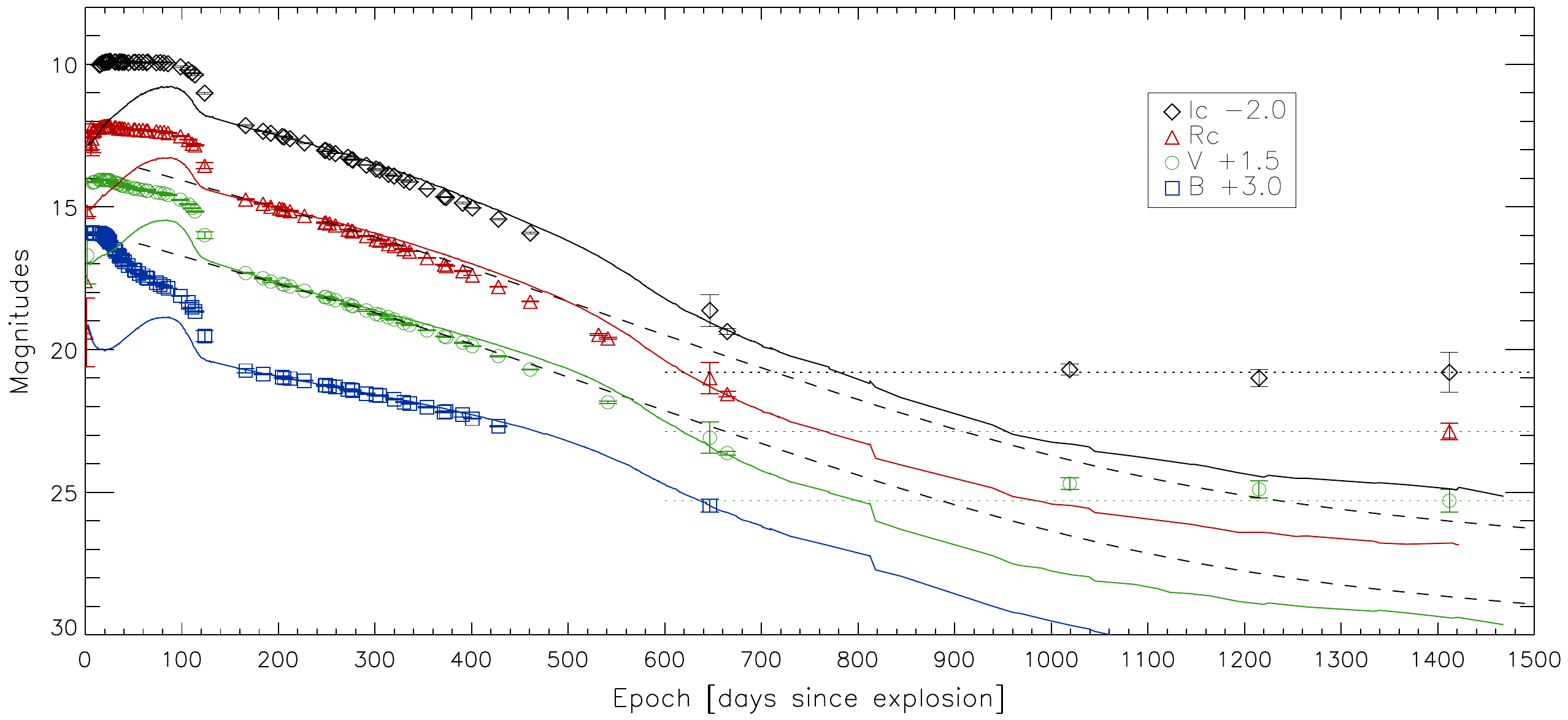} \vspace{1ex}
       {\caption[$BVRI$ light curves of SN~2004et.]{$BVRI$
       light curves of SN~2004et, based on the data of Sahu et al.
       (2006) and the SEEDS photometry listed in 
       {\tablename}~\ref{tab:04et_opt_mag}.
       For clarity, the plots have been
       vertically shifted by the amounts shown in the legend. The
       black dashed lines are the expected light curves based on the
       decay of $^{56}$Co and other isotopes \citep{woosley89}
       normalised to the $V$ and $R$ band magnitudes of SN~2004et
       during the early nebular phase (between \simjo 160--260 days).
       The solid lines are the corresponding light curves of SN~1987A
       \citep{hamuy90,walker91} normalised to those of SN~2004et at
       \simjo\,200 days.  Dotted lines are only to guide the eye along
       the horizontal or ``flat'' parts of the light curve. These
       curves are typical for every SN that shows a light echo, in
       that the light curves level off or ``flatten'' once the SN
       flux drops below that of the
       echo.}\label{fig:04et_BVRI_lcurves}}
\end{figure*}

\begin{figure*}\centering
   \includegraphics[scale=0.65,angle=0,clip=true]
   {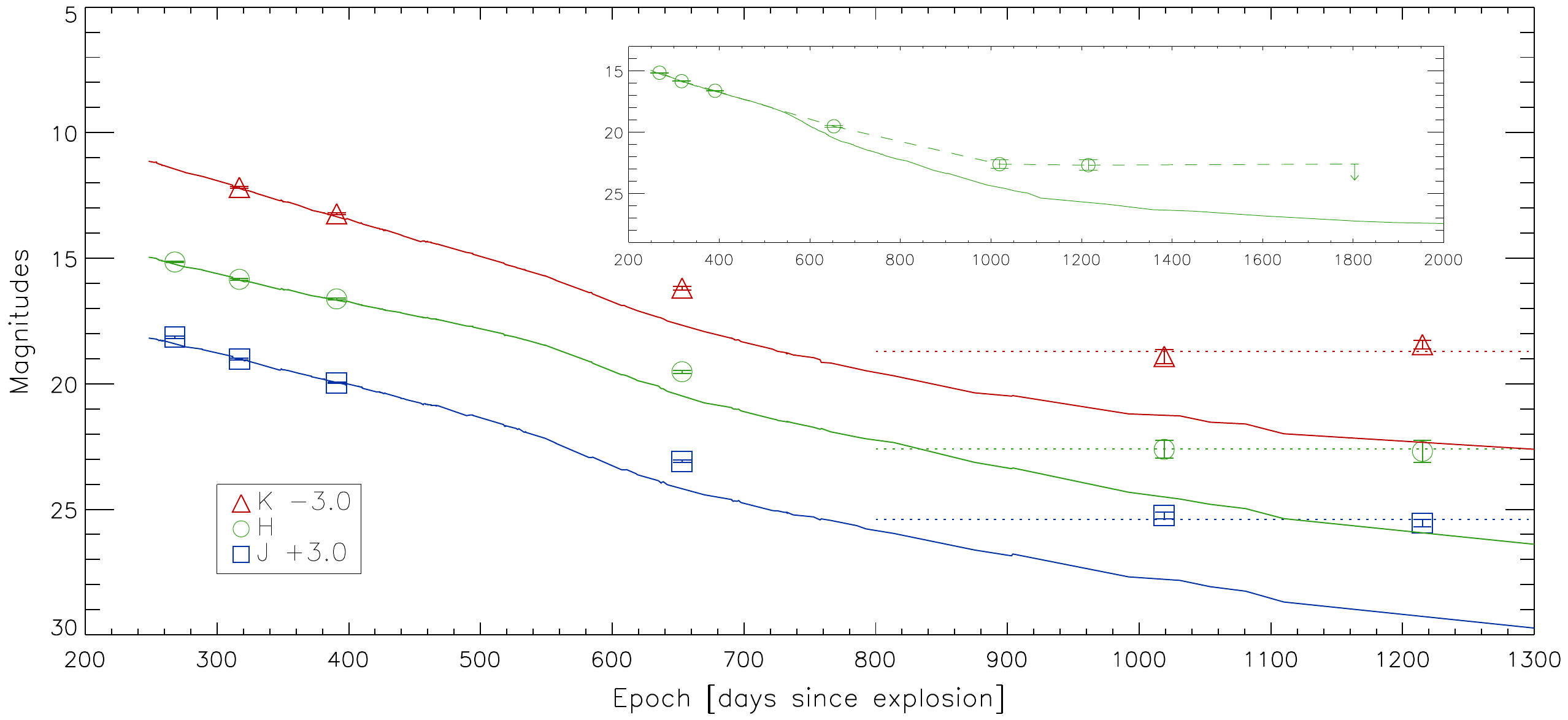} \vspace{1ex}
    {\caption[Late-time NIR light curves of
       SN~2004et.]{Late-time NIR light curves of SN~2004et,
       based on the SEEDS photometry listed in 
       {\tablename}~\ref{tab:04et_nir_mag}.
       For clarity, the plots have been vertically shifted by the
       amounts shown in the legend. The solid lines are the
       corresponding light curves of SN~1987A \citep{bouchet93}
       normalised to the light curves of SN~2004et at day 317, during
       the early nebular phase. Dotted lines are only to guide the eye
       along the horizontal or ``flat'' parts of the light curve. The
       inset shows the $H$-band light curve only for the extended
       period to day 1803, the final SEEDS NIR observation, taken with
       the WIYN telescope in August 2009. Contamination from neighbouring
       sources observed in the high-resolution NICMOS images at days
       1019 and 1215 resulted in an ambiguous detection of the SN with
       the WHRIC at day 1803, for which an upper limit (downward
       pointing arrow) has been derived.
     }\label{fig:04et_nir_lcurves}}
 \end{figure*}

The final optical $VRI$ magnitudes for SN~2004et are presented in
{\tablename}~\ref{tab:04et_opt_mag}, including those of neighbouring
star\,2 discussed previously. The magnitudes for epochs from day 646
onwards have been corrected for the contribution from star\,2, but
this was not necessary for earlier epochs when the SN brightness
dominated.  The $B$-band magnitude at day 646, corrected for the
contribution from star\,2, is detailed in the notes to the table. The
NIR $JHK$ magnitudes of the supernova are given in
{\tablename}~\ref{tab:04et_nir_mag}, although the magnitudes measured
from the NICMOS images at days 1019 and 1215 are \hst\ Vegamags
(Appendix~A2). The optical and NIR light curves are presented in
Figures \ref{fig:04et_BVRI_lcurves} and \ref{fig:04et_nir_lcurves}
respectively.

\begin{table*}\centering
  \parbox[t]{10.5cm}{\protect\caption[Optical photometry of 
SN~2004et]{Optical photometry of SN~2004et.\label{tab:04et_opt_mag}}}
\setlength{\tabcolsep}{1.5mm}
{\small
\begin{threeparttable}
\begin{tabular}{lccccc}
  \hline\hline
  & & & & & \\    
 UT date & Age & \mcc3{Magnitudes} & Source \\
                & [days] & $V$ & $R_c$ & $I_c$ & \\
  & & & & & \\ \hline
  & & & & & \\
2004-12-10 &  79  & 13.09 $\pm$ 0.04 & 12.38 $\pm$ 0.03 & 11.93 $\pm$ 0.03 & Tenagra II \\
2004-12-20 &  89  & 13.21 $\pm$ 0.03 & 12.45 $\pm$ 0.03 & 11.98 $\pm$ 0.03 & Tenagra II\\
2005-08-05 &  317 & 17.35 $\pm$ 0.02 & 16.52 $\pm$ 0.03 & 15.91 $\pm$ 0.04 & Gemini GMOS-N \\
2005-10-31 &  404 & 18.28 $\pm$ 0.04 & 17.69 $\pm$ 0.03 & 16.87 $\pm$ 0.04 & Gemini GMOS-N \\
2006-06-30 &  646\tnote{\astjo} & 21.59 $\pm$ 0.55\tnote{\dag} & 21.00 $\pm$ 0.55\tnote{\dag} & 20.63 $\pm$ 0.56\tnote{\dag} & Subaru FOCAS \\
2006-07-18 &  664 & 22.13 $\pm$ 0.06\tnote{\dag} & 21.56 $\pm$ 0.09\tnote{\dag} & 21.37 $\pm$ 0.11\tnote{\dag} & Gemini GMOS-N \\
2007-07-08 & 1019 & 23.20 $\pm$ 0.20\tnote{\ddag} & $\cdots$ & 22.70 $\pm$ 0.20\tnote{\ddag} & \hst\ WFPC2 \\
2008-01-20 & 1215 & 23.40 $\pm$ 0.30\tnote{\ddag} & $\cdots$ & 23.00 $\pm$ 0.30\tnote{\ddag} & \hst\ WFPC2 \\
2008-08-04 & 1412 & 23.80 $\pm$ 0.40\tnote{\dag} & 22.87 $\pm$ 0.28\tnote{\dag} & 22.80 $\pm$ 0.70\tnote{\dag} & Gemini GMOS-N \\ \hline
Star 2     &      & 24.2 $\pm$ 0.3 &  23.5 $\pm$ 0.5 & 22.9 $\pm$ 0.4  & \hst\ WFPC2 \\ \hline
\mcl2{Zero-magnitude flux [Jy]} &  3670.3 & 2972.3 & 2402.1 & \cite{evans93, glass99} \\
\mcl2{$\lambda_{eff}$ [\micron]} & 0.55 & 0.64 & 0.80 & \\ \hline
\end{tabular}

 \begin{tablenotes}
 \scriptsize
\item [\astjo] Optical photometry at day 646 was estimated from an
  archival Subaru-FOCAS spectrum by integrating over the $BVRI$ filter
  transmission curves. The $B$ band magnitude of SN~2004et at this
  time was 22.47 $\pm$ 0.22, corrected for an estimated $B$-band
  contribution from star\,2 of 25.46 $\pm$ 0.50 (see text).
\item [\dag] The SN magnitudes at these late epochs have been
  corrected for contamination by star 2 (whose magnitudes measured
  from the high-resolution \hst\ data are listed in the final table
  entry).
\item [\ddag] Since the SN and star 2 were resolved in the day 1019 and
  day 1215 \hst\ images, the magnitudes given for these epochs are for
  the SN alone.
 \end{tablenotes}
 \end{threeparttable}
}
\end{table*}

{\figurename}~\ref{fig:04et_BVRI_lcurves} combines the $BVRI$ data of
\cite{sahu06} with the SEEDS data described above to provide optical
light curves (open symbols) from a few days after the explosion to
almost 4 years later. The light curves are well sampled until around
day 400. Few Type~II SNe have been observed beyond \simjo\,500 days.
However, SN~1987A, the closest supernova to have occurred in the past
century, has been well studied for over 2 decades and provides
detailed light curves for comparison. The broad-band $BVRI$ light
curves of SN~1987A \citep{hamuy90,walker91}\footnote{\cite{hamuy90}
  data downloaded from the NOAO archive:
  ftp://ftp.noao.edu/sn1987a/ubvri.txt} are plotted as solid curves in
{\figurename}~\ref{fig:04et_BVRI_lcurves}, normalised to the data of
SN~2004et at around 200 days.  The light curve evolution of both
supernovae is quite similar from the early nebular phase (\simjo\,160
days) to around day 650, although beyond this time their evolution is
markedly different. The brightness of SN~2004et clearly levels off
from around day 1000 in the $VRI$ bands (the last $B$-band measurement
was at day 646), while SN~1987A continued to fade.  Interestingly, the
slope of the radioactive decay curves resemble those of SN~1987A
between days \simjo\,720 and 800, around the time when dust production
for SN~1987A was assumed to have ended \citep[day 775;][]{wooden93}.

For the first few years during the nebular phase, the light curve of
Type II SNe is predominantly powered by $\gamma$-rays from the
radioactive decay of $^{56}$Co to $^{56}$Fe, at a rate corresponding
to the $e$-folding time of the $^{56}$Co decay ($\tau_{56}=111.3$
days).  For example, the $R$-band photometry of the Type II SN~1990E
\citep{benetti94} closely follows this evolution through to
\simjo\,540 days post-explosion, suggesting that simple $^{56}$Co
decay provides a good estimate of the unextinguished $R$-band light
curve for at least that long. The expected decay rate is $\gamma$ (mag
per 100 days) = 0.98 for complete $\gamma$-ray trapping
\citep{patat94}. For SN~2004et, \cite{sahu06} found that the decline
of the broad-band $BVRI$ magnitudes during the early nebular phase
(180--310 days) was linear, with decay rates of $\gamma_B$
\simjo\,0.64, $\gamma_V$ \simjo\,1.04, $\gamma_R$ \simjo\,1.01 and
$\gamma_I$ \simjo\,1.07. \cite{maguire10} found similar results from
their own data (\simjo\, 136--300 days), with $\gamma_B = 0.64 \pm
0.02$, $\gamma_V = 1.02 \pm 0.01$, $\gamma_R = 0.92 \pm 0.01$ and
$\gamma_I = 1.09 \pm 0.01$. With the exception of the $B$ band, the
decay rates were close to that of $^{56}$Co decay, suggesting that
$\gamma$-ray trapping was efficient during this phase.

\begin{table*}\centering
  \parbox[t]{14.5cm}{\protect\caption[Near-infrared photometry of SN~2004et]{Near-infrared photometry of SN~2004et.\label{tab:04et_nir_mag}}}
\setlength{\tabcolsep}{1.5mm}
{\small
\begin{threeparttable}
\begin{tabular}{lccccc}
  \hline\hline

  & & & & & \\    
 UT date & Age & \mcc3{Magnitudes} & Source \\
                & [days] & $J$ & $H$ & $K$ & \\

  & & & & & \\ \hline
  & & & & & \\
2005-06-17 & 268 & 15.14 $\pm$ 0.05 & 15.15 $\pm$ 0.03 & $\cdots$\tnote{a} & {\footnotesize Steward/Bok IR Camera} \\
2005-08-05 & 317 & 16.01 $\pm$ 0.03 & 15.84 $\pm$ 0.03 & 15.18 $\pm$ 0.04 & Gemini NIRI \\
2005-10-17 & 390 & 16.96 $\pm$ 0.03 & 16.62 $\pm$ 0.03 & 16.23 $\pm$ 0.04 & Gemini NIRI \\
2006-07-06 & 652  & 20.09 $\pm$ 0.05 & 19.52 $\pm$ 0.06 & 19.19 $\pm$ 0.07 & Gemini NIRI \\
2007-07-08 & 1019 & 22.25 $\pm$ 0.13\tnote{b} & 22.61 $\pm$ 0.36\tnote{b} & 21.91 $\pm$ 0.28\tnote{b} & \hst\ NICMOS \\
2008-01-20 & 1215 & 22.55 $\pm$ 0.14\tnote{b} & 22.69 $\pm$ 0.43\tnote{b} & 21.44 $\pm$ 0.18\tnote{b} & \hst\ NICMOS \\
2009-08-30 & 1803 & $\cdots$ &  $\leq$ 22.6 & $\cdots$ & WIYN WHIRC \\\hline
\mcl2{Zero-magnitude flux [Jy]} & 1656.3 & 1070.9 & 672.8 & \cite{glass99} \\
\mcl2{$\lambda_{eff}$ [\micron]} & 1.25 & 1.65 & 2.20 & \\ \hline
\end{tabular}
 \begin{tablenotes}
 \scriptsize
 \item [a] There is no $K$ band magnitude at day 268 as unusual image artifacts compromised the photometry. 
 \item [b] \hst\ Vegamags in NICMOS2 filters $F110W$ ($\simeq$ $J$
   band), $F160W$ ($\simeq$ $H$ band) and $F205W$ ($\simeq$ $K$ band)
   for days 1019 and 1215.  For each filter and epoch, the measured
   count rate ($CR$, in units of DN\,s$^{-1}$) at the position of the
   SN was converted to flux by multiplication with the $PHOTFNU$
   (Jy\,s\,DN$^{-1}$) conversion factor given in the fits header,
   where $PHOTFNU$ is the bandpass-averaged flux density for a source
   that would produce a count rate of 1\,DN$^{-1}$. $PHOTFNU$ = $1.21
   \times 10^{-6}$, $1.50 \times 10^{-6}$ and $9.69 \times 10^{-7}$
   Jy\,s\,DN$^{-1}$ for $F110W$, $F160W$ and $F205W$ respectively.
 \end{tablenotes}
 \end{threeparttable}
}
\end{table*}

However, both authors found that the optical decay rates of SN~2004et
steepened beyond \simjo\,300 days, suggesting that either the
supernova had become transparent to $\gamma$-rays and hence
$\gamma$-rays as a source of energy were escaping, or that dust was
forming within the ejecta and causing localised optical extinction, or
possibly was due to both phenomena.

To investigate this further we can look again at the radioactive
deposition.  As the ejecta expand, their
opacity to $\gamma$-rays is expected to decrease, which results in a
modified light curve of the form \citep{woosley89}:

\begin{equation}
\label{co56}
L_{56}^{\gamma}(t)\; \propto\;
e^{-t/\tau_{56}}\;[1-e^{-\kappa_{56}\; \phi_{0}\; (t_{0}/t)^{2}}],
\end{equation}

\noindent where the term in the brackets is the deposition function,
\ie\ the fraction of $\gamma$-rays deposited in the envelope;
$\kappa_{56,\gamma}$ = 0.033 cm$^{2}$\,g$^{-1}$ is the average opacity
to $^{56}$Co-decay $\gamma$-rays, and $\phi_{0}$ = 7$\times$10$^{4}$ g
cm$^{-2}$ is the column depth at the fiducial time $t_{0}$ = 11.6 days
chosen to match the bolometric light curve of SN~1987A.

After this first source of decay energy has become sufficiently weak,
other energy sources which could become important in powering the very
late time light curves are $\gamma$-rays, positrons and
electrons from the radioactive decay of $^{57}$Co,
$^{44}$Ti and $^{22}$Na. The equations that describe the energies from
all these isotopes, including $^{56}$Co, were summarised by
\cite{lih93}, following the work of \cite{woosley89}, to describe the
deposition behaviour of SN~1987A. Adopting the same deposition
behaviour for SN~2004et, the radioactive decay curve attributable to
the energy sources from these isotopes, including a term to account
for the decrease in opacity to $\gamma$-rays as the ejecta expands,
has been plotted in {\figurename}~\ref{fig:04et_BVRI_lcurves} (dashed
line) over both the $V$ and $R$ band magnitudes of SN~2004et,
normalising to the early nebular phase data (\simjo 160--260 days). As
expected, with the exception of the $B$ band, the decay rates during
the early nebular phase closely follow those of the radioactive decay
deposition. However, from about 400 days the $R$ band light curve has
clearly begun to decline more rapidly than the expected light curve
from radioactive decay deposition. The $I$ band follows a
similar trend, whereas the steepening of the decline rate appears to
occur slightly later in the $V$ band, having clearly begun sometime
between 460 and 540 days. In comparison with the expected radioactive
decay deposition behaviour of SN~1987A, as modelled by \cite{lih93},
there is evidence for a steepening decline of the light
curves, indicative of dust formation in the ejecta of SN~2004et from
around 400 days, and possibly earlier \citep{sahu06, maguire10}.
From the $V$-band light curve ({\figurename}~\ref{fig:04et_BVRI_lcurves}), 
we estimate that the difference between the 
observed and predicted light curves was 0.8 magnitudes by day 690,
if allowance is made for the effective opacity term for 
$^{56}$Co $\gamma$-rays, or 1.5 magnitudes if this term is neglected.

It is clear that the elevated brightness of SN~2004et after 1000 days 
cannot be
explained by the inclusion of isotope decays, such as $^{57}$Co and
$^{44}$Ti, which could be important at these late times. The
plateauing of the optical light curves above the expected radioactive
decay suggests an additional energy source has come into play by at
least day 1000 and is consistent with the late rise observed in the
mid-IR observations after this time. Similar phenomena observed for
other SNe have been attributed to light echoes \citep[\eg\
SN~2007od;][]{andrews10}.

{\figurename}~\ref{fig:04et_nir_lcurves} presents the late-time NIR
light curves (open symbols) of SN~2004et based on the SEEDS data
described previously. The $JHK$ light curves of SN~1987A \citep[solid
lines;][]{bouchet93} have been arbitrarily scaled to the early nebular
phase data of SN~2004et for comparison. The NIR light curves of
SN~2004et are not well-sampled but clearly deviate from those of
SN~1987A by day 646, after which time SN~2004et is systematically 
brighter. The NIR light curve evolution reflects that of the optical, with 
a relative plateau in brightness occurring from around 1000 days,
consistent with a light echo hypothesis. 

However, by day 1803 (see inset in
{\figurename}~\ref{fig:04et_nir_lcurves}), the $H$-band brightness has
faded to beyond a clear detection with the WHRIC detector on the 3.5-m
WIYN telescope. The derived upper limit of 22.6 mag accounts for
contamination by neighbouring stars (as described in
{\secname}~\ref{ssec:04et_opt_nir_reduce}) and suggests the SN faded
sometime after day 1215.

\section{Spectral energy distribution analysis based on optical, NIR and 
mid-IR photometry}\label{sec:04et_seds}

The mid-infrared photometry listed in
{\tablename}~\ref{tab:04et_midir_fluxes_psffit_diff}, along with the
optical and IR photometry in {\tablename}~\ref{tab:04et_opt_mag}
and {\tablename}~\ref{tab:04et_nir_mag} and the optical
photometry of \cite{sahu06}, have been used to investigate the
spectral energy distribution of SN~2004et at each of the IRAC 
observation epochs from days 64 to 1395.

\subsection{Blackbody fitting}
\label{ssec:04et_bbfits}

To investigate the physical processes that determine the observed
optical and infrared continuum emission and their evolution,
blackbodies were matched to the SEDs at each of the IRAC epochs from
day 64 to day 1395. Where necessary, the light curves for the optical,
NIR, PUI 16-$\mu$m and MIPS 24-$\mu$m data were used to
interpolate their measured flux densities to the epochs of the IRAC data.
It should be noted that to extrapolate the last $B$ magnitude, obtained
on day 646 to the closest IRAC epoch at day 690, the better-sampled
$V$-band decline rate during this period was adopted in order to
account for the gradual flattening of the light curve. 
Similarly, the $JHK$ magnitudes from day 652
were extrapolated to the closest IRAC epoch of day 690 assuming the
decline rate observed in the $I_{c}$-band during this period.  Gemini
Michelle flux densities at 11.2~$\mu$m were not interpolated due to
insufficient data but, where available, they are compared with the
closest IRAC epoch.

The interpolated optical data were converted from the standard
Johnson-Cousins $BVR_{c}I_{c}$ magnitudes to flux densities using the
zero-magnitude flux densities of \cite{evans93} and \cite{glass99}. 
The interpolated Steward
and Gemini NIR data were converted from standard $JHK$
magnitudes to flux densities using the zero-magnitude flux densities of
\cite{glass99}. The late-time \hst-NICMOS flux densities were obtained by
multiplying the count rate measured for the SN by the $PHOTFLAM$
(erg\,cm$^{-2}$\,$\AA^{-1}$\,DN$^{-1}$) conversion factor from the
fits image headers, where $PHOTFLAM$ is the bandpass-averaged flux
density in $F_{\lambda}$ for a source that would produce a count rate
of 1 DN\,s$^{-1}$. All flux densities were de-reddened using 
\textit{E}(\textit{B\,--\,V})~=~0.41 mag \citep{zwitter04} and assuming 
the extinction law of \cite{cardelli89} with $R_{V} = 3.1$, corresponding 
to $A_{V}=1.27 \pm 0.22$ mag.

\subsubsection{Day 64}

\begin{figure*}
\centering
\subfigure{\includegraphics[scale=0.3]{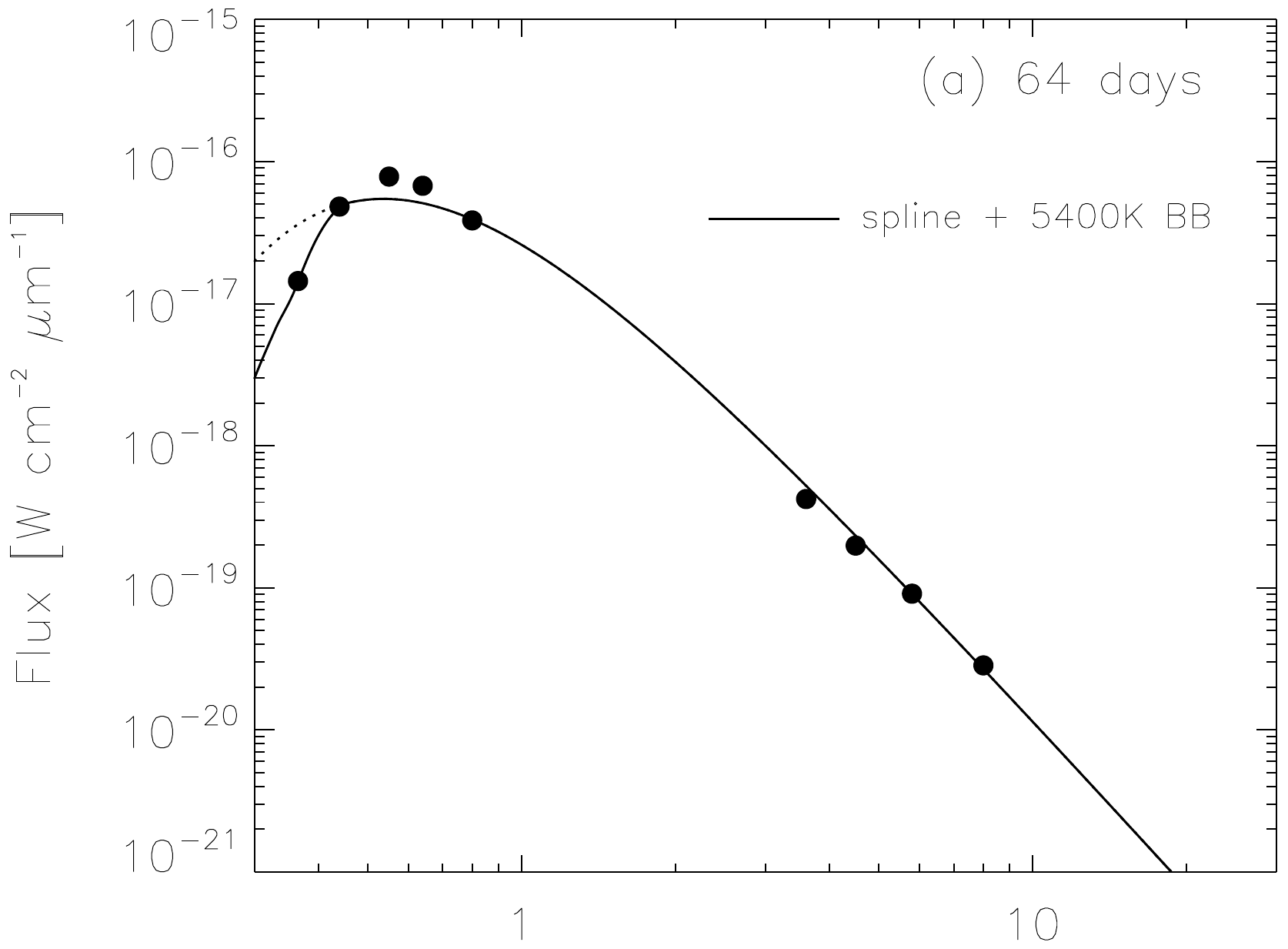}}
\subfigure{\includegraphics[scale=0.3]{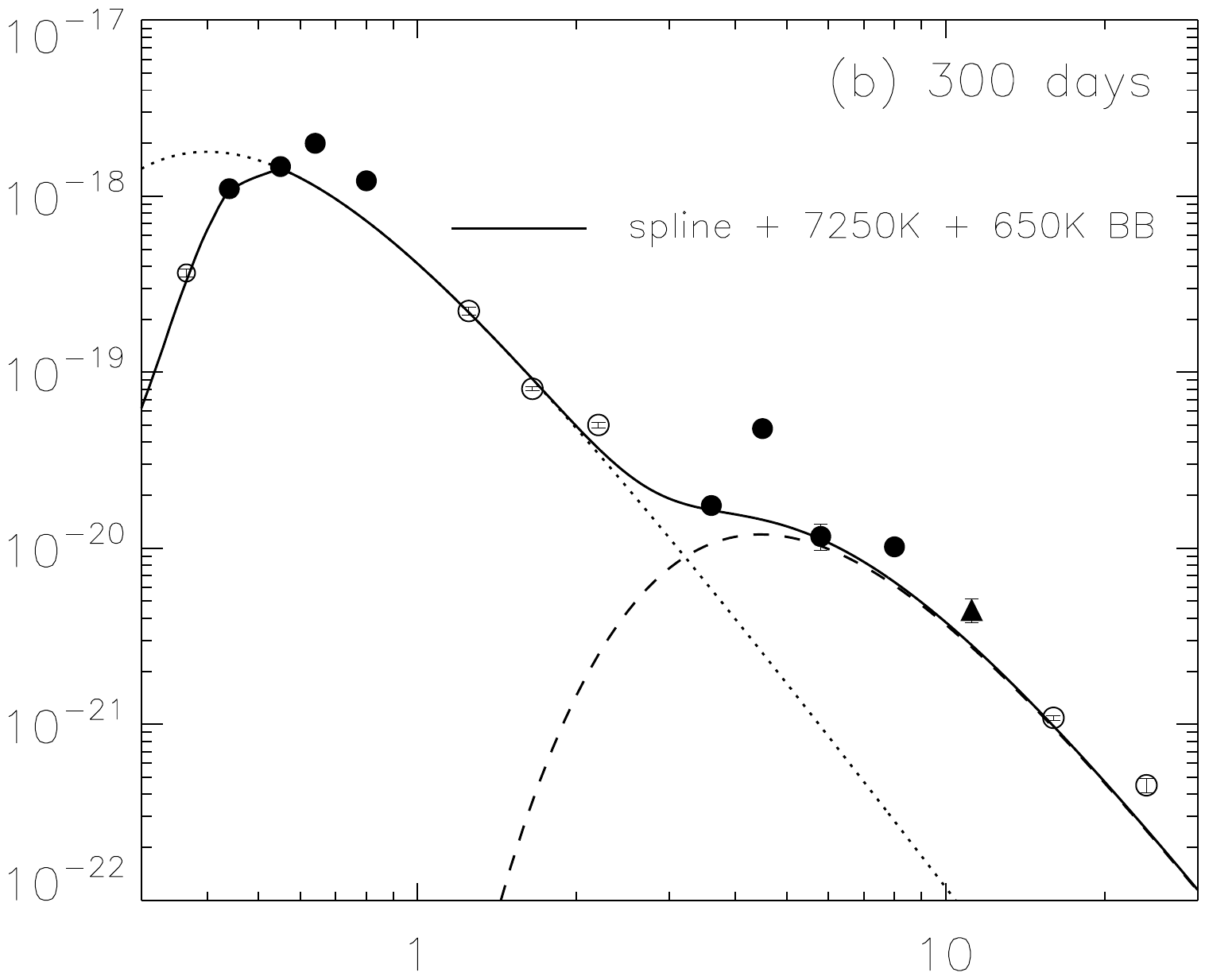}}
\subfigure{\includegraphics[scale=0.3]{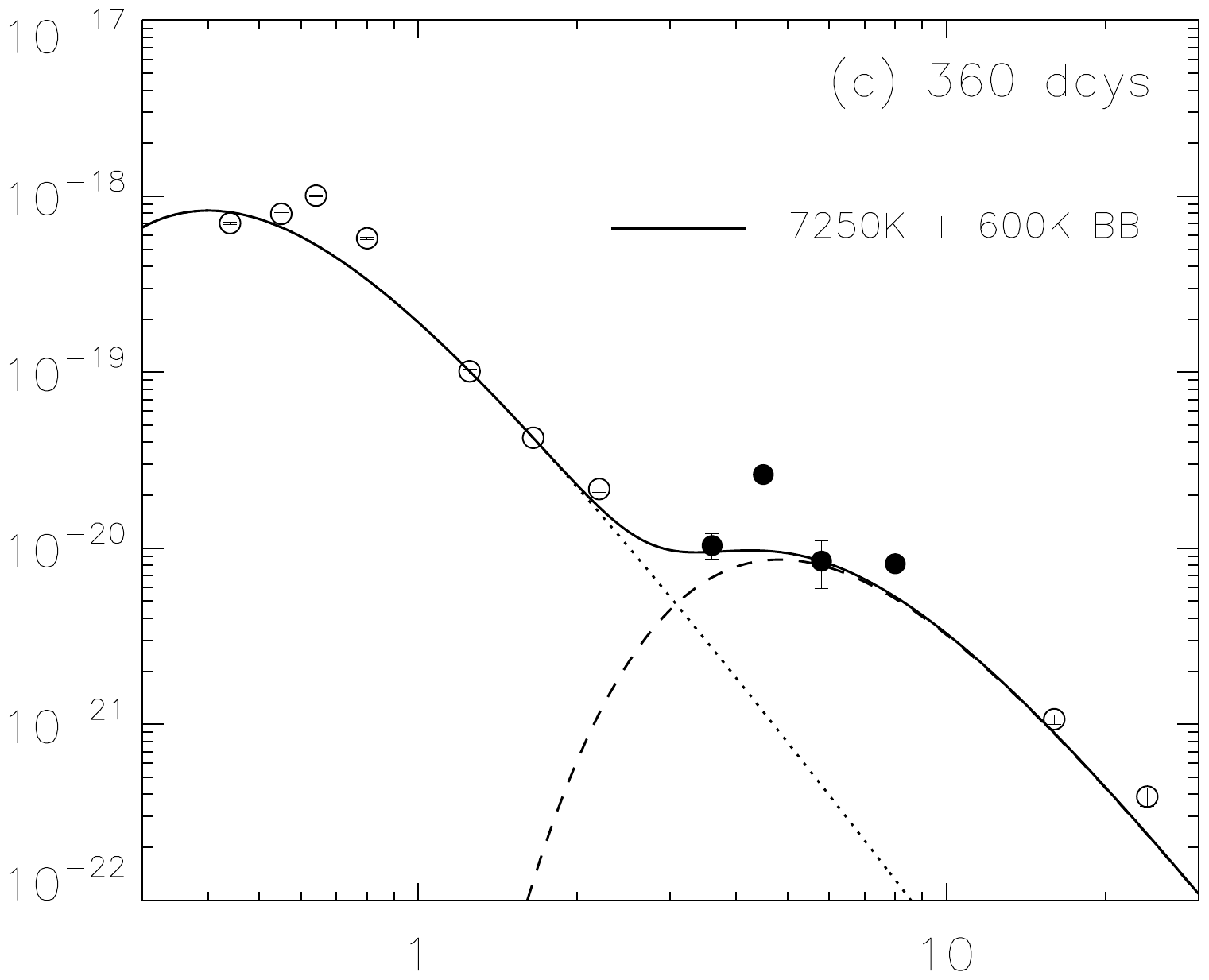}}
\subfigure{\includegraphics[scale=0.3]{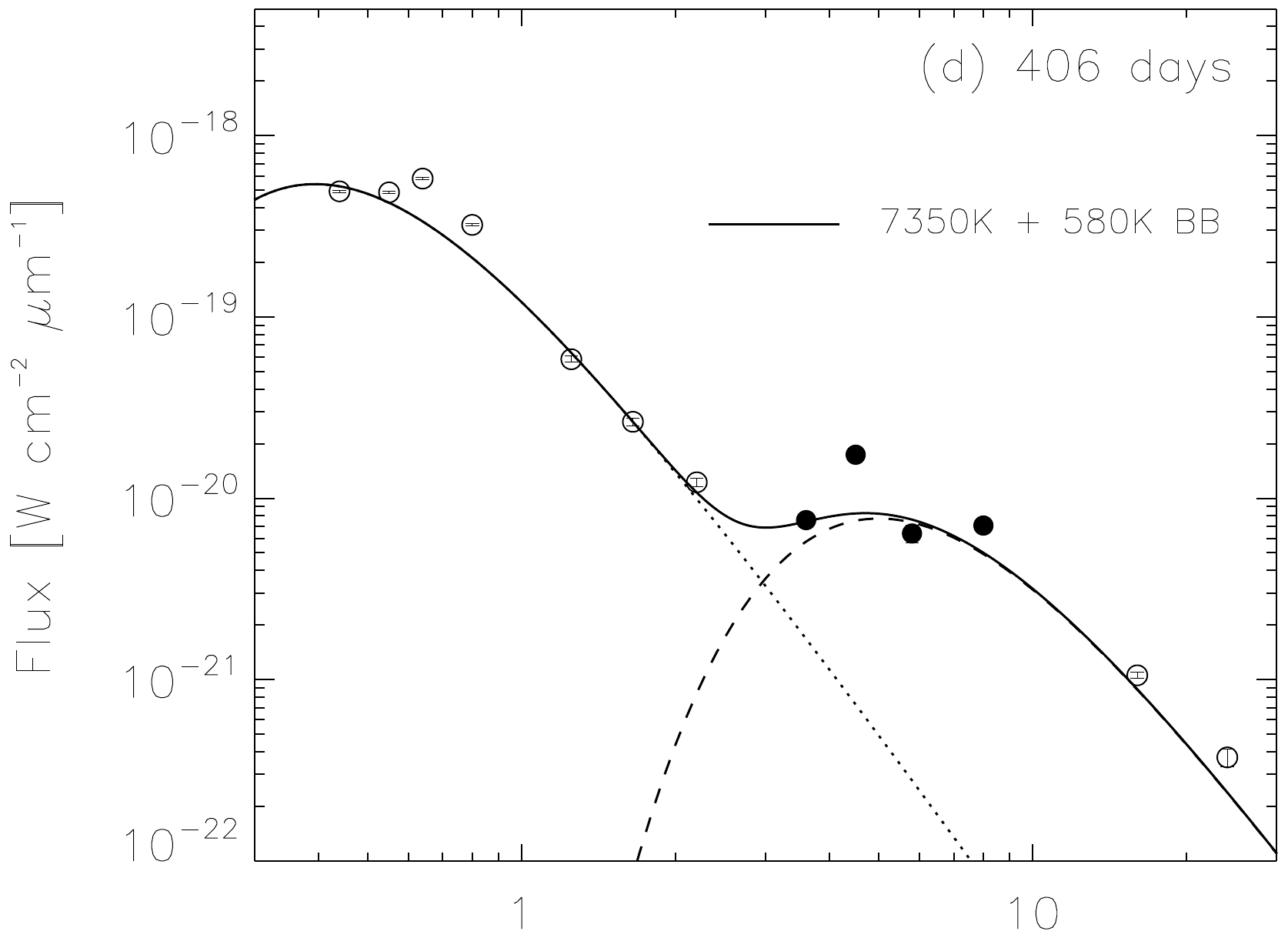}}
\subfigure{\includegraphics[scale=0.3]{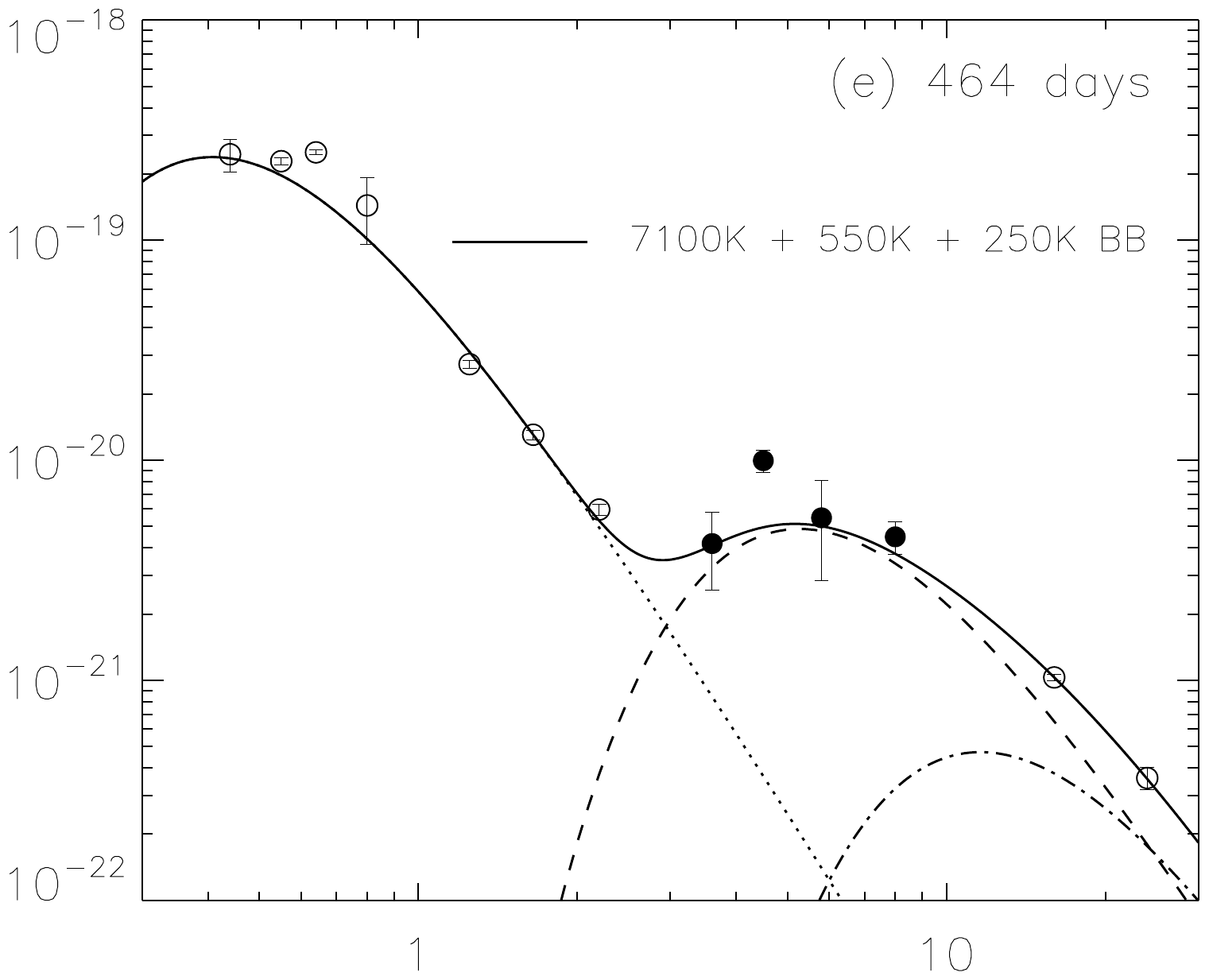}}
\subfigure{\includegraphics[scale=0.3]{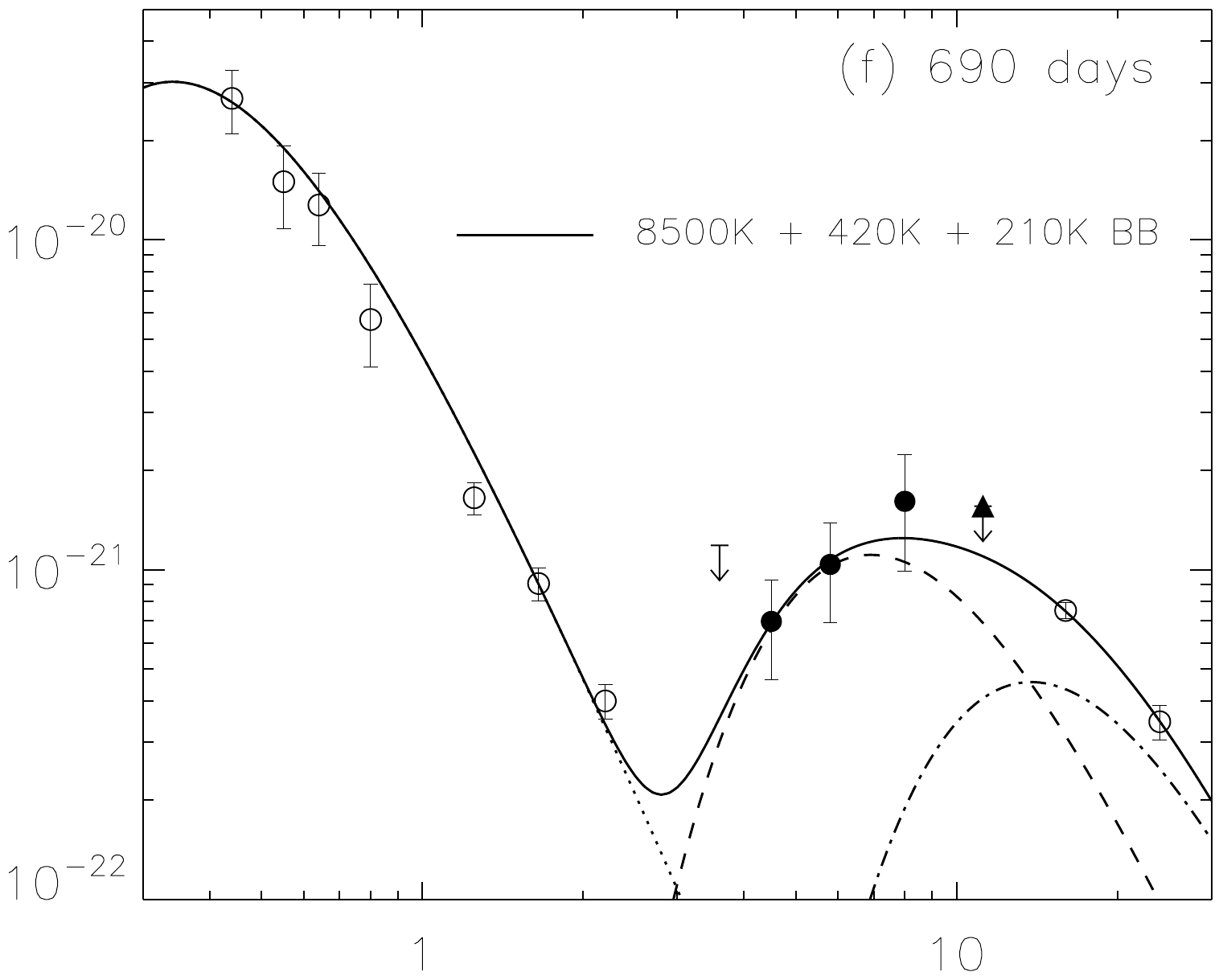}}
\subfigure{\includegraphics[scale=0.3]{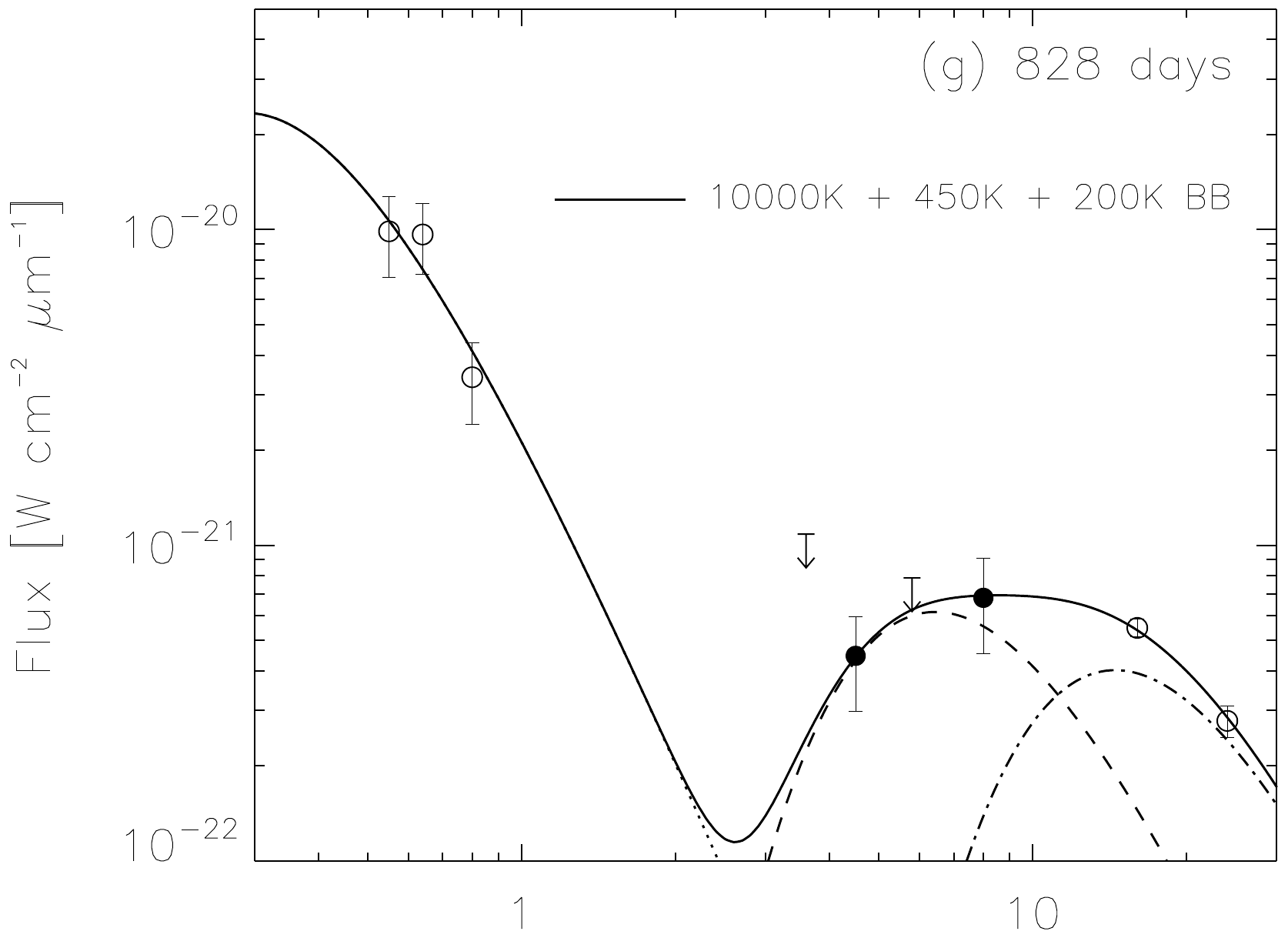}}
\subfigure{\includegraphics[scale=0.3]{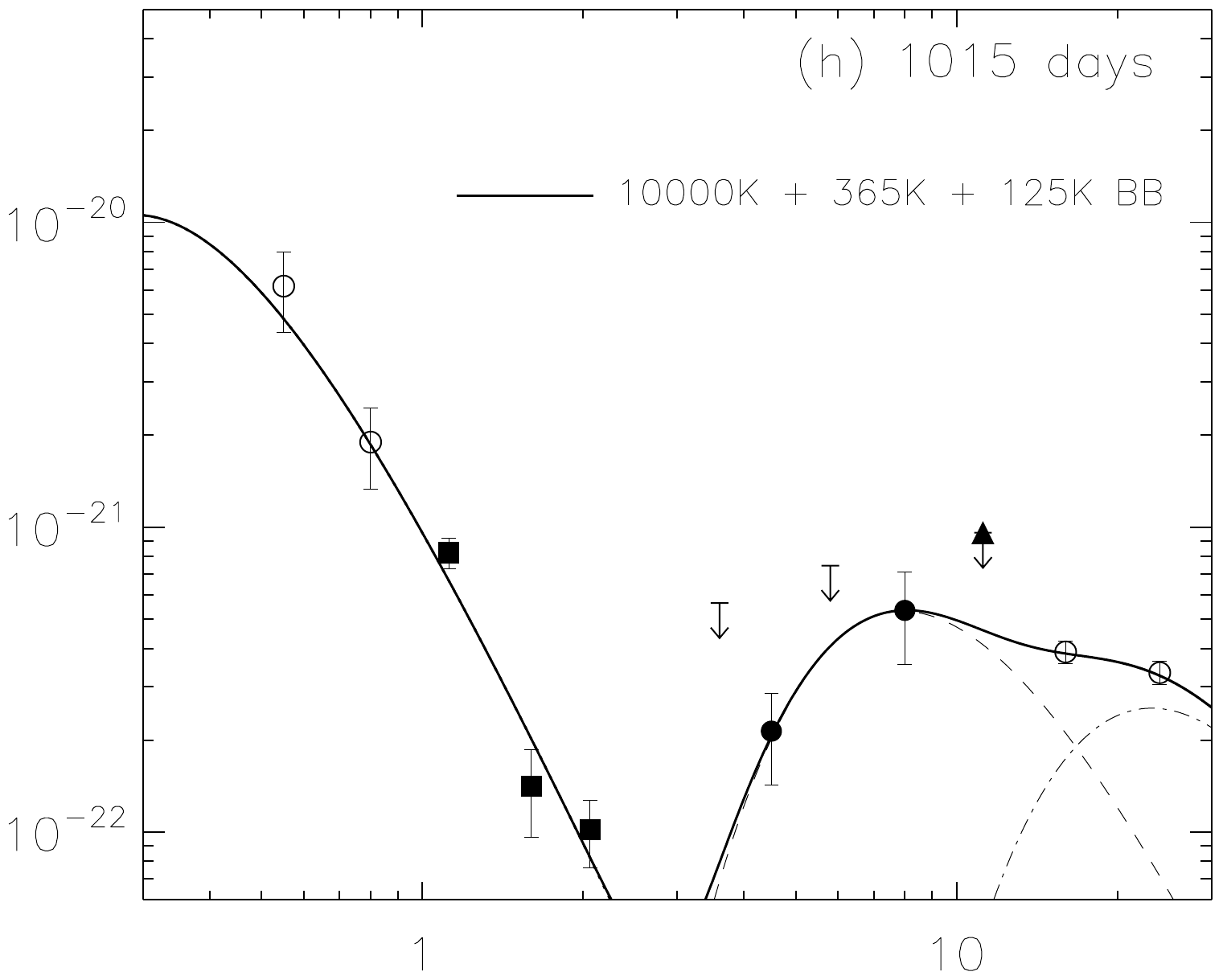}}
\subfigure{\includegraphics[scale=0.3]{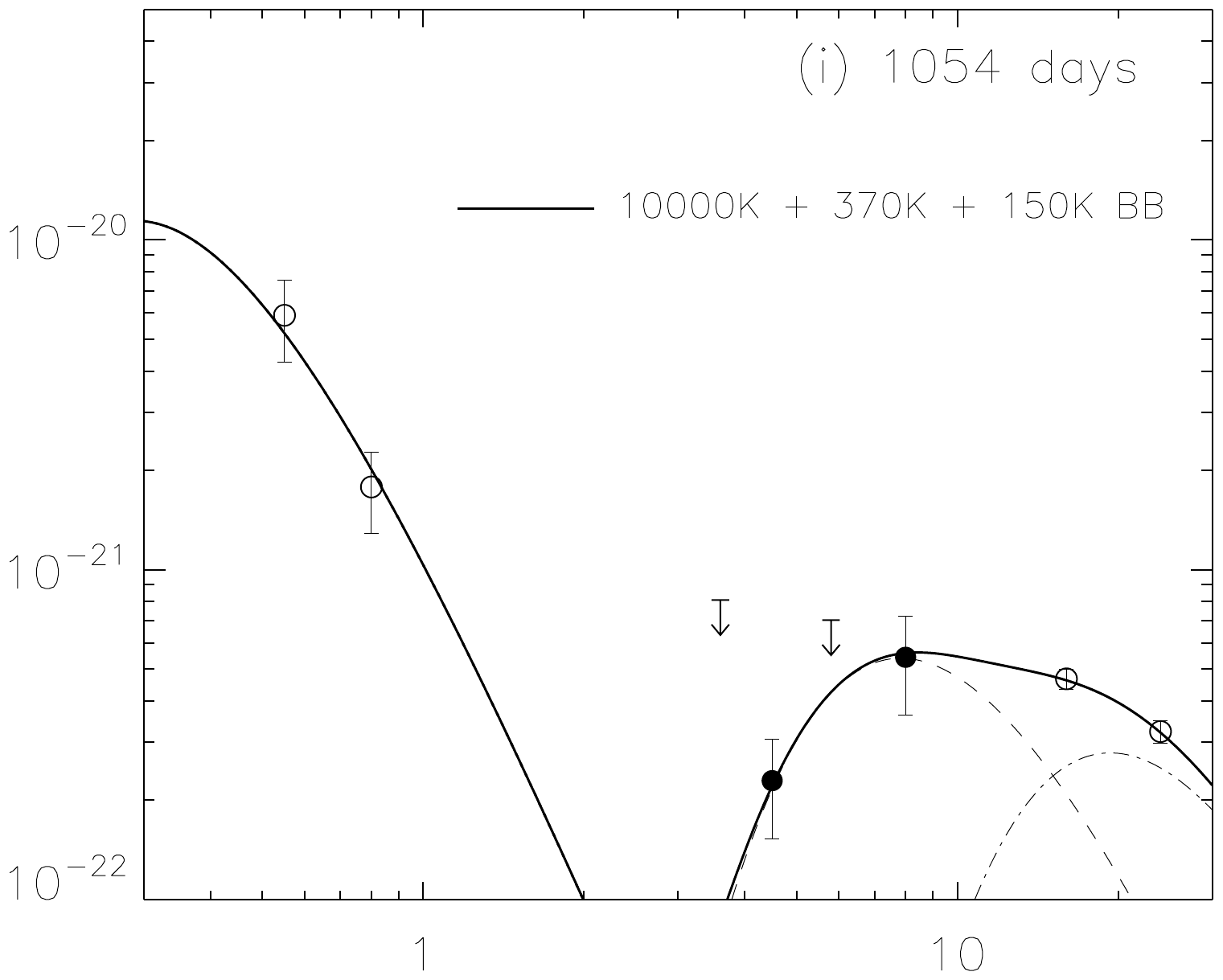}}
\subfigure{\includegraphics[scale=0.3]{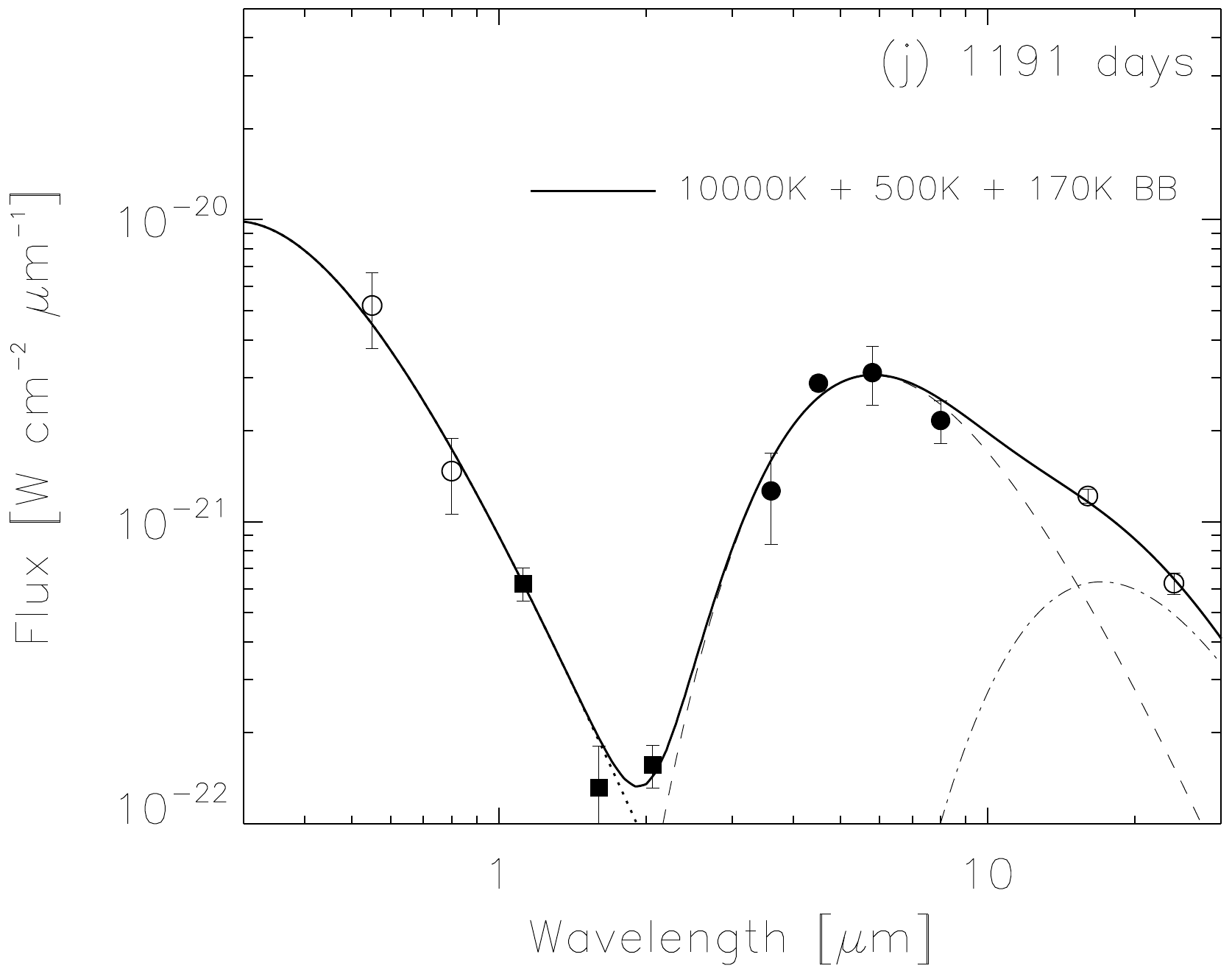}}
\subfigure{\includegraphics[scale=0.3]{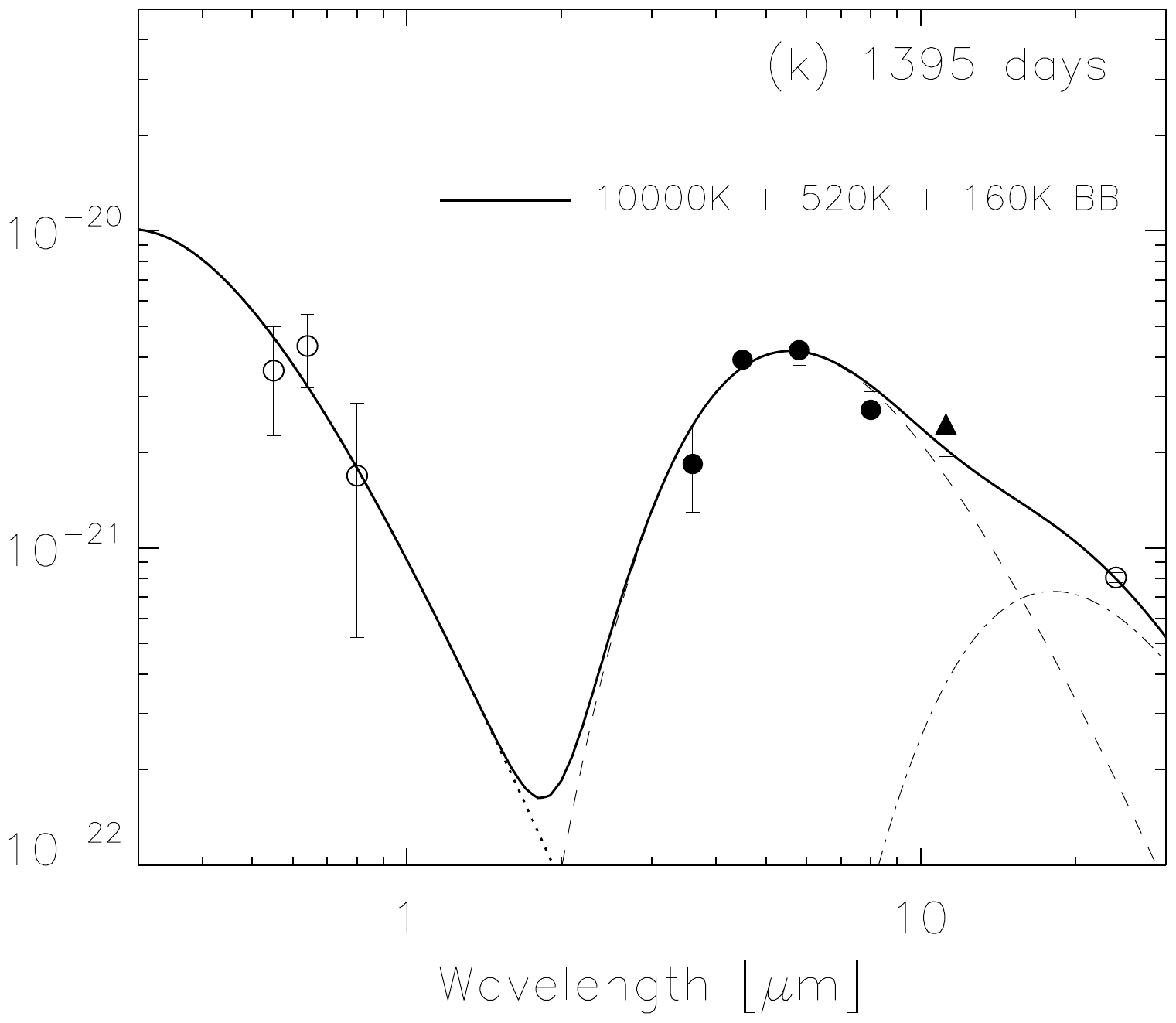}}
\caption[Blackbody fits to the SEDs of SN~2004et - days 300 to 1395]
{Blackbody fits to the day 64 to 1395 SEDs of SN~2004et as defined by the 
optical-IR observations (wavelength range plotted: 0.3--30\,\micron). 
Filled circles indicate fluxes observed at the epoch of the IRAC 
observations. Open circles indicate optical, NIR and mid-IR photometry 
which has been interpolated or extrapolated to the epochs of the IRAC 
observations.  Filled squares indicate the closest epoch (\ie\ not 
interpolated) \hst-NICMOS photometry and filled triangles indicate the 
closest epoch Gemini-Michelle N\primejo-band photometry (see 
{\tablename}~\ref{tab:04et_midir_obs} for epochs of the Michelle and 
NICMOS observations). Upper flux limits for non-detections are indicated 
by the downward-pointing arrows. Where error bars are not shown, 
uncertainties are smaller than the symbol size. All flux densities were 
de-reddened using \textit{E}(\textit{B\,--\,V})~=~0.41 mag 
\citep{zwitter04} and the extinction law of \cite{cardelli89} with $R_{V} 
= 3.1$.}
\label{fig:04et_bbfit_seds_A}
\end{figure*}

Panel (a) of {\figurename}~\ref{fig:04et_bbfit_seds_A} shows a fit
to the SED at day 64, during the photospheric plateau phase of
SN~2004et. The fit, which uses a 5400\,K, blackbody normalised to the IRAC 
5.8-$\mu$m flux density, under-estimates the $V$ and $R$ band flux 
densities, which we attribute to the presence of strong emission lines 
such as H$\alpha$ in the 5400--7000{\,\AA} wavelength region
\citep{sahu06, kotak09}, while it significantly 
overestimates the $U$ band flux density. A spline curve fitted to the 
$UB$ photometry was therefore used to estimate the total flux shortwards 
of $B$. This spline curve was combined 
with the 5400~K blackbody, truncated at wavelengths $\leq B$ (4400\,\AA),
to yield a total integrated flux of 4.5$\times 10^{-13}$\,W\,m$^{-2}$
(with a corresponding luminosity of 4.9$\times 10^{8}$\,\lsun),
about 93\% of the total integrated flux/luminosity from the hot
blackbody alone. At wavelengths $\leq B$, the total flux corresponding to 
the spline fit constituted about 49\% of that from the 5400~K blackbody 
fit. In panel (a) of {\figurename}~\ref{fig:04et_bbfit_seds_A} the solid line
represents the combined spline and truncated blackbody fit, whilst
{\tablename}~\ref{tab:04et_bbfits} lists the parameters for the
blackbody fits.

It is
well known that all Type I SNe show a pronounced early-time deficit at
ultraviolet (UV) wavelengths relative to a blackbody fitted at longer
wavelengths \citep[\eg][]{panagia03}.  This has been interpreted as
due to strong line blanketing by many low excitation lines
of Fe\,II and other lines shortwards of \simjo\,4000{\,\AA}
\citep[\eg][]{branch86}. The situation for Type II SNe seems to be
less clear.  In his review of optical spectra of supernovae,
\cite{filippenko97} noted that most Type II SNe do not show this
feature, with the early-time spectra approximating a
single-temperature Planck function from UV through to IR wavelengths,
and occasionally even showing a slight UV excess. However,
\cite{fransson87}, from their studies of the peculiar Type II
SN~1987A, concluded that supernova atmospheres with a normal (solar)
chemical composition can give rise to line blanketing effects, such as
those seen in the UV spectra of SN~1987A. They proposed that the
differences in UV spectra of supernovae may instead be due to
differences in the density of the CSM. The earliest IUE
(1150--3200\,\AA) spectra of SN~1987A (\simjo\,6 days after explosion)
showed a strong UV deficit in the wavelength range
\simjo\,1250--3200\,{\,\AA}, relative to the 6000\,K blackbody curve
defined at optical and infrared wavelengths \citep{danziger87}.  This
was still present at day 60, and possibly as late as day 260, as shown
by the best-fit SEDs of \cite{wooden93}.
The apparent UV drop-off relative to the blackbody fit to the
photometric SED of SN~2004et at day 64 suggests that a similar
effect is present in this Type II-P SN. \cite{li05b} noted from a
spectrum of SN~2004et at day 9 that ``there is a peculiar decline
blueward of 4000~\AA\ not commonly observed in the spectra of normal SNe
II-P''.

\subsubsection{Days 300-828}

The blackbody fits to the SEDs at days 300 to 828 are shown in
{\figurename}~\ref{fig:04et_bbfit_seds_A}. The corresponding blackbody
parameters are listed in {\tablename}~\ref{tab:04et_bbfits}. 
For each epoch, more than
one component was required to obtain a reasonable fit to the optical,
NIR and mid-IR photometry. These were comprised of (i) a hot
blackbody, with temperatures during the period 300--828 days ranging
from 7100--10000\,K, representing the optical and NIR
continuum emission from the optically thick hot gas of the ejecta,
(ii) a warm blackbody (420--650\,K) representing the
emission at mid-infrared wavelengths, and (iii) from day 464 onwards,
a cooler (200--250\,K) blackbody to account for the observed emission at 
the two longest mid-IR wavelengths (16 and 24~$\mu$m). Excess emission
is also present at 24~$\mu$m on days 300--406, but at a relatively
constant level, see {\figurename}~\ref{fig:04et_bbfit_seds_A}.

From days 300 to 464, the mid-IR emission demonstrated a clear excess
at 4.5~$\mu$m, as noted by \cite{kotak09}. A similar feature has
been seen in mid-IR photometry of other Type II SNe, such as the Type
II-P SN~2003gd at day 499 \citep{sugerman06}, and SN~2007it at day 340
\citep{andrews10}.  This can be attributed to emission from the carbon
monoxide (CO) fundamental band at 4.65~$\mu$m, which was directly
observed in the infrared spectra of the Type IIpec SN~1987A from as
early as 100 days after explosion \citep{suntzeff90} and stayed
visible until at least day 615 \citep{wooden93}. The red wing of the
same emission line was detected in the \spitzer-IRS spectra of the
Type II-P SN~2004dj at days 109 and 129 \citep{kotak05}.  Together
with the detection of the first overtone of CO at \simjo\,2.3~$\mu$m
in the NIR spectrum of SN~2004et \citep{maguire10}, as well as in the 
spectra of several other Type II SNe \citep[and references
therein]{gerardy02}, these observations indicate that strong CO
emission is common in Type II SNe.

\cite{kotak09} also noted an excess at 8.0~$\mu$m compared to
blackbodies matched to the mid-IR continuum emission from days 300 to
690. They found this to be consistent with a broad emission feature
between 8--14~$\mu$m seen clearly in \spitzer-IRS spectra until at
least days 450--481 and attributed this to silicate emission, with a
contribution from the silicon oxide (SiO) fundamental band in the
7.7--9.5-$\mu$m region.  Accordingly, our warm blackbody component was
normalised to the 3.6 or 5.8-$\mu$m fluxes as best representing the
continuum emission during these times. For the best-fit SED at day 300
(panel (b) in {\figurename}~\ref{fig:04et_bbfit_seds_A}), the Gemini
11.2-$\mu$m flux from day 311 is also underestimated by the warm
blackbody component. This excess is attributed to the broad silicate
emission feature seen at its strongest in the IRS spectra at days 294
and 349 presented by \cite{kotak09}.  The 8.0-$\mu$m excess is no
longer evident after day 690, consistent with the IRS spectrum at day
823 presented by \cite{kotak09}. Panel (g) of
{\figurename}~\ref{fig:04et_bbfit_seds_A} shows that by day 828 the
observed 8.0-$\mu$m flux is well-matched by the blackbody fits.

\subsubsection{Days 1015--1395}

Best fits to the SEDs at epochs of days 1015, 1054, 1191 and 1395 are also 
shown in {\figurename}~\ref{fig:04et_bbfit_seds_A} and the corresponding 
blackbody parameters are listed in {\tablename}~\ref{tab:04et_bbfits}.
Hot (10000\,K), warm (350--520\,K) and cool
(120--170\,K) blackbody components were matched to the optical, NIR
and mid-IR photometry.
For days 1015 and 1191, the adopted NIR
fluxes were those measured in the {\em F110W}, {\em F160W} and {\em F205W} 
NICMOS2 filters at the reasonably contemporary epochs of days
1019 and 1215.

\begin{table*}\centering
\scriptsize
  \setlength{\tabcolsep}{0.3mm}
\parbox[t]{21cm}{\protect\caption[Blackbody-fitted parameters to SN 2004et SEDs]{~~Blackbody-fitted parameters to the SEDs of SN 2004et.$^1$ \label{tab:04et_bbfits}}}
\begin{threeparttable}
\begin{tabular}{lccccc|ccccc|ccccc|cc}
  \hline\hline
  & & & & & & & & & & & & & & & & & \\    
  \mcc1{Epoch} & \mcc1{$T_{hot}$} & \mcc1{$F_{hot}$ {\scriptsize[10$^{-15}$}}  & \mcc1{$R_{hot}$} & \mcc1{$v_{hot}$} & \multicolumn{1}{c|}{$L_{hot}$} & 
   \mcc1{$T_{warm}$} & \mcc1{$F_{warm}$ {\scriptsize[10$^{-15}$}} & \mcc1{$R_{warm}$} & \mcc1{$v_{warm}$} & \multicolumn{1}{c|}{$L_{warm}$} & 
    \mcc1{$T_{cool}$} & \mcc1{$F_{cool}$ {\scriptsize[10$^{-15}$}} & \mcc1{$R_{cool}$} & \mcc1{$v_{cool}$} & \multicolumn{1}{c|}{$L_{cool}$} & 
    \mcc1{$F_{tot}$ {\scriptsize[10$^{-15}$}} & \mcc1{$L_{tot}$} \\

  \mcc1{[days]} & \mcc1{[K]} & \mcc1{{\scriptsize W\,m$^{-2}$]}} & \mcc1{[10$^{15}$ cm]} & \mcc1{[\kms]}
                                                                        & \multicolumn{1}{c|}{[10$^{6}$ \lsun]} & 
   \mcc1{[K]} &  \mcc1{{\scriptsize W\,m$^{-2}$]}} & \mcc1{[10$^{15}$ cm]} & \mcc1{[\kms]}
                                                                        & \multicolumn{1}{c|}{[10$^{6}$ \lsun]} & 
    \mcc1{[K]} &  \mcc1{{\scriptsize W\,m$^{-2}$]}} & \mcc1{[10$^{15}$ cm]} & \mcc1{[\kms]}
                                                                        & \multicolumn{1}{c|}{[10$^{6}$ \lsun]} &   
    \mcc1{{\scriptsize W\,m$^{-2}$]}} & \mcc1{[10$^{6}$ \lsun]} \\
  & & & & & & & & & & & & & & & & & \\ \hline
  & & & & & & & & & & & & & & & & & \\
64\tnote{2}&5400&446&1.75&3167&485&$\cdots$ &$\cdots$ &$\cdots$ &$\cdots$ &$\cdots$ &$\cdots$ &$\cdots$ &$\cdots$ &$\cdots$ &$\cdots$ &446&487\tnote{2}\\
300\tnote{3}&7250&10.9&0.15&59&11.9&650&0.81&5.16&1989&0.89&$\cdots$ &$\cdots$ &$\cdots$ &$\cdots$ &$\cdots$ &11.7&12.8\tnote{3}\\
360&7250&5.04&0.10&33&5.50&600&0.63&5.34&1718&0.69&$\cdots$ &$\cdots$ &$\cdots$ &$\cdots$ &$\cdots$ &5.67&6.19\\
406&7350&3.24&0.081&23&3.54&580&0.59&5.51&1570&0.68&$\cdots$ &$\cdots$ &$\cdots$ &$\cdots$ &$\cdots$ &3.83&4.18\\
464&7100&1.48&0.058&15&1.62&550&0.39&4.99&1246&0.42&250&0.083&11.15&2782&0.091&1.96&2.13\\
690&8500&0.16&0.013&2.2&0.17&420&0.12&4.67&784&0.13&210&0.096&16.96&2845&0.10&0.37&0.40\\
828&10000&0.10&0.0078&1.1&0.11&450&0.060&2.93&409&0.065&200&0.089&17.99&2515&0.096&0.25&0.27\\
1015&10000&0.047&0.0052&0.60&0.051&365&0.064&4.59&523&0.069&125&0.090&46.36&5286&0.098&0.20&0.22\\
1054&10000&0.050&0.0054&0.60&0.055&370&0.064&4.47&491&0.070&150&0.082&30.72&3374&0.089&0.20&0.21\\
1191&10000&0.044&0.0050&0.49&0.047&500&0.27&5.02&487&0.29&170&0.16&33.87&3291&0.18&0.48&0.52\\
1395&10000&0.045&0.051&0.42&0.049&520&0.35&5.32&441&0.39&160&0.20&42.33&3512&0.22&0.60&0.65\\
  & & & & & & & & & & & & & & & & & \\ \hline

\end{tabular}
\begin{tablenotes}
\scriptsize
\item[1] Up to three components -- hot, warm and cool -- were used to fit the continuum emission of the SEDs (see text). 
For each component, $R$ is the blackbody radius corresponding to the best fit temperature, $T$; $v$ is the expansion velocity 
corresponding to the radius $R$; and $L$ is the luminosity for an adopted distance to the SN of 5.9\,Mpc.
\item[2] The day 64 5400\,K blackbody fit over-estimated the
  $U$ band flux. A spline curve fit to the {\em UB} fluxes and
  extrapolated to shorter wavelengths was combined with the 5400\,K
  blackbody fit to the longer wavelength data, where the blackbody was
  truncated at wavelengths shorter than the $B$-band -- see panel (a) of
  {\figurename}~\ref{fig:04et_bbfit_seds_A}.  The total integrated
  flux and corresponding luminosity of this ``spline $+$ 5400\,K
  blackbody'' fit are about 93\% of those values for the pure 5400\,K
  blackbody fit shown in this table.
\item[3] At day 300, it was found that the hot component blackbody
  still over-estimated the fluxes at the shortest wavelengths. A
  spline curve fit to the {\em UBV} fluxes and extrapolated to shorter
  wavelengths was combined with the two-component blackbody fit to the
  longer wavelength data, as shown in panel (b) of
  {\figurename}~\ref{fig:04et_bbfit_seds_A}, where the 7250K blackbody
  component was truncated at wavelengths shorter than the V-band. The
  total integrated flux and corresponding luminosity of this ``spline
  $+$ 7250K $+$ 650K blackbody'' fit are about 73\% of those values
  for the combined 7250K $+$ 650K blackbody fit shown in this table.
\end{tablenotes}
\end{threeparttable}
\end{table*}

\subsection{Discussion of results from blackbody fitting}
\label{ssec:04et_bbfit_discuss}

\begin{figure}\centering
  \includegraphics[scale=0.5,angle=0,clip=true]
  {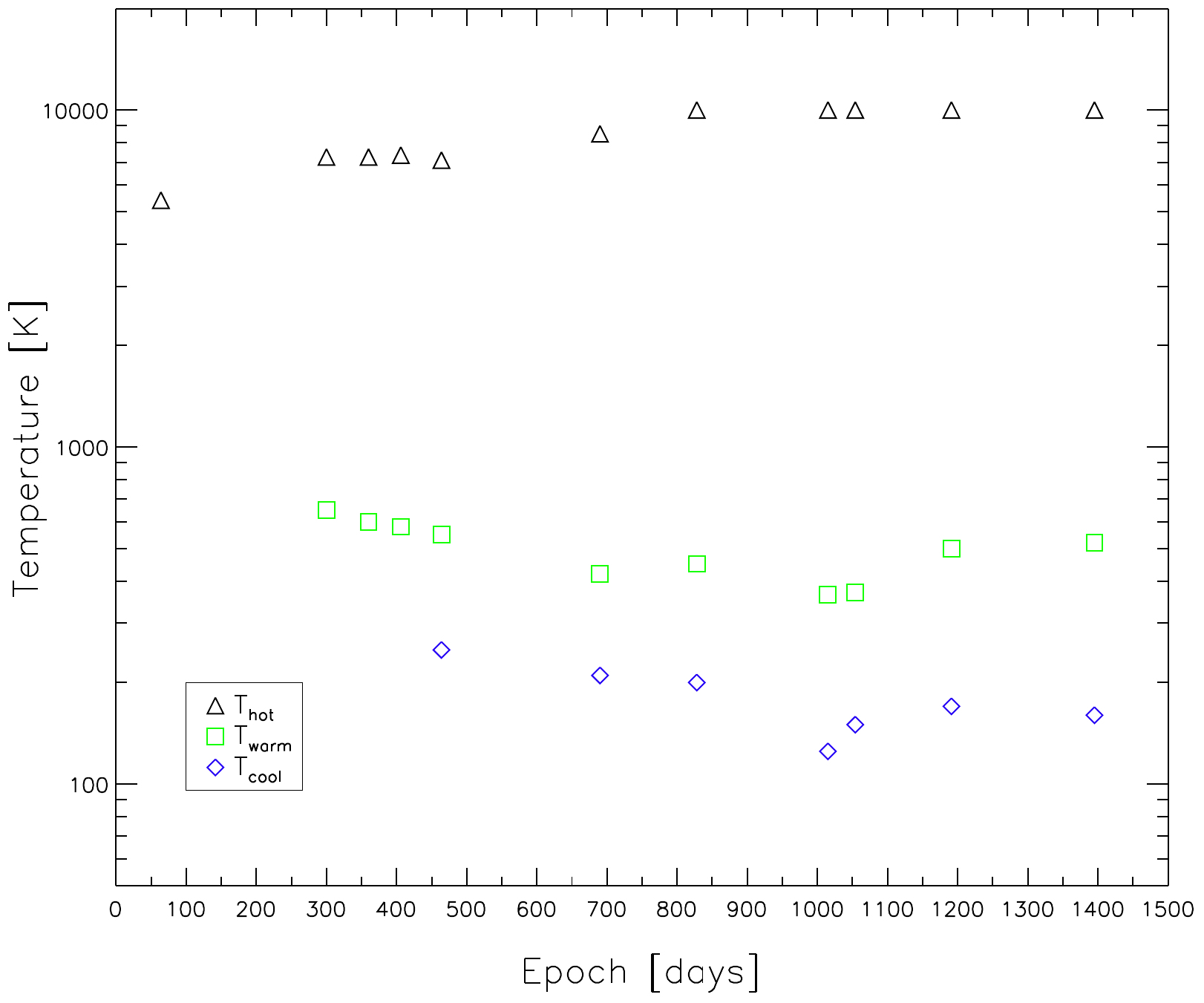} \vspace{2ex}
    {\caption[Temperature evolution from multi-component
    blackbody fitting.]{The temperature evolution of the
    blackbody components that were fit to the SED of 
    SN~2004et at different epochs. Hot component temperatures ranged from 
    5400~K to 10000~K, warm component temperatures ranged from 370~K to 
    650~K and cool component temperatures ranged from 125~K to 
    250~K (see Table~7).}\label{fig:04et_bbfits_temp}} \end{figure}

The complete set of parameters from the blackbody fits 
to the photometry from days 64 to
1395 are listed in {\tablename}~\ref{tab:04et_bbfits}. The temperature
evolution of the best fit multi-component blackbodies is shown in
{\figurename}~\ref{fig:04et_bbfits_temp}.

The left-hand panel of 
{\figurename}~\ref{fig:04et_bbfits_lum} shows the evolution of the
luminosities of each of the blackbody components, and their sum,
compared with the theoretical luminosity
due to the radioactive deposition of $^{56}$Co, $^{57}$Co, and other
isotopes, adopting the deposition behaviour of SN~1987A as modelled by
\cite{lih93} and \cite{woosley89}. The radioactive decay deposition
curve for SN~1987A was scaled by a factor of 0.69 to normalise it to
the total luminosity of SN~2004et between days 300--464. It should be
noted that the blackbody fits to the photometry mainly trace the
continuum emission of the SN (known optical and mid-IR emission line
features were deliberately not matched by the blackbodies as discussed
in the previous sections). As a consequence, the luminosities listed
in {\tablename}~\ref{tab:04et_bbfits} slightly underestimate the total
luminosities.

\subsubsection{Evolution of the hot component}

The day 64 photometry can be fitted by a 5400\,K blackbody, consistent 
with an origin from the optically-thick emission of the ejecta
photosphere, with the Rayleigh-Jeans tail of
the corresponding blackbody extending into the infrared. As the
ejected envelope adiabatically expands and cools, the hydrogen ionised
by the initial SN shock approaches the temperature for recombination
(\simjo\,5000\,K) and a recombination wave recedes through the
envelope. The recombination front defines the photosphere and as such
the temperature of the photosphere is characterised by the
recombination temperature of hydrogen. \cite{wooden93} found that a
hot component of 5000-5500\,K was a good match to the SED of SN~1987A
from days 60 to 777, which they found to be in agreement with
observations of other Type II supernovae by \cite{kirshner73}. With a
similar temperature, the 5400\,K blackbody fit shown in panel (a) of
{\figurename}~\ref{fig:04et_bbfit_seds_A} is 
representative of the photospheric continuum emission at this time.
The ejecta
velocity of 3167\,\kms\ implied by the emitting radius of the 5400\,K
blackbody is in reasonable agreement with the velocity of
\simjo\,3500\,\kms\ found by \cite{sahu06} for SN~2004et in the
plateau phase, estimated from the minimum of weak, unblended
absorption lines of \Feii\ at 4924, 5018 and 5169{\,\AA}. 
\cite{kotak09} obtained a reasonable fit to the day 64 SED of
SN~2004et with a single 5300\,K blackbody, concluding
there was little sign of thermal emission from dust.

As expected, the corresponding luminosity at day 64 exceeds that
from radioactive decay deposition, which only begins to dominate the
light curve during the nebular phase (from \simjo\,130 days,
\citealt{maguire10}), following the sharp decline from the plateau at
\simjo\,110 days \citep{sahu06}.  For the 5400\,K blackbody fit the
day 64 luminosity exceeds that from radioactive deposition by a 
factor of 4.9, or by a factor of 4.5 if considering the spline plus 
truncated
blackbody fit which is a better match to the $U$ band data.  This
compares to the factor of 3.8 found by \cite{kotak09}.

\begin{figure*}
  \includegraphics[scale=0.47,angle=0,clip=true]
  {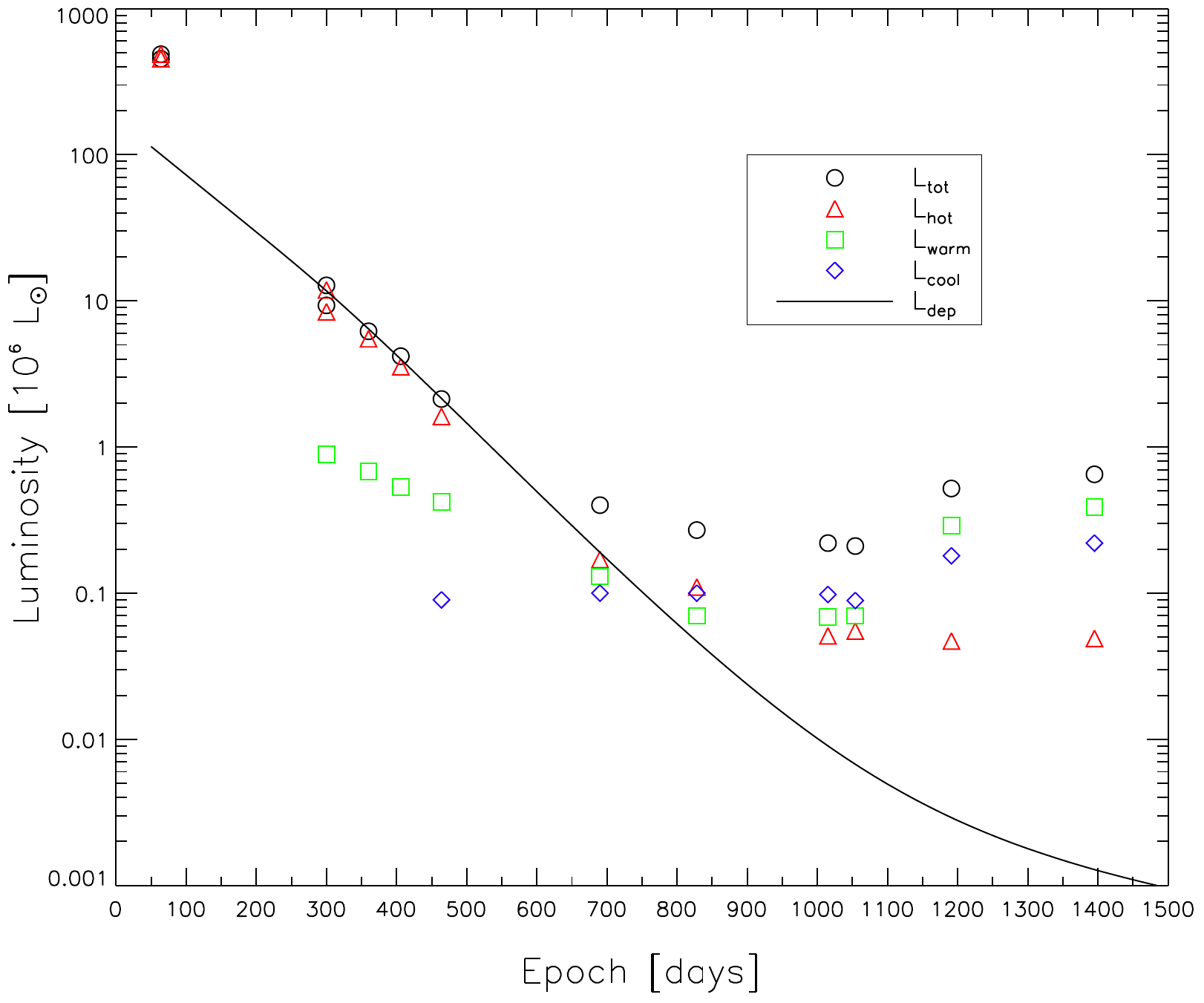} 
  \includegraphics[scale=0.40,angle=0,clip=true]
  {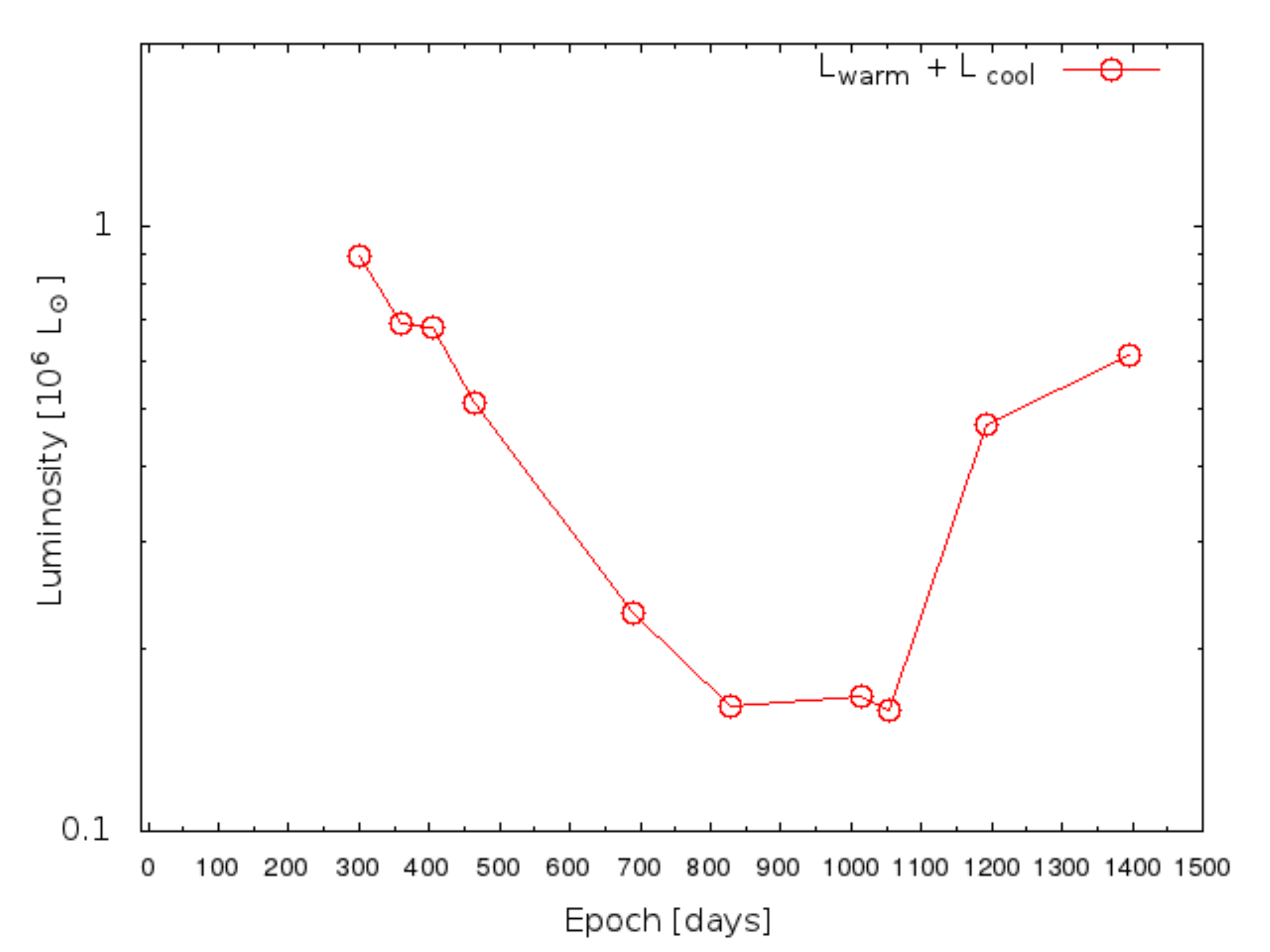} \vspace{2ex}
    {\caption[Luminosity evolution from multi-component
    blackbody fitting.]{The luminosity evolution of SN~2004et, from 
    multi-component blackbody fits to the SED at each epoch (see 
    Table~7). The 
    temperature evolution of the individual components is shown in Fig.~9.  
    Left: The luminosity evolution of the individual
    components (hot, warm and cool) is shown at each epoch, together with 
    that of the corresponding total luminosity. The solid curve  
    shows the predicted luminosity due to radioactive deposition, as 
    scaled from the models of \cite{lih93} (see text for details).
    Right: the sum at each epoch of the luminosities of the two 
    components, warm and cool, attributed to dust emission, showing a 
    steady decline up to day 
    828, followed by a subsequent rise.}\label{fig:04et_bbfits_lum}}
\end{figure*}

By day 300, the estimated temperature of the hot blackbody component had
increased to 7250\,K and then remained relatively constant
($\pm$\,\simjo\,150\,K) until day 464
({\figurename}~\ref{fig:04et_bbfits_temp}), whilst its luminosity faded
quite rapidly (by a factor of \simjo\,7) during this time
({\figurename}~\ref{fig:04et_bbfits_lum}). Note that
at day 300, the blackbody fit significantly over-estimates the fluxes
at the shortest wavelengths in the $U$ and $B$ bands, as also seen at day
64 (see previous discussion).  By day 690 the temperature had
increased again to 8500\,K and to 10,000\,K by day 828, whilst the
luminosity continued to decrease (by a factor of \simjo\,15). For
epochs beyond 1000 days, the temperature of the hot component appears
to remain constant.

\subsubsection{Evolution of the warm and cool dust components}

The warm dust component cooled monotonically from 650--420\,K between 
days 300 and 690 ({\figurename}~\ref{fig:04et_bbfits_temp}). 
This is consistent with the fading of SN~2004et observed in
the mid-IR during this time. The sum of the luminosities of the 
hot component and warm component is less
than or comparable to the predicted radioactive deposition up to day
690. On day 828 however, the sum of the hot and warm component 
luminosities was a factor of 1.7 higher than the predicted radioactive 
deposition luminosity.

The presence of a mid-IR excess from day 300, demonstrated by the
requirement of a warm component to match the SEDs from this time, and the
evolution of this warm component from 300 to 690 days, are consistent with
emission from dust freshly synthesised in the supernova ejecta, in 
agreement with the results of \cite{kotak09} from their interpretation of 
similar data.

The approximately constant luminosity of
the warm component between days 828 and 1054 corresponded to the lowest
level reached, since at day 1191 the warm component luminosity
had increased by a factor of
4.3.  From day 1015 the warm component luminosity clearly exceeds that
of the radioactive deposition (by a factor of \simjo\,8) and at day
1191 by a factor of over 100. 

The cool dust component, required to fit the mid-IR fluxes longwards of 
16~$\mu$m from day 464 onwards, showed a monotonic decline in temperature 
from 250\,K at day 464 to \simjo\,120\,K at day 1015, but then increased 
slightly and remained around 160\,K between days 1054 and 1395. Its 
luminosity stayed roughly constant from days 464 to 1054, but had 
increased by a factor of \simjo\,2 by days 1191--1395. The velocities 
implied by the minimum emitting radii of these cooler blackbodies ranged 
from \simjo\,2500--6000\,\kms.

\begin{table*}
 \parbox[t]{14.5cm}{\protect\caption[Dust density distribution models]{Input parameters for $R^{-2}$ density distribution dust models.\label{tab:04et_RTmodels}}}
\centering
\small
\begin{tabular}{cccccccccccc}
\hline\hline
   & & &\multicolumn{2}{c}{diffuse source}& & & &\multicolumn{2}{c}{smooth}&\multicolumn{2}{c}{clumpy}\\    
\\
\cline{4-5}
\cline{9-10}
\cline{11-12}
Epoch &$R_{in}$&$R_{out}$/$R_{in}$&L&T&AmC:Sil&$a_{min}$--$a_{max}$&n($a$)$\propto$$a^{-p}$&$\tau_{0.55}$&$M_{d}$&$\tau_{0.55}$&$M_{d}$\\
(day) &(10$^{15}$ cm) &&(10$^{6}$ L$_{\odot}$)&(K)&($\%$)&($\mu$m)&$p$=&&(10$^{-5}$ M$_{\odot}$)&&(10$^{-5}$ M$_{\odot}$)\\
\hline
300 &7.0 &4.0 &12.8 &8000 &20:80 &0.1--1.0 &3.5 &0.11 &3.8&0.07 &7.6\\
360 &6.0 &4.5 &6.19 &7000 &20:80  &0.1--1.0 &3.5 &0.20 &5.6&0.12 &10.0\\
406 &7.0 &4.0 &4.18 &8000 &20:80  &0.1--1.0 &3.5 &0.18 &6.5&0.24 &22.4\\
464 &8.0 &4.0 &2.13 &8000 &20:80  &0.1--1.0 &3.5 &0.38 &11.1&0.39 &50.0\\
690 &6.0 &3.5 &0.40 &8000 &20:80   &0.1--1.0 &3.5 &1.30 &43.9&1.31 &150\\
\hline
\end{tabular}
\end{table*}

\cite{kotak09}, on the other hand, found that the temperature of their 
cool component remained approximately constant, at 120\,$\pm$\,10\,K for 
all epochs from day 300--1395, with minimum blackbody radii corresponding 
to velocities as large as 12000\,\kms, so they ruled out ejecta-condensed 
dust as a source of the cool dust emission. The differences between our 
results for the cool component and those of \cite{kotak09} are most likely 
due to the differences in the photometry at the longest wavelengths, 16 
and 24~$\mu$m, as discussed in {\secname}~\ref{sec:04et_midir_results}. 
Our 16 and 24-$\mu$m fluxes required generally higher temperature 
blackbodies to fit them than did those of \cite{kotak09}. Since our 
luminosities for the cool component are generally consistent with those of 
\cite{kotak09}, the higher blackbody temperatures that we obtain resulted 
in lower minimum radii and therefore lower minimum outflow velocities.

The right-hand panel of {\figurename}~\ref{fig:04et_bbfits_lum} shows
the evolution of the sum of the luminosities of the two dust components
(warm plus cool). It shows a steady decline until day 828, a period 
of constant luminosity until day 1054 and a steep rise
thereafter. The evolution
beyond day 828 implies the presence of an additional energy source.
\cite{kotak09} found similar results and
interpreted the late rise in the mid-IR flux as due to ejecta-CSM 
interaction and the subsequent formation of a cool dense shell of dust 
behind the reverse shock of the supernova. 
An alternative explanation in terms of a light echo from pre-existing 
extended CSM dust will be presented by Sugerman et al. (in preparation).

\section{Radiative transfer modelling: days 300--690}
\label{sec:04et_rtmodels}

In agreement with previous investigators \citep{sahu06, kotak09}, the
development of a red-blue asymmetry in optical emission line profiles,
the evidence from optical light curves for additional extinction by
dust in the ejecta, and the development of a mid-IR excess
attributable to dust emission all support the inference that dust
formed within the ejecta of SN~2004et from about 300 days after
explosion. As discussed above, by day 828 the total luminosity
exceeded the estimated radioactive deposition luminosity by a factor
of four ({\figurename}~\ref{fig:04et_bbfits_lum}), indicating that an
additional component dominated by that date.  Consequently, epochs
later than day 690, when the thermal IR emission can no longer be
solely attributed to internally heated newly-formed dust in the
ejecta, will not be modeled here - Sugerman et al. (in preparation)
will present a circumstellar light echo model for these later epochs.

To investigate the time evolution of dust formation in the 
ejecta of SN~2004et and to estimate the mass of dust present, a number of 
dust shell models were built to match the observed SEDs at the epochs 
between days 300 and 690 ({\tablename}~\ref{tab:04et_RTmodels}).
The models were calculated using the 
three-dimensional Monte Carlo radiative transfer (RT) code MOCASSIN 
\citep{ercolano03, ercolano05} which accounts for the primary and 
secondary components of the radiation field in a fully self-consistent 
manner through absorption, re-emission and scattering of photons. The 
photon paths are followed from a specified source through a given 
composition, grain-size distribution, density and geometry of dust. The 
particular choices of these parameters are either constrained a priori or 
are varied until the model emission and extinction match the observations.

For the day 300--690 models we assumed that the observed IR
emission originated from dust formed in the SN ejecta.
Heating is due to $\gamma$-rays from the decay of $^{56}$Co, which are
reprocessed to optical and UV wavelengths through interaction with the
gas. We assume that this leads to a local radiation field whose
strength is proportional to the local ejecta density. Based on these 
assumptions, and following the previous
modelling of SN~2003gd by \cite{sugerman06} and of SN~1987A by
\cite{ercolano07}, the RT models were constructed such that the dust
and source luminosity were mixed within a spherical expanding shell of
inner radius $R_{\rm in}$ and outer radius $R_{\rm out}=YR_{\rm in}$.
For the dust density distribution, we considered
two cases: (i) a smooth distribution following an $r^{-2}$
density profile (smooth model), and (ii) a clumpy model
where dense clumps exist in a less-dense interclump medium (ICM), where 
the ICM follows a smooth $r^{-2}$ density distribution, with the
local heating source located only in the ICM. For each case, we
compared the observed SEDs with those reproduced by the models to
determine which gave the best fit to the observations.

\begin{table}
\centering
\caption{The contribution from line emission to the
$VRI$-bands.\label{tab:linecontribs}}
\begin{tabular}{cccc}
\hline\hline
Epoch &$V$&$R$&$I$\\
(days)&($\%$)&($\%$)&($\%$)\\
\hline
336 &21 &75&68\\
417 &20 &58&60\\
454 &26 &58&57\\
649 &26 &60&27\\
\hline
\end{tabular}
\end{table}


Before running the models, we estimated the contribution from line 
emission to the $VRI$-bands. We used day 417 and 454 optical spectra 
downloaded from $SUSPECT$ plus day 336 TNG and day 646 Subaru spectra, 
along  with the Subaru/FOCAS $VRI$-band filter transmission 
curves\footnote{See 
http://www.naoj.org/Observing/Instruments/FOCAS/camera/filters.html} because 
these have similar band centres and widths to the standard Johnson 
filters. The percentage line contributions in each band are listed in 
{\tablename}~\ref{tab:linecontribs}.
When evaluating the fitting accuracy of the SED modeling, we excluded 
the $R$ and $I$ bands on days 300, 360, 406, and 464 because both bands 
were dominated by line emission at  those epochs. We also omitted
the IRAC 4.5-$\mu$m data-points from the fitting, due to the potentially 
very large CO line emission contributions in that band \citep{kotak09}.
Although our modelling took into account potential emission from the 
broad silicate 10-$\mu$m band, the SiO fundamental vibrational band
can also contribute to the IRAC 8-$\mu$m band -- from {\em Spitzer}
IRS spectra its contribution was deduced by \cite{kotak09} to be 
significant on days 300-464.

\subsection{Smooth dust distribution models}

Both amorphous carbon (AC) and silicate dust grains were considered.
Optical constants were taken from \cite{zubko96} for amorphous carbon
(their ACH2),
and from \cite{draine84} for the silicates. To investigate the dust
composition, models were run with amorphous carbon:silicate mixtures of
(100-x):x\,\%, for x=0,20,40,60,80,100. It was found that the case
of 20\,\% amorphous carbon and 80\,\% silicate (by mass) best
matched the observed SEDs at all epochs, and this composition was
adopted for all subsequent models.

A standard MRN $a^{-3.5}$ distribution \citep{mathis77} 
with $a_{\rm min}$=0.005~$\mu$m and $a_{\rm max}$=0.25~$\mu$m could not
reproduce the steepness of the observed SED in the $JHK$-bands before day
690, nor the 16- and 24-$\mu$m flux densities. An improved
fit was found using an MRN distribution with $a_{\rm min}$=0.1~$\mu$m 
and $a_{\rm max}$=1.0~$\mu$m and this was subsequently adopted for all 
epochs.

\begin{figure*}
\centering
\subfigure{\includegraphics[scale=0.45]{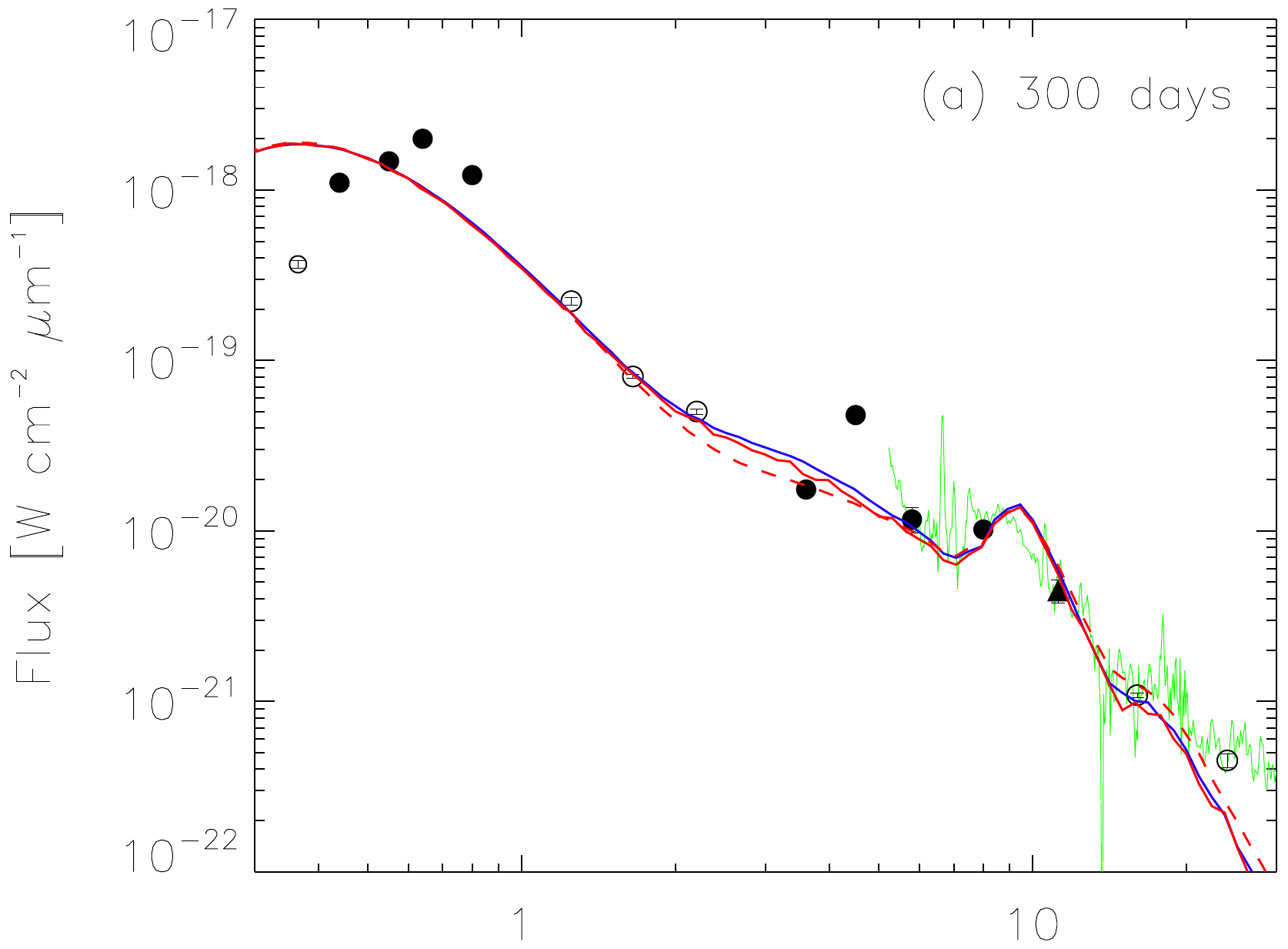}}
\subfigure{\includegraphics[scale=0.45]{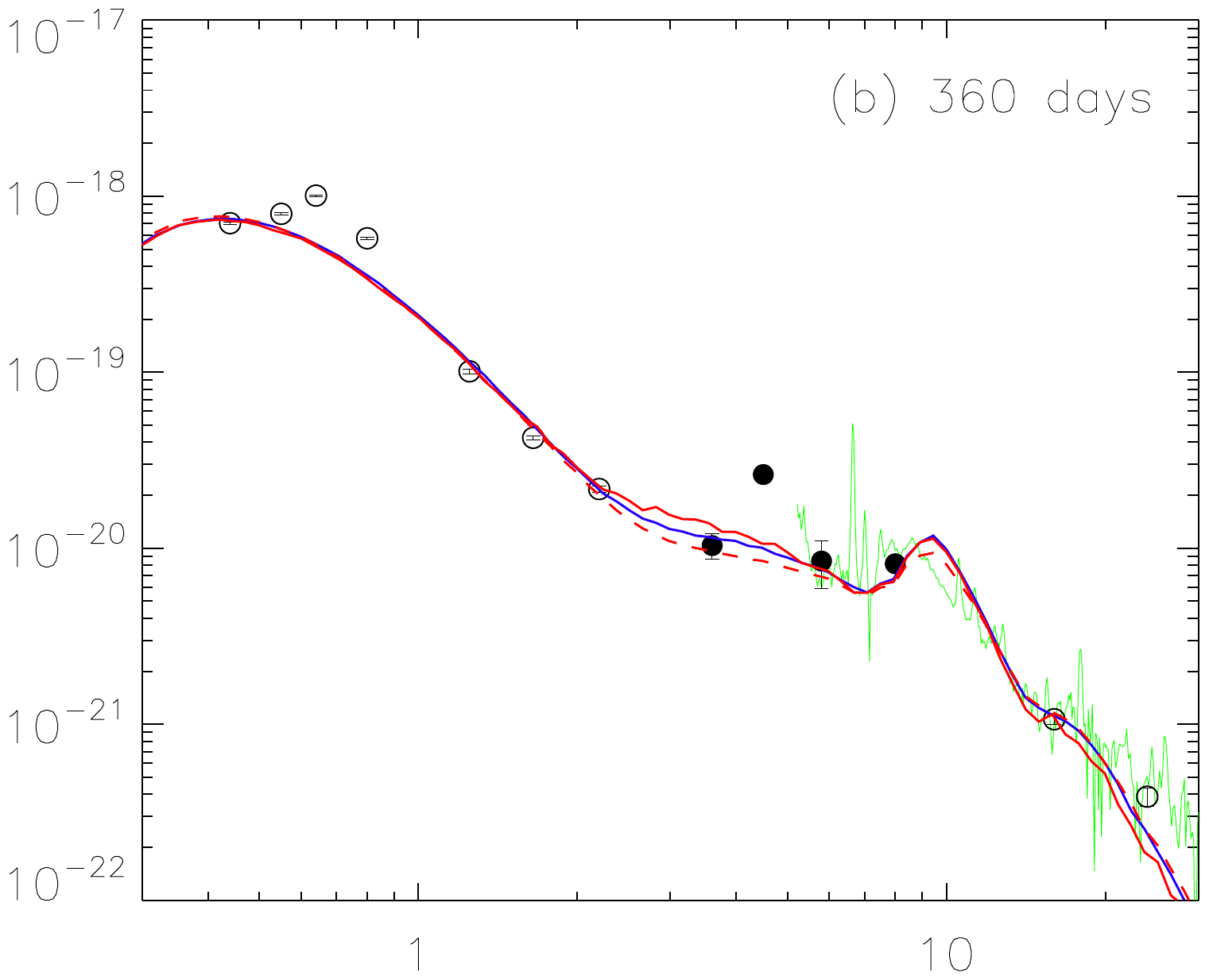}}
\subfigure{\includegraphics[scale=0.45]{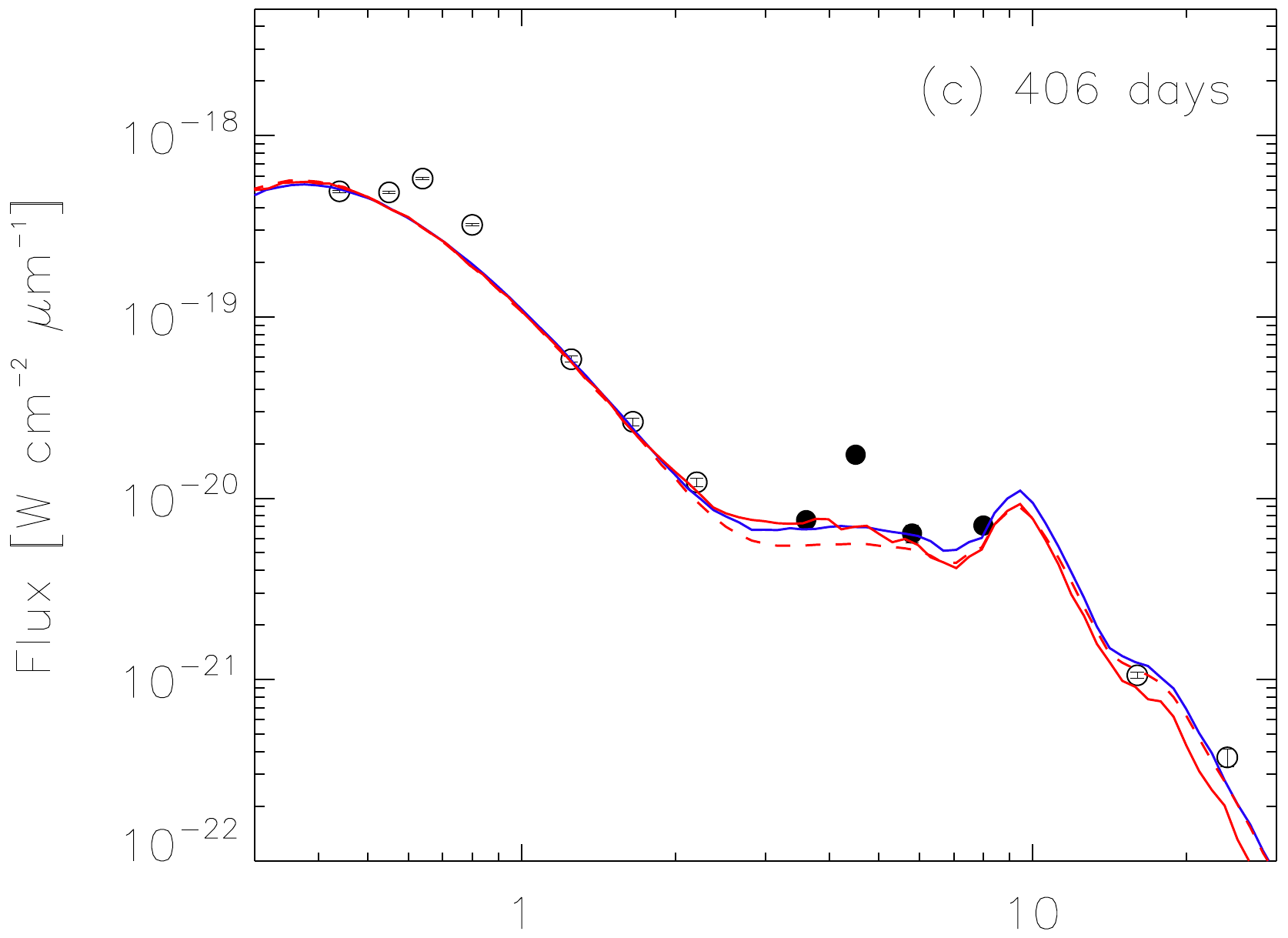}}
\subfigure{\includegraphics[scale=0.45]{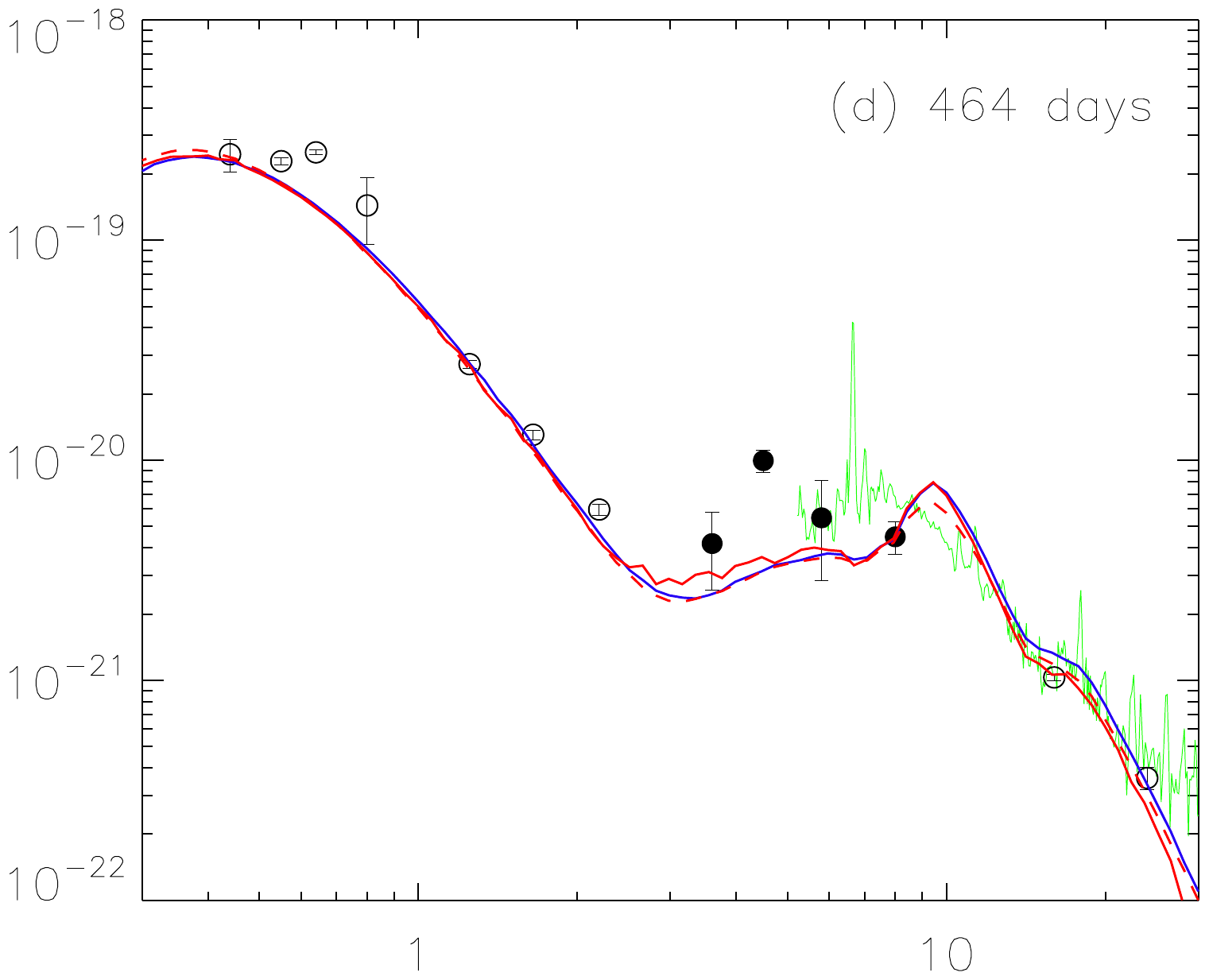}}
\subfigure{\includegraphics[scale=0.45]{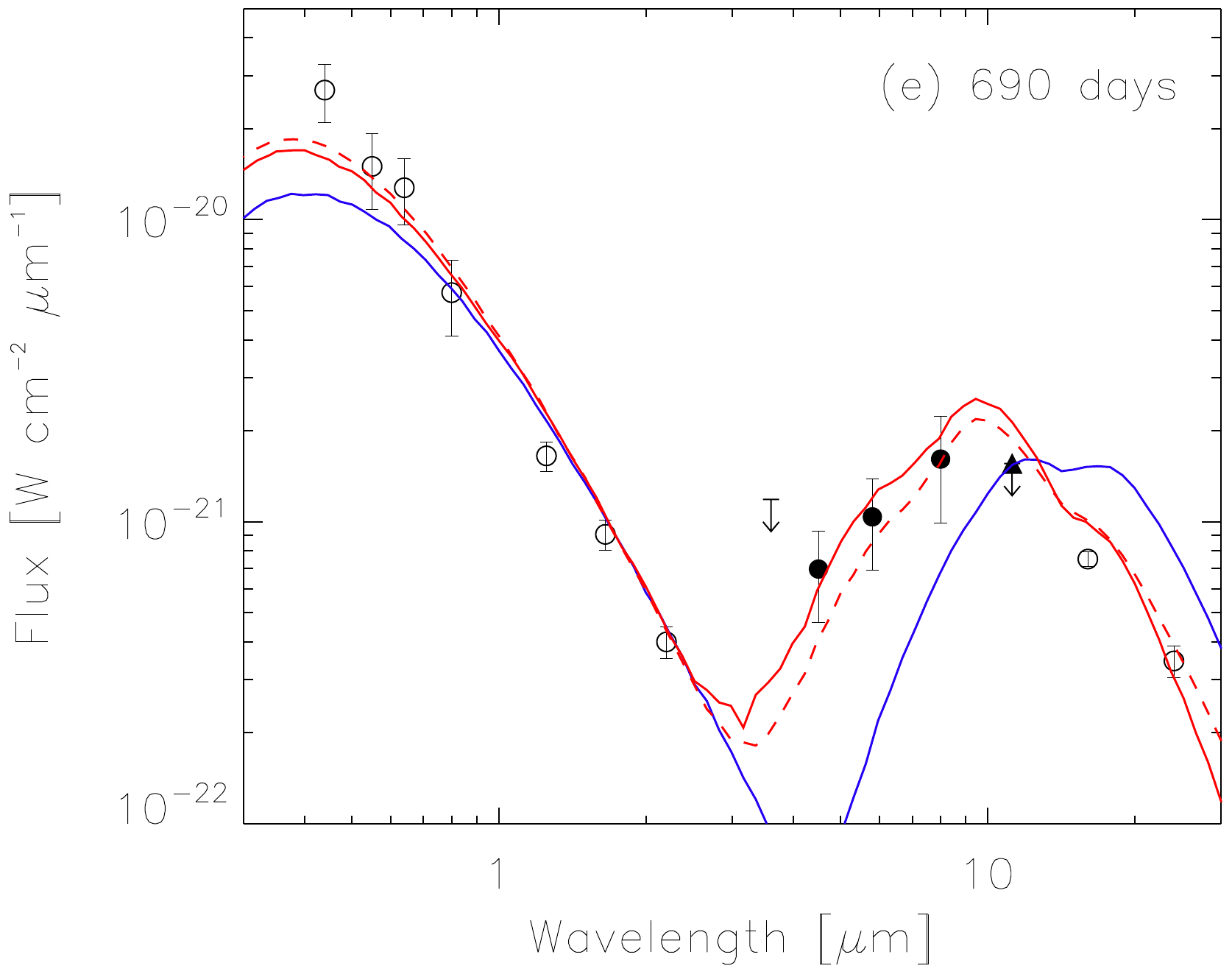}}
\caption[Monte Carlo radiative transfer dust model fits to the
optical--IR SEDs of SN~2004et between days 300 and 690]{MOCASSIN
  dust model fits to the optical--IR
  (0.3--30\,\micron) SEDs of SN~2004et between days 300 and 690.
  Filled circles indicate fluxes observed at the epoch of the IRAC
  observations, whereas open circles indicate optical, NIR and mid-IR
  photometry which has been interpolated/extrapolated to the epochs of
  the IRAC observations.  The filled triangles indicate the
  Gemini-Michelle N\primejo-band observations, and upper limits to the
  flux densities are shown as downward-pointing arrows.  \spitzer\ IRS
  spectra obtained at epochs close to the IRAC observations (on days 294, 
  348 and 480, respectively: green solid lines)
  are in good agreement with the mid-IR photometry.  Model fits to the
  data are indicated by the black solid curves (smooth dust
  distribution, fixed Y = 4.0), red solid curves (smooth dust distribution,
  varying Y) and the red dashed curves (clumpy dust distribution,
  varying Y). All models adopted a composition of 20:80 per cent
  amorphous carbon:silicates by mass.  See text for further details.
  \label{fig:04et_dust_models}}
\end{figure*}

The density distribution in SN ejecta can range from approximately 
flat, to very steep in the layers that had formed the photosphere of the 
progenitor. Here, we adopt an $r^{-2}$ law. We found that distributions 
steeper than $r^{-3}$ led to too much emission in the 1.6-3.6-$\mu$m 
spectral region at the earlier epochs, as a result of the high 
densities and high heating rates 
at the inner edge of the ejecta. As a first guess, we adopted initial
values for $R_{in}$ from our blackbody fitting, and then varied them 
to match the observations. We calculated models with shell size 
scaling parameters $Y = R_{out}/R_{in}$ = 3.0, 3.2, 3.5, 4.0, 4.5. 
The adopted heating luminosities were the total luminosities measured from 
the blackbody fitting at each epoch, as listed in the final column of 
{\tablename}~\ref{tab:04et_bbfits}. 
{\tablename}~\ref{tab:04et_RTmodels} lists the parameters
of the best-fitting models for each epoch, including
derived dust masses, $M_{d}$, and visual optical depths
$\tau_{0.55}$. The variation in derived dust masses when
parameters such as $R_{in}$, $Y$, $L$ and $T$ are 
varied from their best-fitting value allows us to
estimate that the uncertainty in the derived
dust mass is $\leq$50\%. 
{\figurename}~\ref{fig:04et_dust_models} 
plots the emergent SEDs (red solid lines) and the observations, 
where the plotted flux densities have been corrected for foreground
extinction. 

The observed SEDs were best fitted using diffuse field
radiation temperatures of 7000-8000~K, a little higher
than the best-fitting hot blackbody temperatures listed in 
{\tablename}~\ref{tab:04et_bbfits}. This might be attributable to the 
effects of internal dust extinction on the emergent optical
energy distribution in the radiative transfer models.
The best-fitting models for the different epochs were not 
homologous, in that the $R_{\rm in}$ and $R_{\rm out}$ values did not 
increase linearly with time ({\tablename}~\ref{tab:04et_RTmodels}).
Since the derived dust masses increased by a factor
of four between days 300 and 464, indicating ongoing dust formation,
the dominant dust-emitting regions could change with time.

However, we also investigated homologous models for the different epochs, 
taking the day 300 value of Y = 4.0 from 
{\tablename}~\ref{tab:04et_RTmodels}
and keeping it the same for subsequent epochs, adjusting only the total 
dust mass to obtain a best fit. These models are plotted as the back solid 
lines in {\figurename}~\ref{fig:04et_dust_models}. They produced 
slightly improved fits to the day 406 and 464 24-$\mu$m data-points,
although they failed to match the observed SED on day 690
({\figurename}~\ref{fig:04et_dust_models}). 

{\tablename}~\ref{tab:04et_RTmodels} shows that the smooth model dust 
masses increased from 0.4$\times10^{-4}$~M$_{\odot}$ on day 300 to 
1.1$\times10^{-4}$~M$_{\odot}$ on day 464 and 
4.4$\times10^{-4}$~M$_{\odot}$ on day 690. The inner and outer radii of 
the day 300 model correspond to expansion velocities of 2700 and 
10800~km~s$^{-1}$, respectively, while those for the day 406 and 464 
models correspond to expansion velocities of 2000 and 8000~km~s$^{-1}$ 
respectively. The above inner radii velocities are consistent with line 
absorption minimum velocities measured in optical spectra obtained at 
these epochs, e.g. \cite{sahu06} measured Fe~{\sc ii} absorption minimum 
expnasion velocities of $\sim$2000~km~s$^{-1}$ after day 150. Since 
absorption line optical depths scale as $\int{n}dr$, where $n$ is the 
density, then for $n \propto r^{-2}$ or steeper density distributions, 
line optical depths are strongly weighted to the inner radii, where 
velocities are lowest, as are emission lines, whose emissivities typically 
scale as $\int{4\pi r^2n^2}dr$. At the earliest epochs however, when 
ejecta densities are much higher, line optical depths of unity do not 
penetrate very deep into the outermost layers of the ejecta, where 
expansion velocities are much higher, e.g. the day 25 spectrum of 
\cite{sahu06} showed H$\alpha$ and H$\beta$ absorption minima at expansion 
velocities of $\sim$8000~km~s$^{-1}$ (their Fig.~9), consistent with the 
outer radius expansion velocities of our dust models, while H$\alpha$ 
absorption was even detectable out to $-$14,500~km~s$^{-1}$ on day 25.

\subsection{Clumpy dust models}

For the Type~II SN~2003gd, \cite{sugerman06} demonstrated that smooth dust
models could underestimate the dust mass by an order of magnitude or more
compared to models that allow for clumping. \cite{ercolano07} showed that
both smooth and clumpy dust models could fit the observed SEDs of SN~1987A
at late epochs, with clumpy models able to accommodate significantly
larger dust masses. We have constructed clumpy models for SN~2004et,
employing a similar modelling strategy to that used for SN~1987A and
SN~2003gd.

For the clumpy models, we assume that dense homogeneous clumps are
embedded in a less-dense interclump medium (ICM) with an $r^{-2}$ density
distribution. The clumps have radius $\delta$$\times$$R_{out}$ and a
volume filling factor, $f$. We adopted the same $\delta=1/30$ and $f$ =
0.01 for all epochs. The density contrast between the clumps and the
smooth ICM is defined by $\alpha$ =
$N_{clump}$($R_{in}$)/$N_{smooth}$($R_{in}$), where $N_{clump}$($R_{in}$)
and $N_{smooth}$($R_{in}$) are the densities of the clumps and the smooth
ICM at the inner radius, respectively. We set $\alpha$ = 55
for all epochs. For the other parameters, we adopted the same source
luminosity, temperature, dust composition and size distribution as used
for the smooth dust models. Similar SEDs to those from the smooth dust
models were obtained for the clumpy models (red dashed curves in
{\figurename}~\ref{fig:04et_dust_models}). Compared to the variable-Y
smooth dust models, the counterpart clumpy models produced an improved fit
to the 3.6-$\mu$m photometry on days 300-360 and to the 24-$\mu$m
photometry on days 300-406. 

Compared to the smooth dust distribution models, the clumped dust models
were able to accommodate two to five times larger dust masses without
increasing the effective dust optical depths in the visible region of the
spectrum. The smooth and clumpy dust models for day 690 both predict an
effective optical depth of 1.3 in the V-band, consistent with our estimate
from the observed light curve of 0.8-1.5 magnitudes of internal extinction
at this epoch ({\secname}~\ref{ssec:04et_opt_nir_lcurves}). The clumped
dust model for 690 listed in {\tablename}~\ref{tab:04et_RTmodels}
had a total dust mass of 1.5$\times10^{-3}$~M$_{\odot}$, but we found that
up to 4$\times10^{-3}$~M$_{\odot}$ of dust could be accommodated in clumps
at that epoch without seriously reducing the goodness of fit to the
observed SED.

\section{Discussion and conclusions}
\label{sec:04et_discuss}

We have used a range of new and archival optical and infrared 
data, extending from day 64 to day 2151, to investigate the formation 
of dust in the ejecta of the Type II-P SN~2004et, focusing in particular 
on the mid-IR observations obtained with the \spitzerst\ and Gemini-North 
telescope. 

As discussed by \citet{lucy89} for SN~1987A, there are three distinct 
signatures of
dust formation in the ejecta of core-collapse SNe. These are
(1) the appearance of asymmetric blue-shifted emission lines caused
by dust preferentially extinguishing emission from the receding
(redshifted) gas; (2) a drop in visual
brightness due to increased extinction by the newly formed dust;
accompanied by (3) a mid-infrared excess due to thermal dust emission, 
These three phenomena have all been observed
in the case of SN~2004et, as found by previous studies 
\citep{sahu06, kotak09, maguire10}. 
 
From an analysis of new and recalibrated spectra, we showed
that between days 259 and 646 the peak of the H$\alpha$ emission line 
shifted to the blue by 600 km~s$^{-1}$ (Fig.~2).
The optical light curve of SN~2004et declined more rapidly than 
that expected from radioactive deposition, 
with the onset of ejecta dust formation estimated to have
occurred \simjo\,300--400 days after explosion (Fig.~6)
The same light curves allowed us to estimate that by day 690 the 
additional extinction in the $V$-band, attributable to newly formed ejecta 
dust, was between 0.8 and 1.5 magnitudes. 
From day 300 onwards the SN~2004et SEDs exhibit clear excess 
mid-IR emission relative to blackbodies extrapolated from the optical. 
The day 300, 360 and 406 SEDs could each be fitted by two blackbody 
components: (i) a hot component attributed to 
emission from optically thick gas, and (ii) a warm component 
attributed to dust freshly synthesised and radioactively heated in the SN 
ejecta (Fig.~8). While these two-component fits provided adequate
matches to the observed SEDs out to 16~$\mu$m, they did underpredict the 
24-$\mu$m fluxes, which remained relatively constant
between days 315 and 709 (Fig.~4). The day 294, 348 and 480
{\em Spitzer} IRS spectra (Fig.~11) also confirm the presence of
relatively steady excess emission longwards of 20~$\mu$m at these 
epochs. By day 464 a cool third component was definitely required to 
fit an increasing excess seen in the 16-$\mu$m photometry.

For days 300 to 690, our ejecta dust radiative transfer modelling was able 
to match the observed SEDs out to 20~$\mu$m (Fig.~11), longwards of which 
the relatively invariant excess emission discussed above was present.
The observed rise in mid-IR fluxes after 1000 days coincided with a 
flattening of the optical and NIR light curves at around this 
time. This, coupled with the fact that the minimum luminosities estimated 
from blackbody fitting exceeded those expected from radioactive decay 
from {\em c}.\,day 690 onwards, implies that an additional emission source 
is required after that date. One possibility, discussed by 
\cite{kotak09}, is that this emission source was due to the formation
of a cool dense shell of dust as a result of ejecta-CSM interaction.
Another, discussed by Sugerman et al. (in preparation) is that this
emission was due to a light echo from pre-existing CSM dust.

For days 300, 360 and 406, the 80\% silicate, 20\% amorphous carbon dust 
masses derived from our smooth dust models (Table~8)
agree with the silicate dust masses derived by \cite{kotak09} but
our derived dust masses increase faster than their values from day 464
onwards and are a factor of 5 larger by day 690. The clumped dust models
of \cite{kotak09} had similar masses to their unclumped dust models, while
our clumped models have dust masses that are factors of 2-4 larger than
for our smooth dust models (Table~8). The dust mass derived from our day
464 clumped model was 5$\times10^{-4}$~M$_{\odot}$, increasing to
1.5$\times10^{-3}$~M$_{\odot}$ by day 690. This mass of newly formed dust
is similar to, or larger than, values derived for a number of other recent
Type~II-P SNe at similar epochs, e.g. SN~1987A \citep{wooden93,
ercolano07}, SN~2003gd \citep{sugerman06, meikle07}, SN~2004dj
\citep{szalai11, meikle11}, SN2007it \citep{andrews11} and SN2007od
\citep{andrews10}. But this mass is still a factor of 100 or more smaller 
than the CCSN ejecta dust masses estimated to be needed to account for the 
large quantities
of dust observed in some high redshift galaxies \citep{kozasa89, todini01,
nozawa03, bianchi07, dwek07, dwek11} and implies that SNe of this type
cannot make a major contribution to the dust content of galaxies, unless
their dust masses continue to grow at later epochs than have typically
been observed at mid-IR wavelengths.

\section*{Acknowledgments}

J. Fabbri and R. Wesson acknowledge STFC funding support. M. 
Otsuka and M. Meixner acknowledge funding support from NASA JWST grant 
NAG5-12595 and from HST/STScI grant GO-11229.01-A. The work of J. Andrews
and G. C. Clayton has been supported by NSF grant AST-0707691 and 
NASA GSRP grant NNX08AV36H. A portion of the data was obtained at the 
Gemini Observatory, which is operated by the Association of Universities 
for Research in Astronomy (AURA) under a cooperative agreement with the 
NSF on behalf of the Gemini partnership. Some observations were obtained 
with the NASA/ESA Hubble Space Telescope, which is operated by the 
Association of Universities for Research in Astronomy, Inc., under 
a contract with NASA.  This work is based in part on observations made 
with the Spitzer Space Telescope, which is operated by the Jet Propulsion 
Laboratory, California Institute of Technology under a contract with NASA.

\bibliographystyle{mn2e}
\bibliography{mn_sn2004et_21sep11.bib}

\appendix

\section{Photometric data processing}
\label{sec:04et_data_processing}

\subsection{Mid-IR data processing}
\label{ssec:04et_midir_reduce}

The earliest \spitzer\ data for the field around SN~2004et were obtained
by the SINGS Legacy team \citep{kennicutt03}, who produced enhanced data
products which could be downloaded from the SINGS Legacy Data Deliveries
section\footnote{http://ssc.spitzer.caltech.edu/legacy/singshistory.html}
of the SSC website. The MIPS 24\,\micron\ mosaic of NGC~6946 (providing a
pre-explosion image of the field around SN~2004et at day -75) were
downloaded from the SINGS fifth and final data delivery. The
mosaic was created from multiple \spitzer\ images obtained in scan-mapping
mode over two days (75 and 73 days prior to the explosion of SN~2004et)
and processed with the MIPS Data Analysis Tool version 3.06
\citep{gordon05}, along with additional custom processing by the SINGS
team.

However, the enhanced SINGS IRAC data were not used, due to the
unusually long time span between the two observations of NGC~6946
(\simjo~six months, compared to the usual one or two days) which were
combined by the SINGS team to construct the final enhanced mosaic.
Instead, the standard BCD pipeline data for each day (corresponding to
the pre-explosion image at day -104 and the first mid-IR
post-explosion image at day 64) were downloaded from the \spitzer\
archive and processed along with the rest of the \spitzer\
observations as described below.

The basic calibrated data (BCD) from the
\spitzer\ pipeline were combined into final mosaic images using the
\spitzer\ MOPEX software
package \citep{makovoz06}, which includes outlying-pixel rejection,
background matching, and mosaicing with drizzling to increase the
sampling of the point-spread function (PSF).  The IRAC and MIPS data
were re-drizzled to plate scales of 0\farcs75/pixel and
1\farcs5/pixel, respectively (compared to the standard \spitzer\
pipeline products with a spatial sampling of 1\farcs2/pixel for IRAC
and 2\farcs45/pixel for MIPS), following the pixel sizes adopted by
the SINGS team for producing their enhanced data products.  The PUI
data were re-drizzled to a plate scale of 1\farcs2/pixel (compared to
the standard pipeline data with a spatial sampling of 1\farcs8/pixel).
The final mosaic images for all three \spitzer\ instruments were
calibrated in surface brightness units of \mjysr\ during the BCD
pipeline stage.  

For the IRAC and MIPS 24\,\micron\ data, PSF-matched difference images,
which used the pre-explosion SINGS mosaics as the reference images,
were produced as follows. The final mosaics produced by MOPEX were
geometrically registered to a common reference frame using matching
point sources identified within the fields of view, with a 2nd-order
general fit within the IRAF {\tt geotran} task. In all cases,
registration residuals were less than 0.1 pixels RMS in both the $x$
and $y$ dimensions. Once registered, the data were PSF-matched and
differenced using the DIFIMPHOT package \citep{tomaney96} as
implemented and modified by \cite{sugerman05c}. Two approaches were
taken to PSF-matching. In the first, an empirical PSF was built for
each image using uncrowded point sources combined using the {\tt
  daophot} PSF-building tasks \citep{stetson87}. In the second, these
same tasks were run on a single theoretical PSF available for each
\spitzer\ IRAC and MIPS 24\,\micron\ image from the archive, after
that model had been rotated and scaled according to the data's
particular plate scale and position angle. Images were PSF-matched
separately using the empirical and theoretical PSFs, photometrically
scaled by the median brightness of a number of matching point sources,
and then subtracted to yield the final difference images. In general,
the difference images made using the theoretical PSFs were of higher
quality (\ie\ smaller subtraction residuals and less background noise)
since the shapes of \spitzer's PSFs have varied quite little during its
mission.  In practice, these techniques allow the reliable detection
and measurement of changes in point sources that would be considered
below the 1-$\sigma$ level in direct, undifferenced images
\citep{sugerman02}. 

For the photometry from the IRAC and MIPS difference images, uncertainties
were measured with a custom implementation of an optimal photometry code
written for the original version of DIFIMPHOT, which includes the full
noise model of the {\tt daophot}-{\tt allstar} routine as well as correct 
noise contributions from
both input and reference images and the original sky values in each image
prior to pipeline calibration. 

The Gemini Michelle data were downloaded from the Gemini Science
Archive and processed with the Gemini IRAF\,\,{\tt midir}\,\,tasks and
further cleaning procedures. The fluxes, or upper limits, in
counts measured from the final average-combined images were converted
to $F_\nu$ flux density units by multiplying by flux conversion 
factors derived from aperture photometry of standard stars.

\subsection{Optical and near-IR data processing}
\label{ssec:04et_opt_nir_reduce}

\noindent{\bf Gemini/GMOS-N} On days 317, 404, and 664, 60\,s images were
taken with GMOS-N in the $g'$, $r'$, and $i'$ broad-band filters.
Longer exposures of 2 $\times$ 600\,s in the same filters were taken
at day 1412 when the SN was expected to have faded. The images were
reduced using the Gemini IRAF package. Pipeline processed bias and
flat field images were obtained from the Gemini Science Archive.
Object images were trimmed, corrected for overscan and bias, and
flat-fielded using the {\tt gsreduce} task.  Finally, {\tt gmosaic}
was used to mosaic the three GMOS CCDs into a single image. PSF-fitted
photometry was performed on the SN and a sample of standard stars from
the photometric V, R, and I sequence of \cite{pozzo06} to
establish the nightly zeropoint. The GMOS Sloan Digital Sky Survey
(SDSS) magnitudes were transformed into Johnson $V$, $R$, and $I$
magnitudes using the transformation equations given by \cite{welch07}.
These linear transformations were derived from the photometric V, R,
and I sequence presented by \cite{pozzo06}.\vspace{0.5mm}

\noindent{\bf Gemini/NIRI} In order to ensure a consistent sky background
between exposures, individual exposures of 30\,s in each of the
broad-band $JHK$ filters were taken, with the total number being
dictated by the anticipated decline of the SN NIR light.  A
5\,\arcsec\ dither pattern was employed to ensure efficient removal
of point sources while making sky images.  Data reduction for each
night was performed using the standard NIRI routines within the Gemini
IRAF package.  {\tt nprepare} and {\tt niflat} were used to derive the
normalized flat field and the bad pixel mask while {\tt nisky} was
used to create the final sky image.  {\tt nireduce} was used to
subtract this sky image from and apply the flat field correction to
the processed object images.  Finally, the individual images in each
filter were coadded using the GEMTOOLS routine {\tt imcoadd}.
PSF-fitted photometry was performed on SN~2004et and three standard
stars present within the field.  The $JHK$ magnitudes of the standard
stars are contained within the 2MASS catalogue and were
used to derive the nightly zeropoint in each filter.  The photometric
uncertainty is dominated by the standard deviation of the zeropoint
derived from the three standard stars.\vspace{0.5mm}

\noindent{\bf Tenagra and Steward/Bok} These early optical and NIR data
were reduced, calibrated and measured in a similar manner to the
GMOS-N and NIRI observations described above, using standard routines
within IRAF. The data were bias and dark subtracted, flat-fielded, and
the multiple exposures were combined to form the final image.

\noindent{\bf {\em HST} WFPC2 and NICMOS2} For the WFPC2 observations, 
imaging with
the broadband {\em F606W} and {\em F814W} filters was used to measure
the $V$- and $I$-band flux densities. For the NICMOS2 observations,
imaging was carried out with the {\em F110W}, {\em F160W}, and {\em F205W} 
filters, most closely representing the standard $JHK$-bands.
Small-scale dithering was employed for both instruments to improve
S/N, remove cosmic rays and to improve the pixel-scale of the final
images by drizzle techniques.  The WFPC2 observations used a 4-point
dither $\times$ 400\,s exposure for each band. The NICMOS2
observations used a 5-point dither $\times$ 128\,s in the {\em F110W}
and {\em F205W} filters and a 4-point dither $\times$ 128\,s for {\em 
F160W}.

The data were reduced and calibrated using the IRAF external package
{\tt stsdas} (version 3.8) and included the removal of cosmic rays
and other artifacts, as well as linearity corrections. High-resolution images
were created using the {\tt stsdas}/{\tt drizzle} package and
additional distortion correction and alignment was performed using
background stars.  The point spread function of these reference
background stars were fit by Gaussian profiles to obtain accurate
positions, and then instrumental distortions were corrected with the
IRAF tasks {\tt xyxymatch}, {\tt geomap}, and {\tt geotran}.
The resultant pixel scale is \simjo0.02\arcsec\,pixel$^{-1}$ in the WFPC2
images and \simjo0.04\arcsec\,pixel$^{-1}$ in the NICMOS2 images.  The
FWHM of the PSFs was \simjo3 pixel. Flux measurements were performed
using the IRAF {\tt daophot} tasks.  
For the WFPC2 data, \hst\ magnitudes were converted to the
Johnson-Cousins system using the transforms of \cite{dolphin00a,dolphin09},
which include charge transfer efficiency (CTE) corrections.\vspace{0.5mm}

\begin{figure*}\centering
   \includegraphics[scale=0.55,angle=0,clip=true]
   {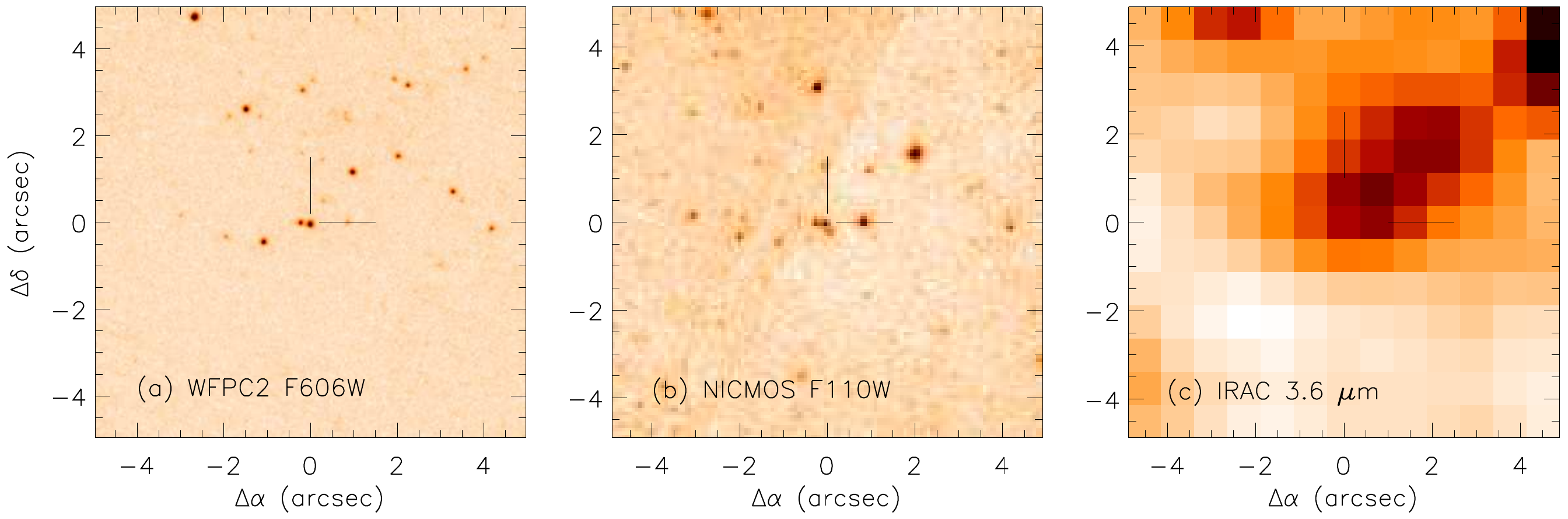} \vspace{1ex}
     {\caption[Late-time high-resolution \hst\ images
     reveal a complex field compared with \spitzer\ IRAC 
     data]{Late-time high-resolution \hst\ images reveal a complex
       field compared with \spitzer\ IRAC data. Panel (a) shows the
       WFPC2 {\em F606W} image at day 1019 with a 5\arcsec$\times$5\arcsec
       FOV centred on the SN position. A source assumed to be the SN
       is located at the centre of the image (indicated by the
       cross-hairs), with a close companion to the east (``star\,2'')
       which is detected in both WFPC2 filters. Panel (b) shows the
       equivalent field in the NICMOS {\em F110W} filter at the same
       epoch. Star\,2 is detected in all 3 NICMOS filters together
       with an additional red object just south of the SN position.
       For comparison, panel (c) shows the \spitzer\ IRAC 
       3.6-$\mu$m image at day 1054.}\label{fig:04et_hst_fovs}}
\end{figure*}

Whilst the broad band NICMOS filters, 
{\em F110W}, {\em F160W} and {\em F205W}, are roughly equivalent to the
$J$, $H$ and $K$ filters respectively, colour transformations are not
well constrained for late-time SN spectra at NIR wavelengths, and
consequently magnitudes were not converted to the standard $JHK$
photometric system.  The late-time \hst-NICMOS magnitudes listed in
{\tablename}~\ref{tab:04et_nir_mag} are for the \hst\ filters in the
Vegamag system, which uses an estimate of the flux density of Vega,
from synthetic spectra integrated over the NICMOS bandpasses, as a
photometric zeropoint. For each filter and epoch, the measured count
rate ($CR$, in units of DN\,s$^{-1}$) at the position of the SN was
converted to flux by multiplication with the $PHOTFNU$
(Jy\,s\,DN$^{-1}$) conversion factor given in the fits header, where
$PHOTFNU$ is the bandpass-averaged flux density for a source that
would produce a count rate of 1\,DN$^{-1}$.  An approximate
Vega-normalised magnitude was then calculated from the following
equation \citep[NICMOS data handbook;][]{nicmosdhb}:

\begin{displaymath}
m = ZP(Vega) - 2.5 \log_{10}(PHOTFNU \times CR \times \langle F_{\nu}(Vega) \rangle^{-1}) 
\end{displaymath}

\noindent where $\langle F_{\nu}(Vega) \rangle$ is the bandpass
averaged flux density (in Jy) for the NICMOS filters using a model
reference spectrum of Vega\footnote{Taken from the NIC2 table of
  Photometric Keywords and Vegamag Zeropoints at\\
  http://www.stsci.edu/hst/nicmos/performance/photometry/postncs\_keywords.html}
\citep{bohlin07} and $ZP(Vega)$ is the magnitude of Vega, which is
defined to be 0.00 mag under the California Institute of Technology
(CIT) infrared photometry scale.\vspace{0.5mm}

\noindent{\bf WIYN/WHIRC}
The WHIRC \citep{meixner10}
has a 2048$\times$2048 HgCdTe VIRGO detector with a pixel
scale of \simjo\,0\farcs1 pixel$^{-1}$. Sky conditions during the
observation were fair, with a seeing of \simjo\,0\farcs8. To minimise
the effects of high background levels and pixel-to-pixel variations on
the array, dithering techniques were employed whereby the source was
offset in each frame of a series of exposures.  The offset images were
used for sky level corrections. Data reduction was carried out using
standard IRAF tasks.  The array linearity correction was performed
using the WHIRC task {\tt wprep} and a distortion correction was
applied using files downloaded from the WIYN-WHIRC web
page\footnote{http://www.noao.edu/wiyn/}. A selection of 2MASS stars
close to SN~2004et were used for final flux calibration, and flux
measurements were performed using the IRAF {\tt daophot} package.

The WIYN $H$-band observation of SN~2004et on day 1803 was the last of our 
NIR observations of the SN, some 5 years after explosion. The SN was not 
clearly detected due to the contribution from neighbouring stars that were 
resolved in the high-resolution \hst-NICMOS observations (on days 1019 and 
1215; see the following section). Consequently, an upper limit to the 
magnitude was derived. The magnitude in an aperture of radius 0\farcs6 
(\simjo\,17\,pc, for the adopted distance of 5.9\,Mpc) was measured using 
the IRAF {\tt daophot} tasks. Estimated contributions from 3 neighbour 
stars resolved in the high-resolution NICMOS $F160W$ filter ($\simeq$ $H$ 
band) were measured from the NICMOS data and subtracted from the WIYN 
magnitude to provide the final upper limit presented in 
{\tablename}~\ref{tab:04et_nir_mag}.

\subsection{Late-time high resolution \hst\ images}
\label{ssec:04et_opt_nir_hst}

The late-time high resolution \hst\ images reveal that the single
point source seen at the SN position in the \spitzer\ images is
actually a complex field comprised of at least 3 sources.
{\figurename}~\ref{fig:04et_hst_fovs} shows example \hst\ WFPC2
$F606W$ ($\approx V$) and NICMOS $F110W$ ($\approx J$) images from
July 2007 (day 1019) compared to the \spitzer\ IRAC 3.6-$\mu$m
image from August 2007 (day 1054). Each field of view is centred on
the position of the SN and a source is located at this position
indicated by the cross-hairs. The star to the east (``star\,2'') was
detected in both WFPC2 filters and all three NICMOS filters at days
1019 and 1215. The third star to the south of the SN position (``star
3'') was detected in all three NICMOS bands but was not detected in
the WFPC2 filters at either epoch.

The magnitude of star\,2 was measured from the WFPC2 images and
transformed to $V$ and $I_{c}$ band magnitudes as described previously
for the SN.

The $V$ and $I_{c}$ magnitudes of star\,2 were
de-reddened using \textit{E}(\textit{B\,--\,V})~=~0.41 mag
\citep{zwitter04} and adopting the extinction law of \cite{cardelli89}
with $R_{V} = 3.1$, corresponding to $A_{V}=1.27 \pm 0.22$ mag. The
intrinsic $(V-I)_{c}$ colour was then used to estimate an $R_{c}$ band
magnitude using a table of intrinsic colours as a function of spectral
type compiled by the Space Telescope Science Institute
(STScI)\footnote{http://www.stsci.edu/~inr/intrins.html} based on the
work of \cite{fitzgerald70} and \cite{ducati01}. The derived colour
index of $(V-I)_{c}=0.8$ indicated star\,2 was probably a K1.0 star with 
an intrinsic $R_{c}$ magnitude of \simjo22.5. The equivalent reddened
$R_{c}$ magnitude of 23.53 $\pm$ 0.50 was in reasonable agreement
(within the errors) with a magnitude estimated from fitting two PSFs
to the blended `SN plus star\,2' in the $R_{c}$-band GMOS-N image at
day 1412.  This PSF-fitted measurement
gave an $R_{c}$ band magnitude for star\,2 of 23.63 $\pm$
0.19. A $B$ band magnitude for star\,2 of 25.46 $\pm$ 0.50 was
estimated in the same way.

The estimated reddened $B$ and $R_{c}$ magnitudes of star\,2, together
with the measured $V$ and $I_{c}$ magnitudes, were subtracted from the
optical magnitudes of SN~2004et obtained from the Subaru-FOCUS and
GMOS-N observations from day 646 onwards. During this time the SN had
faded substantially such that the neighbouring star\,2 made a
significant contribution to the brightness measured at those epochs.
At day 646, the estimated contribution from star\,2 was \simjo\,8\,\%
in the $V$ band and \simjo\,10\,\% in the $I_{c}$ band, whilst by day
1412, the contribution was almost half that of the total flux in both
bands.  For epochs earlier than day 500, the brightness contribution
from star 2 was $\leq$\,1\,\%.

The late-time Gemini-NIRI photometry at day 652
has not been corrected for contamination by
stars 2 and 3. As previously explained,
the NICMOS magnitudes are listed in the \hst\ Vegamag system, and were
not converted to the standard $JHK$ photometric system since colour
transformations are not well constrained for late-time SN spectra at
NIR wavelengths. It is possible that these neighbouring stars made a
small but significant (\simjo\,10\,\%) contribution to the brightness
measured in the NIRI observations at day 652, 
so the uncertainties on the day 690 interpolated flux densities in 
{\tablename}~\ref{tab:04et_nir_mag} have been increased to reflect this, 
by adding a further 10\,\% error in quadrature to the original flux 
uncertainty.

\end{document}